\input eplain

\def\toprule{\vskip1.5pt\hrule height0.8pt\vskip1.5pt}
\def\midrule{\vskip1.5pt\hrule\vskip1.5pt}
\def\bottomrule{\vskip1.5pt\hrule height0.8pt\vskip1.5pt}

\def\nomenclature[#1]#2#3{\line{{#2\hfill} \hbox to0.85\hsize {#3\hfill}}}

\def\abbreviation#1#2{\line{{#1\hfill} \hbox to0.75\hsize {#2\hfill}}}

\newcount\fignumber
\def\figdef#1{\global\advance\fignumber by 1 \definexref{#1}{\number\fignumber}{figure}\ref{#1}}
\def\figdefn#1{\global\advance\fignumber by 1 \definexref{#1}{\number\fignumber}{figure}}
\let\figref=\ref
\let\figrefn=\refn
\let\figrefs=\refs

\newcount\tabnumber
\def\tabdef#1{\global\advance\tabnumber by 1 \definexref{#1}{\number\tabnumber}{table}\ref{#1}}
\def\tabdefn#1{\global\advance\tabnumber by 1 \definexref{#1}{\number\tabnumber}{table}}
\let\tabref=\ref
\let\tabrefn=\refn
\let\tabrefs=\refs

\ifx\pdfoutput\undefined
\input epsf

\def\figscale#1#2{\epsfxsize=#2\epsfbox{#1.eps}}
\else

\def\figscale#1#2{\pdfximage width#2 {#1.pdf}\pdfrefximage\pdflastximage}
\fi

\newcount\scount \scount=0
\newcount\sscount \sscount=0

\makeatletter
\def\section#1\par{
  \vskip\z@ plus.3\vsize\penalty-250
  \vskip\z@ plus-.3\vsize\bigskip\vskip\parskip
  \global\advance\scount by1
  \sscount=0
  \writenumberedtocentry{section}#1{\the\scount}
  \definexref#1{\the\scount}{section}
  \message{#1}
  \noindent\the\scount.\quad{\bf #1}\nobreak\smallskip\noindent}
\makeatother

\def\subsection#1{
  \global\advance\sscount by1
  \smallskip
  \noindent{~~\the\scount.\the\sscount~~{\bf{#1.~}}}}

\input color
\input soul.sty                 %

\centerline{\bf{Skin-Friction and Forced Convection from Rough and Smooth Plates}}
\medskip
\centerline{Aubrey G. Jaffer}
\centerline{e-mail: agj@alum.mit.edu}

\beginsection{Abstract}

{\narrower

  Since the 1930s, theories of skin-friction drag from plates with
  rough surfaces have been based by analogy to turbulent flow in pipes
  with rough interiors.  Failure of this analogy at slow velocities
  has frustrated attempts to create a comprehensive theory.

  Utilizing the concept of a self-similar roughness which disrupts the
  boundary layer at all scales, this investigation derives formulas
  for a rough or smooth plate's skin-friction coefficient and forced
  convection heat transfer given its characteristic length,
  root-mean-squared (RMS) height-of-roughness, isotropic spatial
  period, Reynolds number, and the fluid's Prandtl number.

  This novel theory was tested with 456 heat transfer and friction
  measurements in 32 data-sets from one book, six peer-reviewed
  studies, and the present apparatus.  Compared with the present
  theory, the RMS relative error (RMSRE) values of the 32 data-sets
  span 0.75\% through 8.2\%, with only four data-sets exceeding 6\%.
  Prior work formulas have smaller RMSRE on only four of the
  data-sets.

\par}
\bigskip
{\noindent {\bf Keywords}: skin-friction; forced-convection; height-of-roughness}

\medskip
 This research did not receive any specific grant from funding
 agencies in the public, commercial, or not-for-profit sectors.

\def\Nuz{{N\!u_0}}
\def\Nuzq{{N\!u'_0}}
\def\Nu{{N\!u}}
\def\Nus{{N\!u_\sigma}}
\def\Nut{{N\!u_\tau}}
\def\Nul{{N\!u_\lambda}}
\def\Nur{{N\!u_\rho}}
\def\Nuol{{\overline{N\!u}}}
\def\Nusol{{\overline{N\!u_\sigma}}}
\def\Nufol{{\overline{N\!u_4}}}
\def\Nutol{{\overline{N\!u_\tau}}}
\def\Nuiol{{\overline{N\!u_\iota}}}
\def\NuIol{{\overline{N\!u_{I}}}}
\def\Nurol{{\overline{N\!u_\rho}}}
\def\Nulol{{\overline{N\!u_\lambda}}}
\def\Nuolq{\overline{N\!u'}}
\def\cfol{{\overline{c_f}}}
\def\cf{{c_f}}
\def\Cfol{{\overline{C_f}}}
\def\Cf{{C_f}}
\def\Cft{{C_{\tau}}}
\def\Cftol{{\overline{C_{\tau}}}}
\def\fcol{{\overline{f_c}}}
\def\fciol{{\overline{f_\iota}}}
\def\ffol{{\overline{f_4}}}
\def\fIol{{\overline{f_I}}}

\def\fcrol{{\overline{f_{\rho}}}}
\def\fctol{{\overline{f_{\tau}}}}
\def\fclol{{\overline{f_{\lambda}}}}
\def\fcsol{{\overline{f_{\sigma}}}}

\def\fcw{{f_{\omega}}}
\def\fc{{f_c}}
\def\fcs{{f_\sigma}}
\def\fct{{f_{\tau}}}
\def\fcl{{f_{\lambda}}}
\def\fcr{{f_{\rho}}}
\def\fcb{{f_{\beta}}}
\def\hol{{\overline{h}}}
\def\Pra{{Pr}}
\def\ReI{{Re_I}}
\def\ReW{{Re_W}}
\def\Rey{{Re}}
\def\Rez{{Re_0}}
\def\Rec{{Re_c}}
\def\Reu{{Re_u}}
\def\Rex{{Re_x}}
\def\Rel{{Re_\lambda}}
\def\Rev{{Re_\varepsilon}}
\def\Ret{{\Rey_\tau}}
\def\Ra{{Ra}}

\def\diff{{\rm d}}
\def\W{{\rm W}}
\def\Wz{{\rm W\!}_0}
\def\Ls{{L\!^*}}

\def\gmprm{{\gamma\prime}}
\def\etal{{et~al.~}}

\beginsection{Table of Contents}

\readtocfile

\section{Introduction}

 Fluid flowing along a wall (plate) experiences ``skin-friction drag''
 (or ``resistance'') opposing its flow.  Related to skin-friction,
 ``forced convection heat transfer''
 is the heat transfer to or from a surface induced by fluid flow along
 that surface.
 Skin-friction and forced convection are fundamental processes with
 applications from engineering to geophysics.

\unorderedlist
 \li This investigation seeks to develop formulas to predict the
 skin-friction coefficient and forced convection heat transfer from
 rough and smooth plates.
\endunorderedlist

\subsection{Pipe-Plate Analogy}
 Circa 1930, Prandtl~\cite{Prandtl1932} and
 v{on}~K{\'a}rm{\'a}n~\cite{vonKarman1932} developed theories for
 resistance along (smooth) plates from the results of research on flow
 through pipes; this is the ``pipe-plate analogy''.

 In 1934, Prandtl and Schlichting~\cite{prandtl1934resistance}
 developed a theory of skin-friction resistance for rough plates based
 on their analysis of Nikuradse's~\cite{Nikuradse33lawsof}
 measurements of sand glued inside pipes (``sand-roughness'').  The
 conclusion of the (translated) paper states:

{\narrower ``The resistance law just derived for rough plates has
  chiefly validity for a very specific type of roughness, namely a
  smooth surface to which sand grains have been densely attached and
  where the Nikuradse pipe results have been taken as the basis\dots

  A single roughness parameter (the relative roughness) will in all
  likelihood no longer answer the purpose in continued investigations
  of the roughness problem.''  \par}

 In 1936, Schlichting~\cite{Schlichting1937} investigated the velocity
 profiles and resistance of water flowing through a closed rectangular
 channel having one wall replaced in turn by a series of plates, each
 having an array of identical protrusions attached: spheres, spherical
 caps (bumps), or cones.  The protrusions were positioned on the
 plates in a hexagonal array which was elongated 15\% in the direction
 of flow.

 With significant pressure drop between inflow and outflow of the
 channel, it was not an instance of the isobaric (uniform pressure)
 flow which can occur along external plates.  The similarity of
 channel and pipe flows is well known, but neither supports nor
 refutes treating rough pipe interiors and plates analogously.

 In 1954, Hama~\cite{10010463165} described three challenges of the
 pipe-plate analogy:

{\narrower ``Now there is no obvious reason why pipe flow and
  boundary-layer flow should be identical or even similar.  First, a
  pressure gradient is essential for flow through a pipe but not along
  a plate.  Second, pipe flow is confined and perforce uniform, while
  flow along a plate develops semi-freely and bears no such a priori
  guarantee of displaying similar velocity profiles at successive
  sections.  Finally, the diameter and roughness size are the only
  geometrical dimensions of established flow in pipes, whereas at
  least three linear quantities are necessary to characterize the
  boundary-layer.''  \par}

\subsection{Boundary-Layer}
 Schlichting~\cite{schlichting2014} describes a boundary-layer:
 ``In that thin layer the velocity of the fluid increases from zero at
 the wall (no slip) to its full value which corresponds to external
 frictionless flow.''

 Hama attempted to confirm the rough pipe-plate analogy with
 measurements of wire screens affixed to smooth plates, but concluded
 that it was confirmed only in the fully rough regime (defined below).

\subsection{Sand-Roughness}
 Prior
 works~\cite{prandtl1934resistance,Nikuradse33lawsof,Schlichting1937,10010463165,schlichting2014,A014219,OSFCFRFP,white2006,CHURCHILL1993231}
 specify sand-roughness~$k_S$, the height of ``coarse and tightly
 placed roughness elements such as, for example, coarse sand grains
 glued on the surface'' (Schlichting \cite{schlichting2014}).

 Testing a machined analogue of sand-roughness circa 1975, Pimenta,
 Moffat, and Kays~\cite{A014219} stated that, while agreement with the
 Prandtl-Schlichting model was ``rather good'' in the fully rough
 regime, the apparatus's behavior differed from ``Nikuradse's
 sand-grain pipe flows in the transition region.''

 Modeling the wake component of the velocity profile, in 1985 Mills
 and Hang~\cite{OSFCFRFP} presented a formula improving the match with
 Pimenta \etal data in the rough regime; it did not address other flow
 regimes.

\subsection{Flow Regimes}
 Along with laminar flow, the theory for flow within pipes (and
 channels) distinguishes three turbulent flow regimes: smooth, (fully)
 rough, and transitional.  Smooth regime pipe flow encounters
 viscous resistance varying inversely with the fluid velocity per
 viscosity ratio.  Rough pipe flow encounters resistance varying with
 the height-of-roughness, while largely insensitive to viscosity.  The
 transitional regime describes the range of fluid velocities where both
 viscosity and roughness affect the resistance
 (Colebrook~\cite{doi:10.1680/ijoti.1939.13150}).

\subsection{Reynolds Number}
 The Reynolds number~($\Rey$) represents the bulk fluid velocity
 (far from the plate); $\Rey$ with a subscript represents other fluid
 velocities.  The local Reynolds number $\Rex=x\,\Rey/L$, where $x$ is
 the distance from the leading edge of the plate in the direction of
 flow, and characteristic length~$L$ is the length scale for the
 physical system. Local measurements are made at distance $x$
 from the leading edge.
 Unless stated otherwise, $L$~is the plate length in
 the direction of flow.

\subsection{Skin-Friction Coefficient}
 Skin-friction in prior works is represented by the
 (dimensionless)
 local drag coefficient~$\cf$ or~$\Cf$, a function of the relative
 sand-roughness~$x/k_S$ and $\Rey$ or a subscripted~$\Rey$.

 Prandtl and Schlichting~\cite{prandtl1934resistance} specified the
 boundaries between flow regimes using the ``sand-roughness Reynolds
 number'' $\Rey_k$.  For plates, it assigned the boundaries between
 smooth, transitional, and rough regimes at $\Rey_k=7.08$ and~$70.8$.
 Pimenta \etal \cite{A014219} gave $\Rey_k=65$ as the
 transitional-to-rough boundary.

  \figref{fig:Moody} presents local skin-friction coefficient curves,
  $\Rey_k$ regime boundaries, and measurements from the
  Pimenta \etal \cite{A014219} plate having $k_S=0.794{\rm\,mm}$ at
  $x=0.965{\rm\,m}$.
  \figref{fig:Moody-a} presents the (whole-plate) average
  skin-friction coefficients with $L=2.18$~m.

\vbox{\settabs 2\columns
\+\hfil\figscale{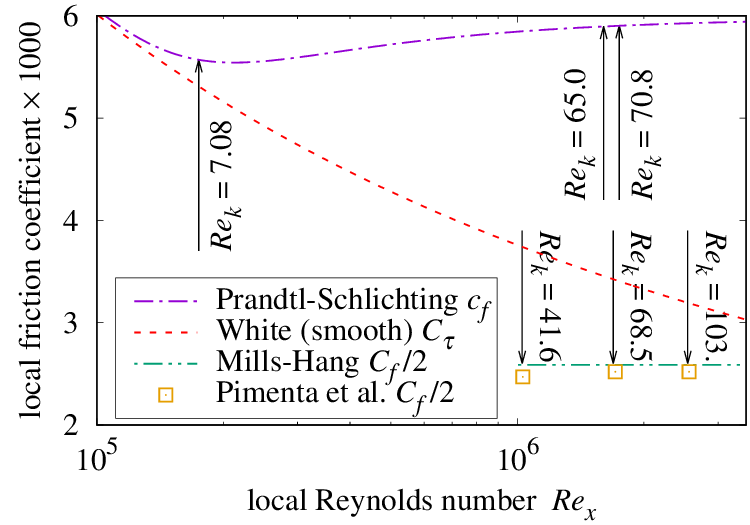}{234pt}&\hfil\figscale{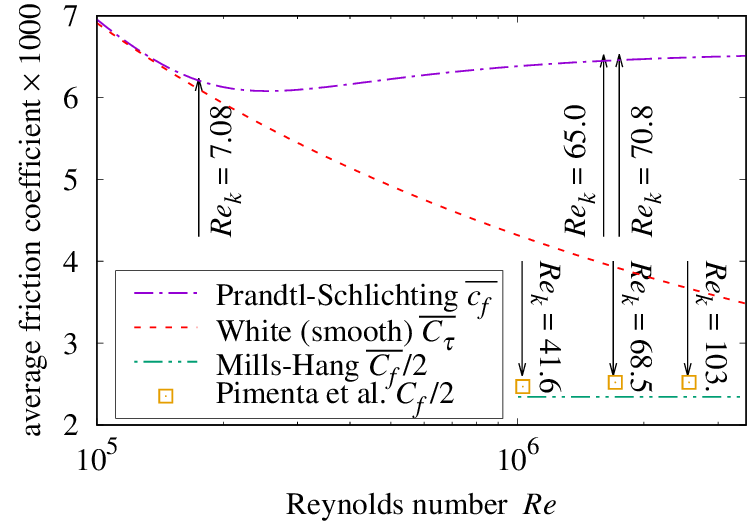}{234pt}&\cr
\+\hfil{\bf\figdef{fig:Moody}\quad Local $\cf$ versus $\Rex$}&
  \hfil{\bf\figdef{fig:Moody-a}\quad Average $\cfol$ versus $\Rey$}&
  \cr}

\subsection{Plate Flow Versus Pipe Flow}
 Schlichting~\cite{schlichting2014} states: ``The resistance to flow
 offered by rough walls [of pipes] is larger than that \dots for
 smooth pipes.''

 The rough pipe-plate analogy holds that this rule also applies to
 rough, external plates.  For example, the
 ``Prandtl-Schlichting~$\cf$'' curve is never less than the ``White
 (smooth)~$\Cft$'' curve in \figref{fig:Moody}.

 The Pimenta \etal measurements are much closer to~$\cf/2$ than they
 are to~$\cf$.\numberedfootnote{The Bergstrom, Akinlade, and
 Tachie~\cite{Bergstrom_2005} local friction coefficient ($\Cf$)
 measurements of woven wire meshes and perforated sheets (present
 work \ref{Rough Skin-Friction Measurements}) are
 also much closer to~$\cf/2$ than~$\cf$.}
 Pimenta \etal \cite{A014219} and Mills and Hang~\cite{OSFCFRFP} both
 designated~$\cf/2$ as the friction coefficient.
 All three measurements in \figref{fig:Moody} are less than the smooth
 regime coefficient~$\Cft$.  If rough friction is never less than
 smooth friction, then these measurements must not be in any turbulent
 flow regime; the remaining alternative is laminar flow.  Laminar flow
 coefficients have a steeper slope than~$\Cft$; yet these measurements
 are near the constant level predicted by Mills-Hang~$\Cf/2$ for the
 rough regime.

\unorderedlist

 \li The pipe-plate analogy fails for roughness because rough
  skin-friction coefficients can be less than
  smooth regime coefficients for external plates, but not inside
  pipes.

\endunorderedlist 

\subsection{More Recent Work}
 With the rough pipe-plate analogy's failure obscured by the factor
 of~2, research based on the pipe-plate analogy continued.  The 2004
 survey article Jim\'enez~\cite{Jimenez_ARFM04} did not question the
 rough pipe-plate analogy, writing: ``The theoretical arguments are
 sound, but the experimental evidence is inconclusive.''

 Circa 2005, Bergstrom, Akinlade, and Tachie~\cite{Bergstrom_2005},
 performed experiments with sandpapers, woven wire meshes, and
 perforated sheets attached onto a flat plate, reporting that:
 $\sqrt{\Cf}=[0.360\pm0.025]\,{\delta^*/\delta}$,
 where $\delta^*$ is the displacement thickness and $\delta$ is the
 99\% velocity boundary-layer thickness.  As a function of a roughness
 metric, this formula has no predictive value because both $\delta^*$
 and $\delta$ must be inferred from velocity measurements along
 the surface under test.  Fortunately, Bergstrom \etal included
 free-stream velocity in their tables, allowing comparisons of their
 skin-friction data with the present theory.

 The 2021 survey article Chung, Hutchins, Schultz, and
 Flack~\cite{doi:10.1146/annurev-fluid-062520-115127} summarizes
 studies for predicting the drag, (boundary layer) velocity profiles,
 or convection from rough walls or pipes in terms of turbulence
 theory.  Several roughness metrics were introduced, however none
 convert to the ``isotropic spatial period'' introduced
 in \ref{Periodic Roughness} of the present work.

 The studies cited by~\cite{doi:10.1146/annurev-fluid-062520-115127}
 generally rely on turbulence theory, sand-roughness, Prandtl and
 Schlichting~\cite{prandtl1934resistance}, and the pipe-plate analogy.
 The present theory relies on none of these.

\unorderedlist

 \li Sand-roughness's lack of generality
 and the failure of the pipe-plate analogy for roughness motivate a
 fresh theoretical analysis of isobaric flow along a rough plate, an
 analysis derived from traceable roughness metrics.

 \li Prior works analyze turbulence in the boundary-layer.  The
 central premise of this investigation is that plate roughness
 disrupts its boundary-layer.  The present theory does not utilize
 turbulence theory.

 \li The present theory is about plates; it has no implications for
 smooth or rough pipe flow.
\endunorderedlist

\subsection{Approach}
 Flow along flat, smooth plates can be laminar or turbulent with a
 continuous boundary-layer.  This investigation uses the term
 ``rough flow'' for disrupted boundary-layer flow from a rough plate.

 \figrefs{fig:Moody} and \figrefn{fig:Moody-a} show that the
 skin-friction from smooth and rough plates
 are substantially different.  Roughness disrupts what would otherwise
 be a viscous sub-layer adjacent to the plate.  Lienhard and
 Lienhard~\cite{ahtt5e} teaches: ``Even a small wall roughness can
 disrupt this thin sublayer, causing a large decrease in the thermal
 resistance (but also a large increase in the wall shear stress).''

 With a sufficiently large roughness, the nascent boundary layers forming
 after each disruption will be smaller than the roughness.  The
 momentum can thus transfer directly between the fluid flow and the
 plate's roughness.  In order for these transfers to constitute most
 of the skin-friction, there must be no significant pressure gradients
 acting on the plate.  Thus, this approach works only with a thin
 plate parallel to an isobaric fluid flow.  It is not applicable to
 pipes, for example, because the pressure decreases as fluid flows
 through a cylindrical pipe.

 Shearing stress can be determined from the free-stream velocity,
 ignoring turbulent velocity perturbations.  Thus, the
 skin-friction can be derived without turbulence theory.

\unorderedlist
 \li While understanding the nature of the flow shed by roughness is
 of theoretical interest, it is not needed for determining the
 skin-friction coefficient from a very rough surface in an isobaric
 flow.

\endunorderedlist

\subsection{Not Empirical}
  Empirical theories derive their coefficients from measurements,
  inheriting the uncertainties from those measurements.  Theories
  developed from first principles derive their coefficients
  mathematically.  For example, Incropera, DeWitt, Bergman, and
  Lavine~\cite{bergman2007fundamentals} gives the thermal conductance
  of one face of a diameter $D$ disk into a stationary, uniform medium
  having thermal conductivity $k$ (${\rm W/(m\cdot K)}$) as
  $8\,k/[\pi\,D]$~${\rm W/(m^2\cdot K)}$.
  The present theory derives from first principles; it is not
  empirical.  Each formula is tied to aspects of the plate geometry,
  fluid, and flow.

\subsection{Mathematics}
  Familiarity with calculus is assumed.  Computational geometry,
  probability, self-similar recurrences, the Lambert $\W$ function,
  vector-space norms, and Fourier transforms are also employed; each
  is briefly introduced or illustrated graphically.
  Differential equations are not explicitly used.

\subsection{Overview}
 After a comparison of roughness metrics in \ref{Roughness Metrics}, a
 thought experiment about flow along self-similar roughness is solved
 using computational geometry in \ref{Flow Over Obstacles}.  The
 resulting skin-friction coefficient formula's range of validity does
 not extend to 0 height-of-roughness.  However, analyzing the case of
 a roughness which induces skin-friction midway between that of a
 self-similar rough surface and that of a smooth surface
 in \ref{Turbulent Friction} produces a formula with unprecedented
 accuracy for smooth plates.

 The self-similar roughness was designed to produce the most
 boundary-layer disruption possible within a given RMS
 height-of-roughness.  A self-similar roughness could be fabricated,
 but measuring its skin-friction could confirm predictions only for
 self-similar roughness.
 Instead, \refs{Spectral Roughness} and \refn{Periodic Roughness} turn
 to isotropic, periodic roughness, which applies in many practical use
 cases.  Fourier analysis and computational geometry result in a
 quantitative test for periodic isotropy which computes the isotropic
 period $L_P$ from a sufficiently large height map of
 roughness.  \ref{Periodic Roughness} computationally tests arrays of
 posts and wells to find the upper/lower area ratio spans which
 qualify as isotropic, periodic roughness.

 \ref{Local Skin-Friction Coefficients} presents formulas for
 converting between local and average friction coefficients, and tests
 them on measurements from Pimenta \etal \cite{A014219},
 Churchill~\cite{CHURCHILL1993231}, and
 \v{Z}ukauskas and \v{S}lan\v{c}iauskas~\cite{Zukauskas1987}.

 \ref{Forced Convection} derives formulas for forced convection heat
 transfer from rough and smooth plates.

 Periodicity enables the derivation of formulas for the onset of rough
  flow from an isotropic, periodic rough surface
 in \ref{Onset of Rough Flow}.

 \ref{Plateau Roughness} finds that the
 skin-friction coefficient of isotropic, periodic roughness is the
 same as for self-similar roughness, except when the periodic
 roughness peaks are all co-planar (at the same elevation) plateaus.
 Plateau roughness is an isotropic, periodic roughness with most of
 its area at its peak elevation.  \ref{Plateau Roughness}
 derives a quantitative test for plateau roughness from a height map
 of roughness.  It also develops formulas for the $\Rex$ thresholds
 separating rough flow and turbulent flow along the
 plate.  These thresholds are tested in
 \refs{Rough Skin-Friction Measurements}
 and \refn{Rough Heat Transfer Measurements}.

 \ref{Combining Transfer Processes} derives formulas for skin-friction
 coefficients and convection heat transfer from a plateau roughness
 shedding rough and turbulent flow from regions of the plate
 separated at the $\Rex$ thresholds developed
 in \ref{Plateau Roughness}

 \ref{Rough Skin-Friction Measurements} tests the
 present theory on measurements from
 Bergstrom \etal \cite{Bergstrom_2005}.

 \ref{Rough Heat Transfer Measurements} tests the present theory on
 measurements from the present apparatus.

 \ref{Smoothness} analyzes fluid flow along isotropic,
 periodic roughness at $\Rey$ smaller than the rough flow
 thresholds found in \ref{Onset of Rough Flow}.  This flow
 strongly resembles the ``oscillations [\dots] discovered in the
 laminar boundary layer along a flat plate'' by Schubauer and
 Skramstad~\cite{Schubauer1948}.  The present work treats them
 identically.  Although measurements of flows over such small
 roughnesses are lacking, smooth plate skin-friction measurements from
 Gebers~\cite{Gebers1908,Gebers1920} and uniform-wall-temperature
 convection measurements by \v{Z}ukauskas
 and \v{S}lan\v{c}iauskas~\cite{Zukauskas1987} support the present
 theory.

\section{Data-Sets and Evaluation}

 \tabrefs{tab:friction-sources} and~\tabrefn{tab:convection-sources}
 list the data-sets to be compared with the present theory.

 Rough surface convection measurements were obtained from an apparatus
 built for this investigation, which measured (whole-plate) average
 convection heat transfer in air at $2300<\Rey<93000$.
 \ref{Appendix A: Apparatus and Measurement Methodology}
 describes this apparatus and its measurement methodology.
 \ref{Rough Heat Transfer Measurements} presents its measurements.

  The Gebers~\cite{Gebers1908,Gebers1920} skin-friction measurements
  were captured from a graph in Schlichting \cite{schlichting2014}
  by measuring the distance from each point to the graph's axes, then
  scaling to the graph's units using the ``Engauge'' software (version 12.1).
  The remaining measurement data-sets were manually entered from
  tables in the cited works.  Several obvious single-digit
  typographical errors were corrected.

  Two non-obvious single digit errors in the text of a prior work are
  detailed in \ref{Rough Skin-Friction Measurements}.

\subsection{RMS Relative Error}
 
 The ``$\pm$'' column of \tabrefs{tab:friction-sources}
 and \tabrefn{tab:convection-sources} lists the estimated measurement
 uncertainties stated by the cited studies.  While essential to
 empirical theories, these estimates are only indicative for
 non-empirical theories.

 Root-mean-squared relative error (RMSRE) provides an objective,
 quantitative evaluation.  It gauges the fit of measurements
 $g(\Rey_j)$ to function $f(\Rey_j)$, giving each of the $n$ samples
 equal weight in Formula~\eqref{eq:RMSRE}.
$$\eqalignno{{\rm RMSRE}=&\sqrt{{1\over n}\sum_{j=1}^n\left|{g(\Rey_j)\over f(\Rey_j)}-1\right|^2}&\eqdef{eq:RMSRE}\cr
                            {\rm bias}={1\over n}\sum_{j=1}^n\left\{{g(\Rey_j)\over f(\Rey_j)}-1\right\}&\qquad
  {\rm scatter}=\sqrt{{1\over n}\sum_{j=1}^n\left|{g(\Rey_j)\over f(\Rey_j)}-1-{\rm bias}\right|^2}&
  \eqdef{eq:bias}\cr}$$

 Along with presenting RMSRE, charts in the present work split RMSRE
 into the bias and scatter components defined in
 Formula~\eqref{eq:bias}.  The root-sum-squared (RSS) of bias and
 scatter is RMSRE.

\vfill\eject

\smallskip
\centerline{\bf\tabdef{tab:friction-sources}\quad{Sources of friction measurements}}
\vbox{\settabs 10\columns
\toprule
\+ \bf Source &&&{\bf Measuring} &&&\hfil $\Pra_\infty$ & ~~$\Rey\ge$ & ~~$\Rey\le$ &$\pm$\hfill\bf Count&\cr
\midrule
\+ \cite{CHURCHILL1993231,smith_walker_1959,spalding_chi_1964} Churchill &&& turbulent average, air &&&\hfil $0.71$ & $1.0\times10^5$ & $1.0\times10^9$ &\hfill 9&\cr
\+ \cite{CHURCHILL1993231,smith_walker_1959,spalding_chi_1964} Churchill &&& turbulent local, air &&&\hfil $0.71$ & $1.0\times10^5$ & $1.0\times10^{10}$ &\hfill 11&\cr
\+ \cite{Zukauskas1987} \v{Z}ukauskas \& \v{S}lan\v{c}iauskas &&& turbulent local, oil   &&&\hfil 55.2 & $3.6\times10^5$ & $1.1\times10^7$ &\hfill 5&\cr
\+ \cite{Zukauskas1987} \v{Z}ukauskas \& \v{S}lan\v{c}iauskas &&& turbulent local, water &&&\hfil 5.42 & $3.6\times10^5$ & $2.4\times10^6$ &\hfill 8&\cr
\+ \cite{Zukauskas1987} \v{Z}ukauskas \& \v{S}lan\v{c}iauskas &&& turbulent local, water &&&\hfil 2.78 & $7.2\times10^5$ & $4.4\times10^6$ &\hfill 8&\cr
\+ \cite{Zukauskas1987} \v{Z}ukauskas \& \v{S}lan\v{c}iauskas &&& turbulent local, air   &&&\hfil 0.71 & $7.6\times10^5$ & $3.2\times10^6$ &\hfill 9&\cr
\+ \cite{Gebers1908,Gebers1920} Gebers &&& transition average &&&\hfil & $7.4\times10^5$ & $3.3\times10^7$ &\hfill 33&\cr
\+ \cite{A014219} Pimenta \etal &&& packed sphere rough local, air &&&\hfil $0.71$ & $3.8\times10^5$ & $5.8\times10^6$ &10\%\hfill 19&\cr
\+ \cite{Bergstrom_2005} Bergstrom \etal &&& turbulent local, air &&&\hfil $0.71$ & $1.6\times10^6$ & $4.6\times10^6$ &~5\%\hfill 4&\cr
\+ \cite{Bergstrom_2005} Bergstrom \etal &&& wire mesh rough local, air &&&\hfil $0.71$ & $1.6\times10^6$ & $4.7\times10^6$ &~9\%\hfill 12&\cr
\+ \cite{Bergstrom_2005} Bergstrom \etal &&& perforated sheet local, air &&&\hfil $0.71$ & $1.6\times10^6$ & $4.6\times10^6$ &~9\%\hfill 12&\cr
\bottomrule
}
\noindent Note: Churchill~\cite{CHURCHILL1993231} extracted its
measurements from Smith and Walker~\cite{smith_walker_1959} and
Spalding and Chi~\cite{spalding_chi_1964}.

\smallskip
\centerline{\bf\tabdef{tab:convection-sources}\quad{Sources of convection measurements}}
\vbox{\settabs 10\columns
\toprule
\+ \bf Source &&&{\bf Measuring} &&&\hfil $\Pra_\infty$ & ~~$\Rey\ge$ & ~~$\Rey\le$ &$\pm$\hfill\bf Count&\cr
\midrule
\+ \cite{KESTIN1961133} Kestin \etal &&& UWT transition local, air &&&\hfil 0.7 & $3.8\times10^4$ & $3.5\times10^5$ &\hfill 13&\cr
\+ \cite{KESTIN1961133} Kestin \etal &&& UWT transition local, air &&&\hfil 0.7 & $4.3\times10^4$ & $2.9\times10^5$ &\hfill ~7&\cr
\+ \cite{Reynolds1958IV} Reynolds \etal &&& UWT transition local, air &&&\hfil 0.71 & $8.2\times10^4$ & $1.1\times10^6$ &~4\%\hfill 22&\cr
\+ \cite{Zukauskas1987} \v{Z}ukauskas \& \v{S}lan\v{c}iauskas &&& UHF transition local, air &&&\hfil 0.71 & $1.1\times10^4$ & $8.2\times10^5$ &~5\%\hfill 10&\cr
\+ \cite{Zukauskas1987} \v{Z}ukauskas \& \v{S}lan\v{c}iauskas &&& UHF transition local, air &&&\hfil 0.71 & $1.1\times10^4$ & $8.2\times10^5$ &~5\%\hfill 10&\cr
\+ \cite{Zukauskas1987} \v{Z}ukauskas \& \v{S}lan\v{c}iauskas &&& UHF transition local, air &&&\hfil 0.71 & $1.1\times10^4$ & $8.2\times10^5$ &~5\%\hfill 10&\cr
\+ \cite{Zukauskas1987} \v{Z}ukauskas \& \v{S}lan\v{c}iauskas &&& UHF transition local, water &&&\hfil 6.57 & $4.0\times10^3$ & $2.2\times10^5$ &10\%\hfill 19&\cr
\+ \cite{Zukauskas1987} \v{Z}ukauskas \& \v{S}lan\v{c}iauskas &&& UHF transition local, water &&&\hfil 6.57 & $5.0\times10^3$ & $2.2\times10^5$ &10\%\hfill 15&\cr
\+ \cite{Zukauskas1987} \v{Z}ukauskas \& \v{S}lan\v{c}iauskas &&& UHF transition local, oil &&&\hfil 108. & $3.0\times10^4$ & $3.0\times10^5$ &~5\%\hfill 17&\cr
\+ \cite{Zukauskas1987} \v{Z}ukauskas \& \v{S}lan\v{c}iauskas &&& UHF transition local, oil &&&\hfil 257. & $1.2\times10^4$ & $1.1\times10^5$ &~5\%\hfill 17&\cr
\+ \cite{Zukauskas1987} \v{Z}ukauskas \& \v{S}lan\v{c}iauskas &&& UWT average, air &&&\hfil 0.71 & $1.1\times10^5$ & $6.3\times10^5$ &~5\%\hfill16&\cr
\+ \cite{Zukauskas1987} \v{Z}ukauskas \& \v{S}lan\v{c}iauskas &&& UWT average, air &&&\hfil 0.71 & $1.7\times10^5$ & $7.5\times10^5$ &~5\%\hfill19&\cr
\+ \cite{Zukauskas1987} \v{Z}ukauskas \& \v{S}lan\v{c}iauskas &&& UWT average, water &&&\hfil 5.8-7.1 & $1.4\times10^6$ & $2.3\times10^6$ &10\%\hfill 5&\cr
\+ \cite{Zukauskas1987} \v{Z}ukauskas \& \v{S}lan\v{c}iauskas &&& UWT average, water &&&\hfil 2.9-7.2 & $2.1\times10^5$ & $6.4\times10^6$ &10\%\hfill 21&\cr
\+ \cite{Zukauskas1987} \v{Z}ukauskas \& \v{S}lan\v{c}iauskas &&& UWT average, water &&&\hfil 2.0-5.8 & $1.8\times10^5$ & $1.4\times10^6$ &10\%\hfill 40&\cr
\+ \cite{Zukauskas1987} \v{Z}ukauskas \& \v{S}lan\v{c}iauskas &&& UWT average, oil &&&\hfil 75-246 & $5.0\times10^4$ & $7.0\times10^5$  &~5\%\hfill 40&\cr
\+ \cite{Zukauskas1987} \v{Z}ukauskas \& \v{S}lan\v{c}iauskas &&& UWT average, oil &&&\hfil 80-205 & $1.1\times10^5$ & $3.6\times10^5$  &~5\%\hfill 11&\cr
\+ \cite{Zukauskas1987} \v{Z}ukauskas \& \v{S}lan\v{c}iauskas &&& UWT average, oil &&&\hfil 92-317 & $2.7\times10^4$ & $7.5\times10^5$  &~5\%\hfill 29&\cr
\+ present apparatus $\varepsilon=3.00$~mm &&& UWT rough average, air &&&\hfil 0.71 & $2.3\times10^3$ & $9.3\times10^4$ &3-7\%\hfill 13&\cr
\+ present apparatus $\varepsilon=1.04$~mm &&& UWT plateau average, air &&&\hfil 0.71 & $2.0\times10^3$ & $6.8\times10^4$ &2-6\%\hfill 14&\cr
\bottomrule
}
\noindent Note: UHF is uniform heat flux; UWT is uniform wall temperature.

\section{Roughness Metrics}

 Two established, traceable roughness metrics are the
 root-mean-squared (RMS) height-of-roughness and the arithmetic-mean
 height-of-roughness.  For an elevation function~$z(x,y)$ defined on
 area~$A$ having a convex perimeter, its mean elevation~$\overline{z}$
 and RMS height-of-roughness~$\varepsilon$ are:
$$\eqalignno{\overline{z}&={\left.\int_A z\,\diff{A}\right/\int_A\diff{A}}&\eqdef{eq:mean-surface-height}\cr
  \varepsilon&=\sqrt{\left.\int_A |z-\overline{z}|^2\,\diff{A}\right/\int_A\diff{A}}&\eqdef{eq:RMS-surface}\cr}$$

 The arithmetic-mean height-of-roughness is defined in terms of the
 same mean elevation~$\overline{z}$ Formula~\eqref{eq:mean-surface-height}:
$${\left.\int_A |z-\overline{z}|\,\diff{A}\right/\int_A\diff{A}}\eqdef{eq:arithmetic-mean}$$

\vfill\eject

\subsection{Sand-Roughness}
Modeling sand-roughness grains as diameter $k_S$ spheres sitting in a
pool of depth~${g}$ glue, the mean elevation~$\overline{z}$ of a cell
of area $A$ containing one sphere is:
$$\overline{z}
  ={g}+{5\,\pi\,k_S^3\over24\,A}-{\pi\,k_S^2\,{g}\over4\,A}\eqdef{eq:grit-z}$$

 With the cell's RMS height-of-roughness~$\varepsilon$ computed from
 Formula~\eqref{eq:RMS-surface},
 \figref{fig:glue-height} shows $k_S/\varepsilon$ versus the ratio of cell
 area to the sphere's shadow area, at six glue-levels between 0\% and
 50\% of~$k_S$.
 \tabref{tab:Nikuradse} lists $k_S$, grain densities, and
 $k_S/\varepsilon$ conversion factors for
 Nikuradse's~\cite{Nikuradse33lawsof} sand coatings, assuming
 $g=0.5\,k_S$.

\medskip
\hbox{
 \vbox{\settabs 2\columns
 \+\hfil\figscale{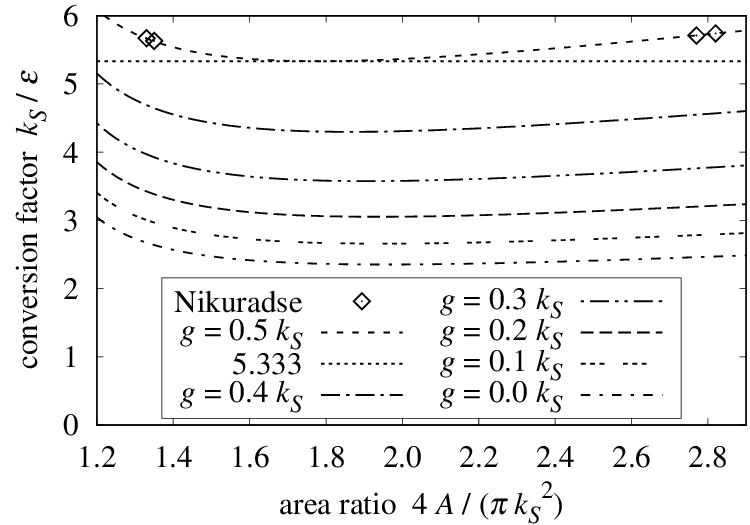}{234pt}&\cr
 \+\hfil{\bf\figdef{fig:glue-height}\quad $k_S/\varepsilon$ versus cell area of sand-roughness}&\cr
 }
\vbox{\settabs 6\columns
\+\hfil{\bf\tabdef{tab:Nikuradse}\quad Nikuradse's sand coatings at $g=0.5\,k_S$}&\cr
\toprule
\+\hfil$k_S$&\hfil grains$/{\rm cm}^2$&\hfil$k_S/\varepsilon$&\cr %
\midrule
\+\hfil.08~cm&\hfil150 &\hfil5.67&\cr %
\+\hfil.04~cm&\hfil590 &\hfil5.63&\cr %
\+\hfil.02~cm&\hfil1130&\hfil5.74&\cr %
\+\hfil.01~cm&\hfil4600&\hfil5.71&\cr %
\bottomrule
\vskip 76pt
}
}

\subsection{Conversions}
 Afzal, Seena, and Bushra~\cite{Afzal2013} fitted 5.333 as the RMS to
 sand-roughness conversion factor $k_S/\varepsilon$, and 6.45 as the
 arithmetic-mean to sand-roughness conversion factor (both in pipes).

 $k_S/\varepsilon=5.333$ is a broad minimum of the $g=0.5\,k_S$ curve
 in \figref{fig:glue-height}.

 The ``$k_S/\varepsilon$'' column values in \tabref{tab:Nikuradse}
 (``Nikuradse'' in \figref{fig:glue-height}) match each other
 within~2\%.
 The tightest spread on \tabref{tab:Nikuradse} data with the
 arithmetic-mean height-of-roughness exceeds 20\%.  Thus,
 sand-roughness correlates an order of magnitude more strongly with
 RMS than arithmetic-mean height-of-roughness.

 Flack, Schultz, Barros, and Kim~\cite{flack2016skin} measured
 skin-friction from grit-blasted surfaces in a duct, writing ``The
 root-mean-square roughness height is shown to be most strongly
 correlated with the equivalent sand-roughness height ($k_S$) for the
 grit-blasted surfaces.''

\unorderedlist
 \li Arithmetic-mean height-of-roughness will not be considered further by
 this investigation.
\endunorderedlist

\subsection{Packed Spheres Roughness}
 The Pimenta \etal \cite{A014219} plate was composed of 11 layers of
 closely packed 1.27~mm diameter metal spheres ``arranged such that
 the surface has a regular array of hemispherical roughness
 elements.''  Joined by brazing, there was no pool of glue surrounding
 the spheres.  Shrinking the cell to the sphere's shadow,
 $\pi\,k_S^2/4$, the RMS height-of-roughness of the top half of the
 1.27~mm sphere is 0.150~mm.  Pimenta \etal gave $k_S=0.794{\rm\,mm}$;
 $k_S/5.333\approx0.149{\rm\,mm}$, which matches 0.150~mm within~1\%.

\unorderedlist
 \li This investigation will use $k_S/\varepsilon=5.333$ as the RMS to
 sand-roughness conversion factor.
\endunorderedlist

\section{Formulas From Prior Works}

 Several prior works gave formulas for skin-friction coefficient in
 the fully rough regime.

\subsection{Prandtl and Schlichting}
 In {\it Boundary-layer theory}~\cite{schlichting2014}, Prandtl and
 Schlichting gave formulas for fully rough local ($\cf$) and plate
 average ($\cfol$) skin-friction coefficient for a rough plate as a
 function of~$x/k_S$ and~$L/k_S$, respectively:
$$\eqalignno{
 \cf&=\left[2.87+1.58\,\log_{10}{x\over k_S}\right]^{-2.5}\qquad x\le L&\eqdef{eq:c'_f}\cr
 \cfol&=\left[1.89+1.62\,\log_{10}{L\over k_S}\right]^{-2.5}\qquad10^2<{L\over k_S}<10^6&\eqdef{eq:c_f}\cr
 }$$

\subsection{Mills and Hang}
 Mills and Hang~\cite{OSFCFRFP} gave a Formula~\eqref{eq:C_f} which is
 more accurate than Formula~\eqref{eq:c'_f} on the local skin-friction
 measurements from Pimenta \etal \cite{A014219}.  Their local ($\Cf$)
 and average ($\Cfol$) coefficient formulas were:
$$\eqalignno{
 \Cf&=\left[3.476+0.707\,\ln{x\over k_S}\right]^{-2.46}
   \qquad750<{x\over k_S}<2750&\eqdef{eq:C_f}\cr
 \Cfol&=\left[2.635+0.618\,\ln{L\over k_S}\right]^{-2.57}&\eqdef{eq:C_D}\cr
 }$$

\subsection{White}
 White~\cite{white2006} gave Formula~\eqref{eq:White} for fully rough
 local skin-friction coefficient:
$$\Cf=\left[1.4+3.7\,\log_{10}{x\over k_S}\right]^{-2}
 \qquad{x\over k_S}>{\Rex\over1000}\eqdef{eq:White}$$

 White is also the source of widely used formulas for turbulent
 skin-friction coefficients of a smooth plate:
$$\Cft(\Rex)={0.455\over\ln^2(0.06\,\Rex)}
  \qquad\Cftol(\Rex)={0.523\over\ln^2(0.06\,\Rex)}
  \eqdef{eq:White-smooth}$$

\subsection{Average Coefficient}
 Mills and Hang~\cite{OSFCFRFP} derived the average
 Formula~\eqref{eq:C_D} from the local Formula~\eqref{eq:C_f} by fitting a
 curve to the result of a numerical integration such as
 Formula~\eqref{eq:integration}:
$$\Cfol\left({L\over k_S}\right)={k_S\over L-L_0}\int_{L_0/k_S}^{L/k_S}\Cf(x)\,\diff{x}\eqdef{eq:integration}$$

 The local formulas \eqref{eq:c'_f}, \eqref{eq:C_f}, and \eqref{eq:White} each
 have a singularity where the expression containing the logarithm
 is~0.  The lower limit of integration ($L_0/k_S$) must be large
 enough to avoid this; but the lower limit is not revealed in the
 prior works.  The averaging Formula~\eqref{eq:integration} is quite
 sensitive to the lower limit because the largest value of the local
 formula occurs there.

 For the Mills-Hang Formula~\eqref{eq:C_f}, with lower bound
 ${L_0/k_S}=1.6$ and initial $\diff{x}/k_S=0.01$, integration of the
 local~$\Cf$ is within $\pm0.5\%$ of the average~$\Cfol$ in
 Formula~\eqref{eq:C_D} over the range $200<x/k_S<2\times10^5$.

 For the Prandtl-Schlichting Formula~\eqref{eq:c'_f}, with lower bound
 ${L_0/k_S}=0.5$ and initial $\diff{x}/k_S=0.5$, integration of the
 local~$\cf$ is within~$\pm0.5\%$ of the average~$\cfol$ in
 Formula~\eqref{eq:c_f} over the range $200<x/k_S<2\times10^5$.

\subsection{Churchill}
 Churchill~\cite{CHURCHILL1993231} compared 8 formulas from diverse
 sources versus the data from Pimenta \etal \cite{A014219}, finding
 none significantly closer to the measurements than the Mills-Hang
 local Formula~\eqref{eq:C_f}.

 \ref{Local Skin-Friction Coefficients} compares the local fully rough regime
 formulas with measurements from Pimenta \etal \cite{A014219}.

\section{Flow Over Obstacles}

 Jim\'enez~\cite{Jimenez_ARFM04} wrote ``In flows with $\delta/k<50$,
 the effect of the roughness extends across the boundary-layer, and is
 also variable. There is little left of the original wall-flow
 dynamics in these flows, which can perhaps be better described as
 flows over obstacles.''

 This investigation focuses first on a case where flow over
 obstacles dominates the dynamics.  It models the shearing stress of
 flow along a roughness which disrupts that flow at a succession of
 scales: $L$, $L/2$, $L/2^2$, $L/2^3$, \dots.  While simpler surfaces
 may produce rough flow, a roughness which disrupts at all these
 scales surely will.

\unorderedlist

 \li This approach departs from prior works because their continuous
 boundary-layer assumption is incompatible with a roughness which
 repeatedly disrupts boundary-layers.

\endunorderedlist

\subsection{Profile Roughness}

 Simpler than surface roughness, profile roughness is nonetheless
 informative.

 Let a ``profile roughness'' be a function~$z(x)$ with $0\le x\le L$;
 its mean elevation~$\overline{z}$ and RMS
 height-of-roughness~$\epsilon$ are computed similarly to surface
 roughness $\varepsilon$:
$$\eqalignno{\overline{z}&={1\over L}\,\int_0^L z(x)\,\diff{x}&\eqdef{eq:z-bar-profile}\cr
  \epsilon&=\sqrt{{1\over L}\,\int_0^L|z(x)-\overline{z}|^2\,\diff{x}}&\eqdef{eq:epsilon-profile}\cr}$$

\subsection{Self-Similar Profile Roughness}
 Let a ``self-similar profile roughness'' be a profile roughness
 function~$z(x)$ such that the RMS height-of-roughness of~$z(x)$ over
 an open interval is twice the RMS height-of-roughness of~$z(x)$ over
 each half of that interval (leaving out the midpoint).

 These $x$ intervals are open (not containing the endpoints); the
 $z(x)$ value for each endpoint contributes to the height-of-roughness
 of its parent interval, but not to any sub-interval.
 This definition is designed so that $z(x)$ will have the following
 property:

\unorderedlist

 \li The RMS height-of-roughness of $z(x)$ over an open interval,
 divided by the length of that interval will be invariant through its
 succession of scales.

\endunorderedlist

\subsection{Ramp Permutation}
 An additional constraint is needed to reduce the uncountable variety
 of possible $z(x)$ functions to a manageable number.  Experience with
 self-similar curves (Jaffer~\cite{2014arXiv1402.1807J})
 suggests a restriction to profile roughnesses which
 are permutations of the linear ramp $z(x)=\varsigma\,x/L$ with
 $0\le{x}\le{L}$.
 Each elevation from 0 to peak height~$\varsigma$ occurs exactly once.

 The only occurrence of $x$ in Formulas~\eqref{eq:z-bar-profile}
 and \eqref{eq:epsilon-profile} is $z(x)$; hence the RMS
 height-of-roughness calculation depends only on the~$z$ values, not
 on their relation to~$x$.  Thus, the height-of-roughness of any
 ramp-permutation is identical to the height-of-roughness of the
 linear ramp:
$$\eqalignno{
  \epsilon&=\sqrt{{1\over L}\,\int_0^{L}\left|{\varsigma\,x\over L}-{\varsigma\over2}\right|^2\,\diff{x}}={{\varsigma}\over\sqrt{12}}&\eqdef{eq:epsilon-permute}\cr
  }$$

\subsection{Self-Similar Ramp Permutation}
 A self-similar integer sequence~$Y(t,w)$ from integers
 $0\le{t}<w=2^q$ allows self-similar behavior to be explored with a
 finite approximation.
 Letting $t=\lfloor{w\,x/L}\rfloor\equiv{\rm floor}(w\,x/L)$
 constructs a profile roughness from a sequence by
 $z(x)=\varsigma\,Y(\lfloor{w\,x/L}\rfloor,w)/w$.

 The following three examples are self-similar ramp-permutation
 sequences.
 Each element of the sequence is generated by calling its recurrence
 function with a sequence index $0\le t\le w$ and $w$, an integer power
 of~2.  Each recursive call divides $w$ by 2, terminating (and
 returning) when $w$ reaches~1.

 Self-similar recurrence~\eqref{eq:Gray} defines
 the integer Gray-code sequence $G(t,w)$ shown in \figref{fig:Gray-rough}.
$$G(t,w)=\cases{
       t,                        & if $w = 1$;\cr
       w+G(w-1-(t\bmod w), w/2), & if $\lfloor t/w\rfloor=1$;\cr
       G(t\bmod w, w/2),         & otherwise.} \eqdef{eq:Gray}$$

 Recurrence~\eqref{eq:wiggly} defines the integer sequence $W(t,w)$
 shown in \figref{fig:wiggly-rough}; it reverses direction at each
 bifurcation, yielding a wiggliest possible self-similar
 ramp-permutation sequence.
$$W(t,w) = \cases{
       t,                                 & if $w = 1$;\cr
       \lfloor t/w\rfloor w+W(w-1-(t\bmod w), w/2), & otherwise.}
 \eqdef{eq:wiggly}$$

 \figref{fig:random-rough} shows a sequence generated by randomly
 reversing or not at each bifurcation in recurrence~\eqref{eq:random}.
 \figref{fig:rand-3d} shows a random reversal, self-similar,
 ramp-permutation surface roughness.
$$Y(t,w)=\cases{
       t,                        & if $w = 1$;\cr
       w+Y(w-1-(t\bmod w), w/2), & with 50\% probability;\cr
       Y(t\bmod w, w/2),         & otherwise.} \eqdef{eq:random}$$

\vbox{\settabs 2\columns
\+\hfill\figscale{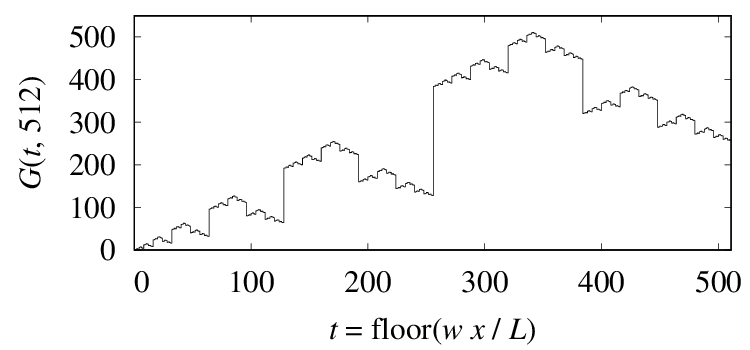}{234pt}\hfill&\hfill\figscale{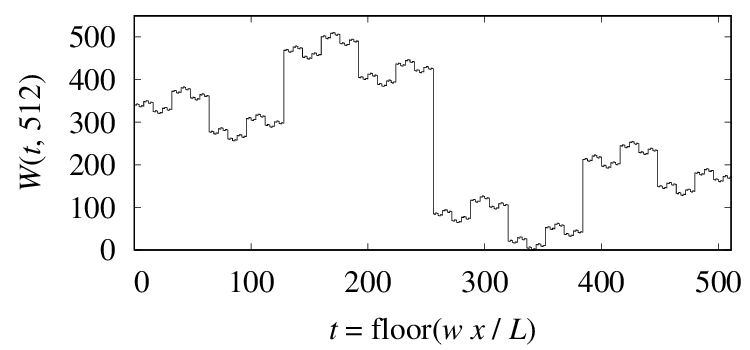}{234pt}\hfill&\cr
\+\hfill{\bf\figdef{fig:Gray-rough}\quad Gray-code profile roughness}\hfill&\hfill{\bf\figdef{fig:wiggly-rough}\quad Wiggliest self-similar roughness}\hfill&\cr
\+\hfill\figscale{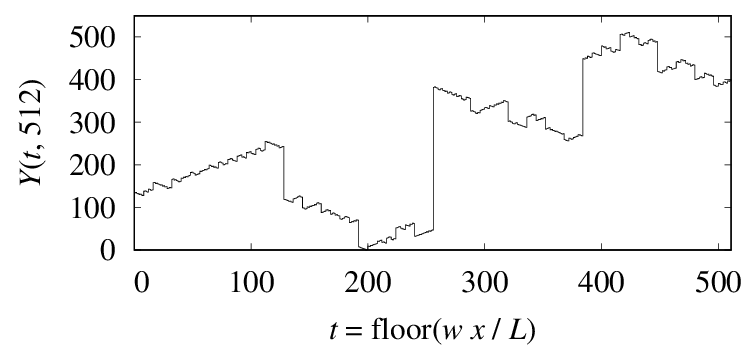}{234pt}\hfill&{\figscale{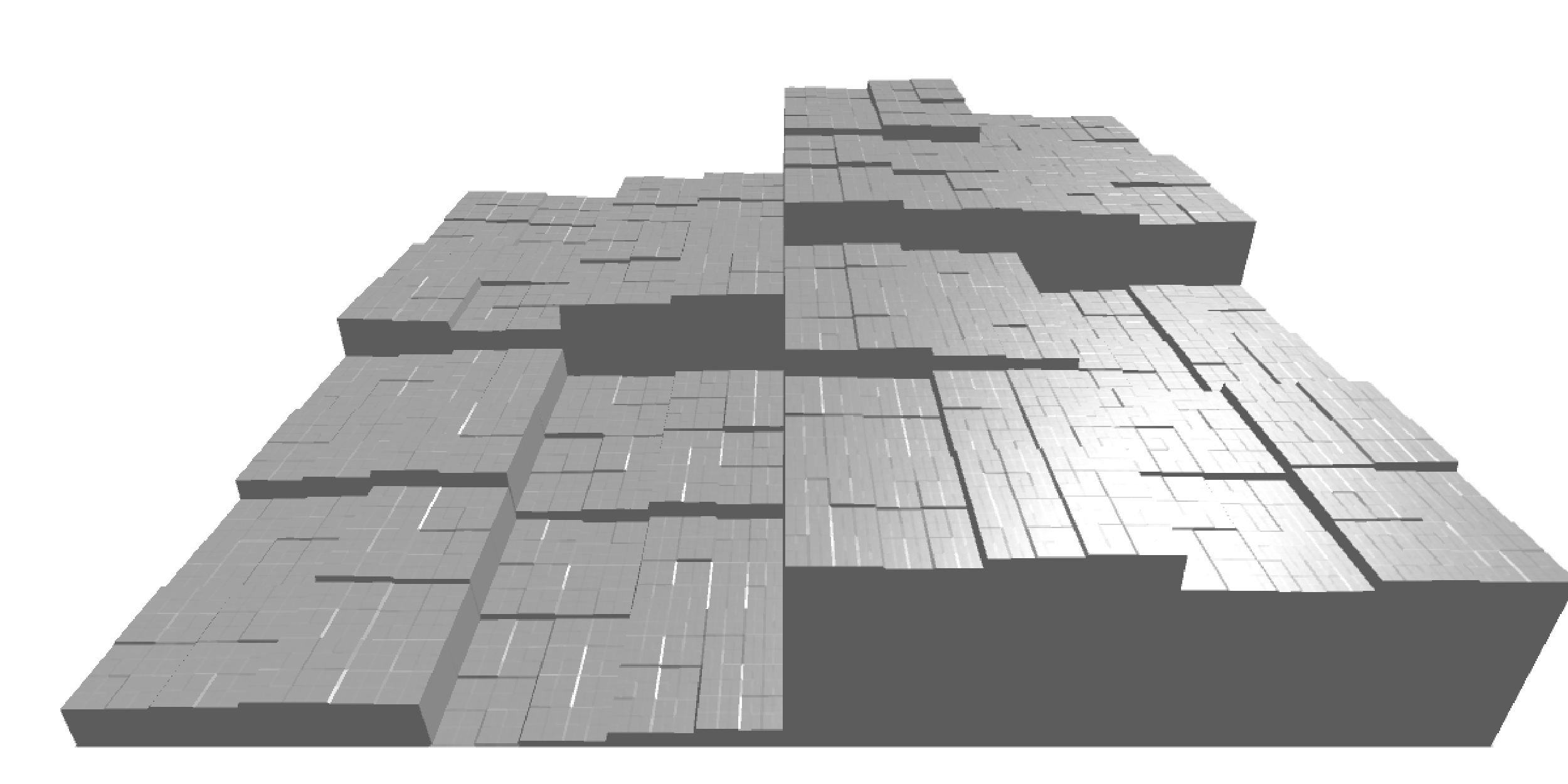}{234pt}}&\cr
\+\hfill{\bf\figdef{fig:random-rough}\quad Random reversal profile roughness}\hfill&\hfill{\bf\figdef{fig:rand-3d}\quad Random reversal ramp surface}\hfill&\cr
}

\subsection{Cardinality}
 The goal is to characterize self-similar roughnesses in general.  The
 theory should work for the vast majority of self-similar roughness
 functions, with few outliers.  For a given number of points
 $w=2^q\ge8$, there are~$2^{w-1}\ge128$ distinct self-similar
 ramp-permutation sequences, of which there are only two distinct
 ramps and two distinct wiggliest sequences.

\subsection{Friction Travel and Velocity}
 When the fluid flow encounters roughness, some particles of fluid
 must move in directions not parallel to the bulk flow.  Such movement
 results from deflections of flow by roughness peaks, pits, ridges,
 and valleys; the extent of deflections should grow with the RMS
 height-of-roughness.

 Let ``run'' be the horizontal axis and ``friction'' be the vertical
 axis of a profile roughness such as in \figref{fig:random-rough}.
 For an integer ramp-permutation sequence~$Y(t,w)$, the sum of the
 (dimensionless) lengths of all its run segments is simply
 $w-1=2^q-1$.  The sum of its friction segment lengths is:
$$\sum_{t=0}^{2^q-2}\bigl|Y(t,2^q)-Y(t+1,2^q)\bigr|\eqdef{eq:sum-R}$$

 If a particle of fluid traces the ramp-permutation sequence $Y(t,w)$
 between $t=0$ and $t=w-1$, then $w-1$ is the run travel,
 while Formula~\eqref{eq:sum-R} is the friction travel.

 \figref{fig:rough-length} shows the friction per run travel ratio
 versus~$q=\log_2(w)$.  The linear ramp trace has slope~0; the Gray-code trace
 has slope~1/2; the random reversal cases have slope of approximately~1/2;
 and the wiggliest roughness trace has slope~2/3.

\smallskip
\vbox{\settabs 1\columns
\+\hfil\figscale{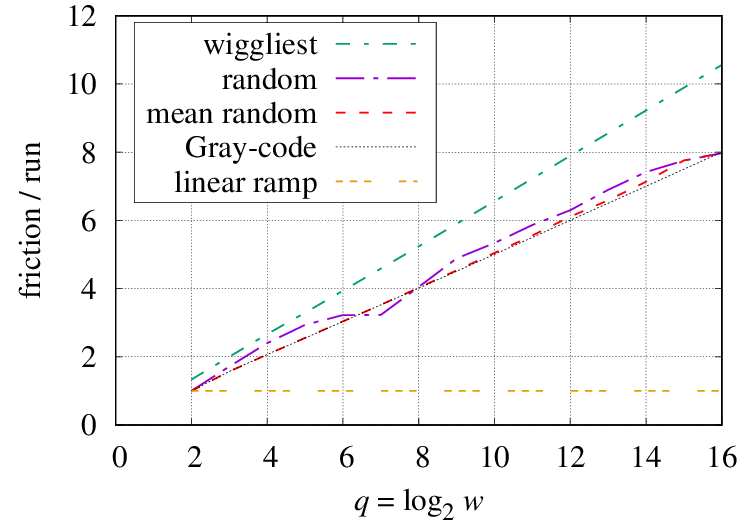}{234pt}&\cr
\+\hfil{\bf\figdef{fig:rough-length}\quad Travel along profile roughness}&\cr
}

\subsection{Roughness Sequence Outliers}
 A wiggliest roughness sequence $W(t,w)$ is an extreme case; it
 reverses friction direction at each increment of run ($t$).  For each
 wiggliest roughness sequence with $w\ge8$ there are $2^{w-1}-2$ other
 random reversal roughness sequences.  In contrast, the linear ramp
 never reverses direction.  For each linear ramp sequence there are
 $2^{w-1}-2$ other random reversal sequences.

\unorderedlist
 \li Being outliers, $W(t,w)$ and linear ramps are excluded from
 further consideration as roughness.
\endunorderedlist

\subsection{Dimensional Analysis}
 Excluding the outliers, \figref{fig:rough-length}'s friction per run
 ratios are about:
$${q\over2}\equiv{\log_2{w}\over2}\eqdef{eq:ratio}$$

 $Y(t,w)$, $t$, and $w$ are dimensionless.  The friction per run
 ratio~\eqref{eq:ratio} needs to be reformulated in terms of
 $\epsilon$~and~$L$, which have length units.  Turning to dimensional
 analysis, the argument to~$\log_2$ must be dimensionless,
 involve~$\epsilon$, and be greater than~1, so that the logarithm will
 be positive.  This friction per run ratio must increase with
 increasing~$\epsilon$.  Thus, $\epsilon$ and the logarithm will be in
 denominators, yielding:
$${2\over\log_2(L/\epsilon)}\eqdef{eq:ratio-inv}$$

 Scaling Formula~\eqref{eq:ratio-inv} by $1/\sqrt{12}$ from
 Formula~\eqref{eq:epsilon-permute} converts it into the RMS friction
 per run travel ratio:
$${1\over\sqrt{12}}\,{2\over\log_2(L/\epsilon)}\equiv{1\over\sqrt{3}\,\log_2(L/\epsilon)}\eqdef{eq:ratio12}$$

\unorderedlist

 \li Considering the run travel and friction travel with respect to
 time lets Formula~\eqref{eq:ratio12} also serve as the friction
 velocity per bulk fluid velocity ratio:~$u_\rho/u$.

\endunorderedlist

\subsection{Isotropy}
 Fluid particles stay within the vertical plane of profile roughness.
 Surface roughness deflects particles in all directions.  Therefore,
 this investigation will use $\varepsilon$ instead of $\epsilon$ and
 restrict its attention to ``isotropic'' roughness, where rotating the
 flow azimuth (direction) in the plane of the rough surface does not
 substantially affect its behavior.  \ref{Periodic Roughness} develops
 a decision procedure for roughness isotropy.

 The friction to run length ratio should not be tied only to
 ramp-permutations based on successive halving.  Instead, use the
 expected value of a continuous random variable~$Z>1$ having a Pareto
 distribution whose probability density function is~$1/Z^2$:
$$\int_1^{L/\varepsilon}{Z\over Z^2}\,\diff{Z}=\ln(L/\varepsilon)
 \qquad{u_\rho\over u}={1\over\sqrt{3}\,\ln(L/\varepsilon)}
 \eqdef{eq:ratio-log}$$

\subsection{Shearing Stress}
 The skin-friction coefficient $\fcol$ is the ratio of the shearing
 stress~$\tau_2$
 per the fluid flow's dynamic pressure (kinetic energy density)
 $\rho\,u^2/2$, where~$\rho$ is the fluid's density:
$$\fcol={\tau_2\over\rho\,u^2/2}\eqdef{eq:f-C-define}$$

 Both~$\tau_2={\rho\,{u_\rho}^2/2}$ and $\rho\,u^2/2$ have units of
 pressure, $\rm{kg/(m\cdot{s^2})}$.
 From Formula~\eqref{eq:ratio-log}:
$${u_\rho}={u\over\sqrt{3}\,\ln(L/\varepsilon)}
  \qquad\tau_2={\rho\,{u_\rho}^2\over2}={\rho\,u^2\over6\,\ln^2(L/\varepsilon)}
  \eqdef{eq:tau}$$

\unorderedlist
 \li Eliminating $\tau_2$ from
 Formulas~\eqref{eq:f-C-define} and \eqref{eq:tau} yields the average
 skin-friction coefficient of an isotropic, self-similar roughness:
$$\fcrol={1\over3\,\ln^2(L/\varepsilon)}\qquad
 {L\over\varepsilon}\gg1\qquad\varepsilon>0\eqdef{eq:f-C}$$

\vbox{\settabs 2\columns
\+\hfill\figscale{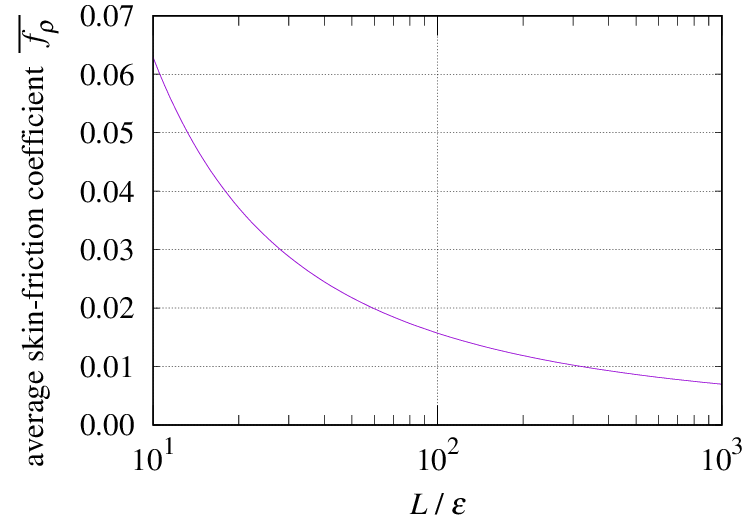}{234pt}\hfill&\hfil\figscale{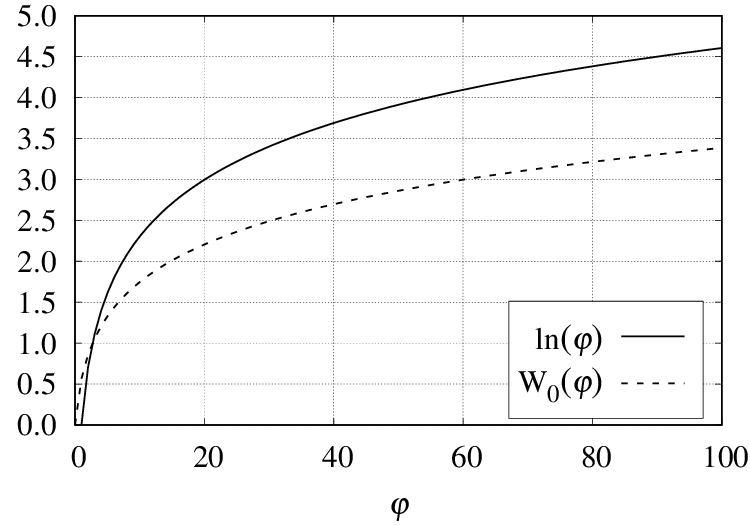}{234pt}&\cr
\+\hfill{\bf\figdef{fig:f-rho}\quad Average $\fcrol$ versus $L/\varepsilon$ of rough plate}\hfill&
  \hfil{\bf\figdef{fig:Lambert-W0}\quad Lambert function $\W_0$}&\cr
}

 \figref{fig:f-rho} plots $\fcrol$ Formula~\eqref{eq:f-C}.
 \refs{Local Skin-Friction Coefficients} and \refn{Rough Skin-Friction Measurements}
 compare Formula~\eqref{eq:f-C} with friction measurements from rough
 surfaces.
\endunorderedlist

 Note that Prandtl and Schlichting~\cite{prandtl1934resistance}
 calculated $\tau={\rho\,{u_\rho}^2}$, not
 $\tau_2={\rho\,{u_\rho}^2/2}$.  As a result, $\cfol\ge2\,\fcrol$ and
 $\cf\ge2\,\fcr$.
 Pimenta \etal \cite{A014219} and Mills and Hang~\cite{OSFCFRFP}
 designated~$\cf/2$ as the friction coefficient.

\section{Turbulent Friction}

 Formula~\eqref{eq:f-C} is not defined for $\varepsilon=0$, a smooth
 plate.

 Given~$\Rey\gg1$ there must be an $L/\varepsilon$ ratio so large that a
 length $L$ plate with a self-similar roughness of RMS
 height~$\varepsilon$ induces skin-friction midway between that of a
 rough surface and that of a smooth surface.

\subsection{Roughness Reynolds Number}
 Let the ``roughness Reynolds number''~$\Rev$ derive from friction
 velocity~$u_\rho$ at scale~$\varepsilon$:
$$\Rev={{u_\rho}\,\varepsilon\over\nu}
  ={\Rey\over\sqrt{3}\,[L/\varepsilon]\,\ln(L/\varepsilon)}\eqdef{eq:roughness-Re}$$
 where $\nu$ is the fluid's kinematic viscosity (with units
 $\rm{m^2/s}$) and $\Rey=L\,u/\nu$.  The~$\Rey$ strength at which
 rough plate friction transitions to smooth plate friction should have
 the same~$\Rev$ value at all $L/\varepsilon\gg1$.  $\Rev=1$ when
 $\varepsilon=u_\rho/\nu$.  Combining $\Rev=1$ with
 Formula~\eqref{eq:roughness-Re} relates $\Rey$ and $L/\varepsilon$ at
 transition:
$$\Rey=\sqrt{3}\,{L\over\varepsilon}\,\ln{L\over\varepsilon}\eqdef{eq:Re-parity}$$

\vfill\eject

 This link between $\Rey$ and $L/\varepsilon$ suggests that the
 turbulent friction coefficient~$\fctol$ can be inferred by combining
 Formulas~\eqref{eq:f-C} and~\eqref{eq:Re-parity}.  However, there
 being no roughness on a smooth plate, the coefficients must be
 different from $\fcrol$~Formula~\eqref{eq:f-C}.  Scaling
 $\fcrol(L/\varepsilon)$ by~${3/2^{5/4}}$,
 and its argument
 by~$1/{\rm e}$:
$$\fctol={3\over2^{5/4}}~\fcrol\!\left({L\over{\rm e}\,\varepsilon}\right)={3\over2^{5/4}}\,\left[{1\over3}\,\ln^{-2}{L\over{\rm e}\,\varepsilon}\right]
  \qquad{2^{-5/4}}\approx{0.4204}\eqdef{eq:f-C-smooth}$$

 Euler's number ${\rm e}=\exp(1)$ is a fixed point of
 $\varphi\,\ln{\varphi}$, which appears in Formula~\eqref{eq:exp-W0}.

\subsection{Lambert $\W$ Function}
 The (natural) logarithm function $\ln$ is the inverse of $\exp(x)$.
 Similarly, the Lambert~$\W$~function is the inverse of $x\,\exp(x)$.
 $L/\varepsilon$ can be eliminated from
 Formulas~\eqref{eq:Re-parity} and \eqref{eq:f-C-smooth} using the
 Lambert~$\W$~function's principal branch~$\Wz$, which is defined by
 equivalence~\eqref{eq:W0} and plotted in \figref{fig:Lambert-W0}.
$$\eqalignno{
 \vartheta=\varphi\,\exp{\varphi}\quad&\Leftrightarrow\quad\varphi=\Wz(\vartheta)&\eqdef{eq:W0}\cr
 \vartheta=\varphi\,\ln{\varphi}\quad&\Leftrightarrow\quad\varphi=\exp\Wz(\vartheta)&\eqdef{eq:exp-W0}}$$

 The related equivalence~\eqref{eq:exp-W0} acting on
 Formula~\eqref{eq:Re-parity} lets ${\exp(\Wz(\Rey/\sqrt{3}))}$
 replace $L/\varepsilon$ in Formula~\eqref{eq:f-C-smooth} when
 $\Rey\gg\sqrt{3}{\rm\,e}\approx4.71$.
$$\fctol={2^{-5/4}}\,\ln^{-2}\left({\exp\left(\Wz\left(\Rey/\sqrt{3}\right)\right)\over{\rm e}}\right)
 ={{2^{-5/4}}\over\left[{\Wz\left(\Rey/\sqrt{3}\right)}-1\right]^2}
  \eqdef{eq:W-smooth}$$

\subsection{Comparison With Measurements}
 Churchill~\cite{CHURCHILL1993231} compared turbulent friction
 formulas from multiple studies with measurements from Smith and
 Walker~\cite{smith_walker_1959}, and Spalding and
 Chi~\cite{spalding_chi_1964}.  \figref{fig:Churchill-smooth} plots
 them and $\fctol$ Formula~\eqref{eq:W-smooth}; the key gives the RMS
 error of the measurements relative to each formula (RMSRE was
 introduced in \ref{Data-Sets and Evaluation}).

\unorderedlist
\li With 0.74\% RMSRE, $\fctol$ Formula~\eqref{eq:W-smooth} has less
 error than any formula evaluated by Churchill.
\endunorderedlist

\vbox{\settabs 1\columns
\+\hfill\figscale{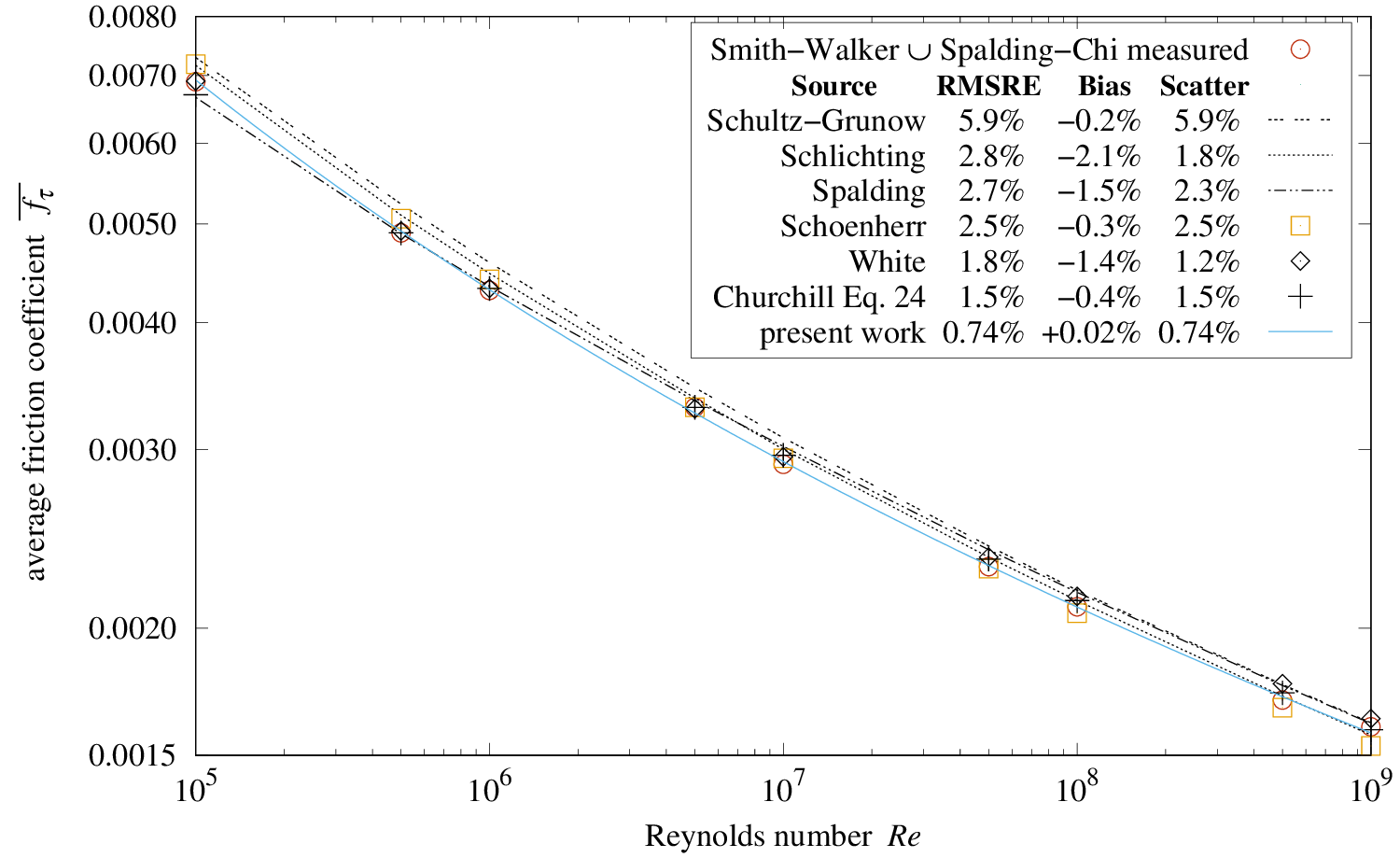}{440pt}\hfill&\cr
\+\hfill{\bf\figdef{fig:Churchill-smooth}\quad Average $\fctol$ versus $\Rey$ of smooth plate}\hfill&\cr
}

\unorderedlist
 \li Self-similar roughness served to establish the skin-friction
 coefficient upper bound for roughness $\varepsilon>0$, and the
 coefficient for $\varepsilon=0$.  Attention now turns to existing
 conventional types of roughness.
\endunorderedlist

\section{Spectral Roughness}

 Several prior
 works~\cite{Nikuradse33lawsof,Schlichting1937,10010463165,OSFCFRFP,doi:10.1680/ijoti.1939.13150}
 use the term ``uniform roughness'' to describe sand-roughness,
 implying that its height-of-roughness is the same at all scales.
 This concept of uniform roughness is incompatible with
 self-similarity; the RMS height-of-roughness of a portion of a
 self-similar surface must shrink with its succession of scales.

 Sand-roughness can be described as ``repeated roughness''.  However,
 a roughness composed of parallel rows of 1000 sand grains spanning
 its length can also be described as having 500 sand grain pairs
 spanning its length.  Needed is an unambiguous method for determining
 the spatial period.

\subsection{Discrete Fourier Transform}
 The discrete Fourier transform~\eqref{eq:Fourier} converts a series
 of equally-spaced samples of a function into a complex-number
 coefficient $X_j$ for each of its harmonic sinusoidal components:
$$X_j=\sum_{t=0}^{w-1}Y(t,w)\,\exp\left({-2\,\pi\,{\rm i}\,j\,t\over w}\right)\eqdef{eq:Fourier}$$

 A complex number consists of two real numbers as $a+b\,{\rm{i}}$,
 where ${\rm{i}}=\sqrt{-1}$; $b$ is called the imaginary part.
 The amplitude of $a+b\,{\rm{i}}$, written $|a+b\,{\rm{i}}|=\sqrt{a^2+b^2}$.

 There is a profound connection between
 $X_j$ and the RMS height-of-roughness~$\epsilon$:
$$\epsilon=\sqrt{{1\over{w}}\,\sum_{j=1}^{w-1}\left|X_j\right|^2}\eqdef{eq:Fourier-epsilon}$$

 $|X_{w-j}|=|X_j|$ because all the $Y(t,w)$ elevations have imaginary
 parts $=0$; hence there are $w/2+1$ distinct~$|X_j|$; only
 $0\le{j}\le{w/2}$ needs to be considered in the developments which
 follow.

\subsection{Dominant Component of Roughness}
 The $X_0=\overline{z}$ term, the mean value of~$Y\!$, is the
 only~$X_j$ term not included in the
 Formula~\eqref{eq:Fourier-epsilon} sum for~$\epsilon$.  Hence, the
 dominant component of roughness will be the~$X_j$ ($0<j\le{w/2}$)
 having the largest amplitude.

\unorderedlist
 \li Let the ``period index'' $j_P$ be the nonzero index~$j$ of
 the~$X_j$ having the largest amplitude~$|X_j|$.

 \li When one $|X_j|$ is dominant, $j_P$ is well-defined and the
 profile roughness's spatial period is $L_P=L/j_P$.
\endunorderedlist

 \figref{fig:Gray-spect} shows the~$|X_j|/w$ amplitude spectrum of the
 Gray-code profile roughness from \figref{fig:Gray-rough}; also the
 mean Fourier spectrum amplitudes from 187 instances of 128-point
 random reversal profiles ($w=128$).
 For both spectra, $X_1$ has the largest amplitude; thus $j_P=1$,
 indicating that neither spectrum is from repeated roughness.

\medskip
\vbox{\settabs 2\columns
\+\figscale{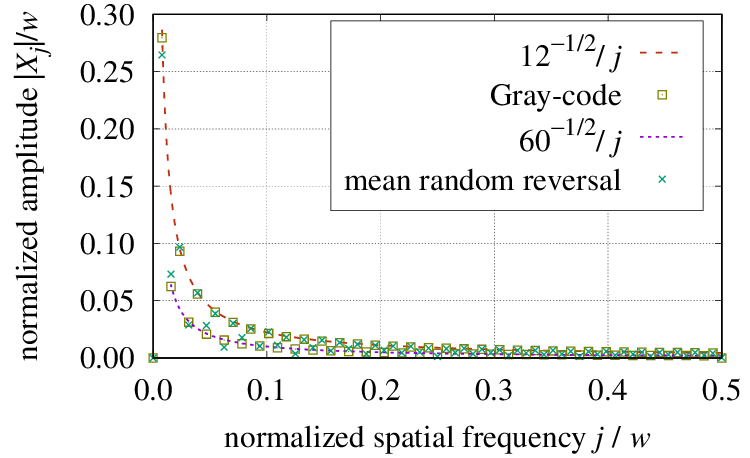}{234pt}\hfill&\hfill\figscale{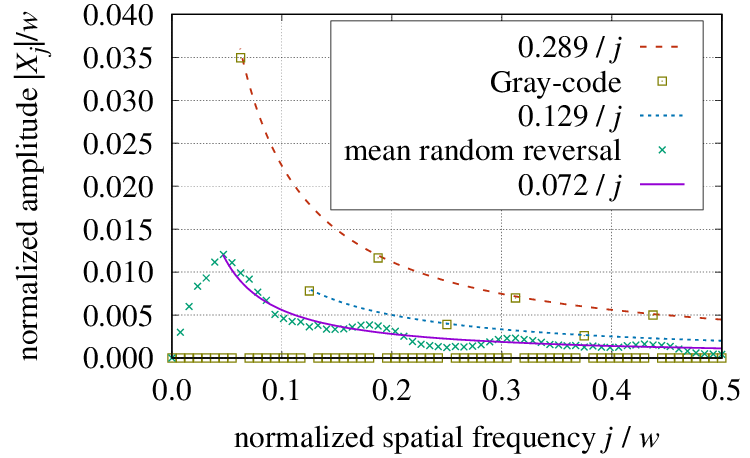}{234pt}&\cr
\+\hfill{\bf\figdef{fig:Gray-spect}\quad Gray and random spectra}\hfill&\hfill{\bf\figdef{fig:Gray8-spect}\quad Gray and random eighths}\hfill&\cr
}

 \figref{fig:Gray8-spect} shows the spectrum of eight concatenated
 repetitions of a Gray-code sequence; also the mean Fourier spectrum
 amplitudes from 187 instances of eight concatenated random reversal
 sequences.  The period index~$j_P$ of the Gray-code eighths is~8, as
 expected; but the random reversal sequences have $j_P=6$ because
 their amplitudes are not correlated between the random eighths.

\unorderedlist
 \li This use of the discrete Fourier transform was able to determine
 the spatial period of profile roughness.
 \ref{Periodic Roughness} generalizes this spatial period metric to isotropic
 surface roughness.
\endunorderedlist

\section{Periodic Roughness}

\unorderedlist
 \li Let a ``periodic roughness'' be a flat surface tiled with many
 isotropic, uniformly sized patches, all sharing the same mean
 elevation~$\overline{z}$ and RMS height-of-roughness~$\varepsilon$.

 \li The mean elevation and RMS height-of-roughness of the entire
 surface will therefore be~$\overline{z}$ and~$\varepsilon$.
\endunorderedlist

\subsection{Discrete Spatial Fourier Transform}
 Let $S_{s,t}$ be a $w\times{w}$ matrix of mean elevations from
 a~$w\times{w}$ square grid of an $L_w\times L_w$ region of a rough
 surface.  Its 2-dimensional discrete spatial Fourier transform is:
$$X_{j,k}=\sum_{t=0}^{w-1}\sum_{s=0}^{w-1}S_{s,t}\,\exp\left(-2\,\pi\,{\rm i}\,{j\,t+k\,s\over w}\right)
  \eqdef{eq:Fourier2}$$

\unorderedlist
 \li Let the 2-dimensional period index $j_P=\sqrt{j^2+k^2}>0$, where
 $0\le{j}\le{w/2}$ and $0\le{k}\le{w/2}$ are the indexes of the
 coefficient~$X_{j,k}$ having the largest amplitude,
 excluding~$X_{0,0}$.

 \li The 2-dimensional spatial period $L_P=L_w/j_P$.
\endunorderedlist

 \figref{fig:dft-0.500} shows the $j_P$ values of a square equal-area
 bi-level surface (regular array of square posts on a flat plate)
 computing~$X_{j,k}$ from a $64\times64$ interpolated sampling of that
 surface with azimuth from $0^\circ$ through $45^\circ$, and which is
 scaled between 9 and 13 cells per side.  At each scale, $j_P$ varies
 within a $\pm1$ range as the azimuth is
 rotated.  \figref{fig:dft-0.250} shows $j_P$ values of a 25\% high,
 75\% low, bi-level surface; some of its~$j_P$ traces have peaks
 outside of the~$\pm1$ range.  This suggests a quantitative criterion
 for surface roughness isotropy:

\unorderedlist
 \li A surface roughness is isotropic if $j_P\gg1$ varies no more
 than~$\pm{2}$ through its full flow azimuth rotation.

 \li  More specifically, using $w\times{w}$ ($w\ge64$) samplings of
 roughness at nine scales over a 2:1 range such that most of the
 calculated $j_P$ values satisfy $8\le j_P\le16$ at each scale, do 56
 sampling trials with randomized offset and randomized azimuth
 rotation per scale.  A roughness is considered isotropic if no more
 than 5 of the $504 (=56\times9)$ trials have $j_P$ varying more than
 $\pm{w/32}$ from its scale.
\endunorderedlist

\vbox{\settabs 2\columns
\+\figscale{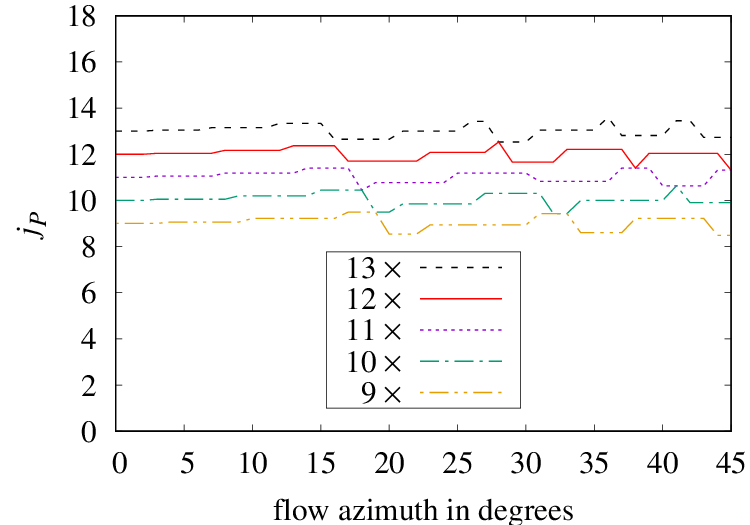}{234pt}\hfill&\hfill\figscale{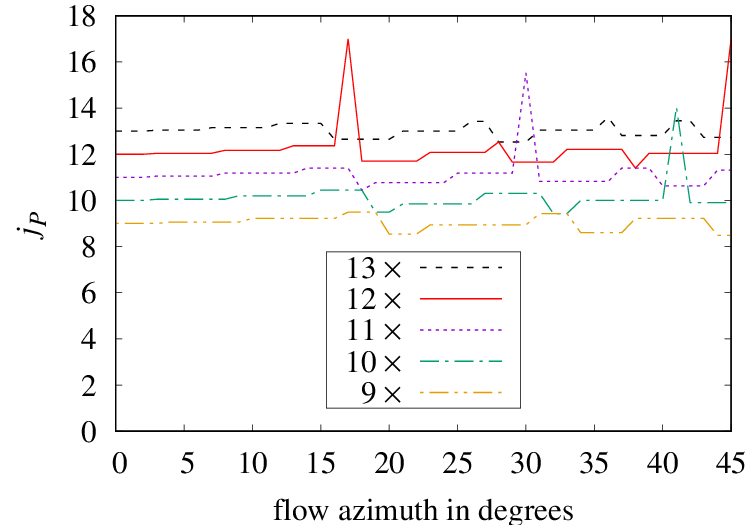}{234pt}&\cr
\+\hfill{\bf\figdef{fig:dft-0.500}\quad Bi-level plate 50\%}\hfill&\hfill{\bf\figdef{fig:dft-0.250}\quad Bi-level plate 25\%}\hfill&\cr
}

\subsection{Exploring Isotropy}
 Tests of this criterion applied to $128\times128$ interpolated
 samplings found the following roughnesses to be isotropic:
\unorderedlist
 \li square arrays of aligned square posts having an upper area
 fraction between~27\% and~76\%;

 \li square arrays of circular columns having an upper area fraction
 between~24\% and~78\%;

 \li hexagonal arrays of circular columns having an upper
 area fraction between~6\% and~75\%;

 \li hexagonal arrays of circular wells (depressions) having an upper
 area fraction between~25\% and~94\%;

 \li hexagonal arrays of aligned square posts having an upper area
 fraction between~4\% and~49\%;

 \li hexagonal arrays of aligned square wells having an upper area
 fraction between~51\% and~96\%;

 \li 15\% elongated hexagonal arrays of cone or bump
 protrusions described in Schlichting~\cite{Schlichting1937}.

\endunorderedlist

\subsection{Visual Isotropy}
This isotropy test is not equivalent to the visual appearance of
isotropy.  Square post arrays having upper area fractions of 20\% fail
the isotropy test, while those with 30\% pass.  Plates from
Schlichting~\cite{Schlichting1937} having elongated hexagonal arrays
of cones are not visually isotropic, yet pass the test.

\section{Local Skin-Friction Coefficients}

 Conversions between local and average skin-friction are needed in
 order to compare prior with present work.

\subsection{Continuous Local to Average Skin-Friction}
 The ratio of average to local skin-friction~$\fcol/\fc>1$ of a
 continuous boundary-layer is calculated from~$\fc$ by
 Formula~\eqref{eq:f-C-from-f-c-continuous}.  Using $\Rez>0$ as the
 integration lower-bound avoids the division-by-zero singularity at the
 leading edge of the plate.
$${\fcol(\Rey)\over \fc(\Rey)}={1\over\Rey-\Rez}\,\int_{\Rez}^\Rey{\fc(\Rex)\over \fc(\Rey)}\,\diff\Rex
 \eqdef{eq:f-C-from-f-c-continuous}$$

 The Blasius laminar model in Schlichting \cite{schlichting2014}
 derives the local drag coefficient:
$$\fcl={0.664\over\sqrt{\Rex}}\eqdef{eq:laminar-friction-local}$$

 Applying transform~\eqref{eq:f-C-from-f-c-continuous} to
 Formula~\eqref{eq:laminar-friction-local} and factoring the
 denominator produces a novel formula for the average laminar friction
 coefficient which lacks the leading-edge singularity:
$${\fclol(\Rey)}=1.328\,{\sqrt{\Rey}-\sqrt{\Rez}\over\Rey-\Rez}={1.328\over\sqrt{\Rey}+\sqrt{\Rez}}\eqdef{eq:laminar-friction}$$

 Lienhard and Lienhard~\cite{ahtt5e} estimates $\Rez=600$, leading
 to $\fclol(0)\approx0.0542$.

\subsection{Disrupted Local to Average Skin-Friction}
 The local skin-friction coefficient is not well-defined for
 self-similar roughness because of its constant $L/\varepsilon$.
 Periodic roughness has varying $L/\varepsilon$; conveniently, it also
 provides $L_P/\varepsilon$ as the lower-bound of integration.

 A crucial distinction between periodic roughness and smooth surfaces
 is that periodic roughness disrupts the boundary-layer repeatedly.
 Thus, the local skin-friction coefficients being averaged are
 independent.  Instead of $\fcrol/\fcr$ scaling linearly,
 it should scale as the square:
$${\fcrol(L/\varepsilon)\over \fcr(L/\varepsilon)}=\left[{\varepsilon\over L-L_P}\int_{L_P/\varepsilon}^{L/\varepsilon}{\fcr(r)\over \fcr(L/\varepsilon)}\,\diff{r}\right]^2
\eqdef{eq:f-C-from-f-c}$$

 Applying Formula~\eqref{eq:f-C-from-f-c} to the Mills-Hang local
 Formula~\eqref{eq:C_f} yields the average skin-friction coefficient:
$${\left.\Cfol^{\,2}\right/\Cf} \eqdef{eq:C_D-from-C_f}$$
 where $\Cf$ and $\Cfol$ are from equations~\eqref{eq:C_f} and~\eqref{eq:C_D},
 respectively.

\subsection{Average Formulas From Prior Work}
 \figref{fig:average} compares fully rough regime average
 skin-friction formulas with $\fcrol$ Formula~\eqref{eq:f-C}.  To the
 right of each ``$\in\pm$'' is the maximum discrepancy from $\fcrol$
 over the Mills-Hang range $750<{L/k_S}<2750$, which is
 $4000<L/\varepsilon<14666$.

 Note that \ref{Plateau Roughness} establishes that $\fcrol$
 Formula~\eqref{eq:f-C} applies to sand-roughness.

\unorderedlist

 \li ``$0.5~\Cfol^{\,2}/\Cf \in\pm\,~2.4\%$'' is 1/2 of
 Formula~\eqref{eq:C_D-from-C_f};
 it matches $\fcrol$ within~$\pm2.4\%$.

 \li ``$0.5~\cfol\qquad\,\,~\in\pm\,~9.3\%$'' is 1/2 of the Prandtl-Schlichting
 average Formula~\eqref{eq:c_f}.

 \li ``$0.5~\Cfol\qquad~\in\pm23.1\%$'' is 1/2 of the Mills-Hang average
 Formula~\eqref{eq:C_D}.
\endunorderedlist

\vbox{\settabs 1\columns
\+\hfill\figscale{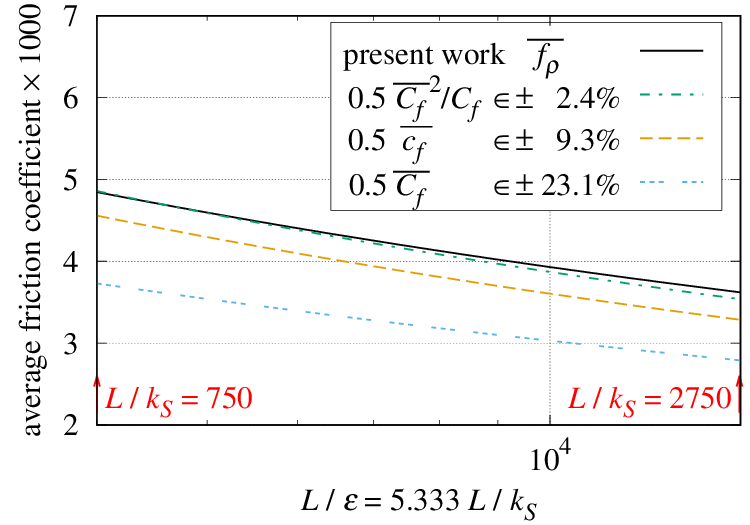}{234pt}\hfill&\cr
\+\hfill{\bf\figdef{fig:average}\quad Average friction coefficient of sand-roughness}\hfill&\cr
}

\subsection{Disrupted Average to Local Skin-Friction}
 For disrupted boundary-layers, the transform for local friction~$\fcr$
 given average friction~$\fcrol$ is:
 $${\fcr(L/\varepsilon)\over \fcrol(L/\varepsilon)}=\left[\left.{\diff{\left([L-L_P]\,\fcrol(L/\varepsilon)\right)}\over \diff{L}}\right/{\fcrol(L/\varepsilon)}\right]^2
  \eqdef{eq:f-c-from-f-C}$$

 Applying Formula~\eqref{eq:f-c-from-f-C} to~$\fcrol$ Formula~\eqref{eq:f-C}
 with ${L}\ge{x}>{L_P}\ge\varepsilon$ yields the local friction
 coefficient:
$$\fcr(x/\varepsilon)={1\over3}{\left[{\ln{(x/\varepsilon)}+2\,\left[L_P/x-1\right]\over\ln^2{(x/\varepsilon)}}\right]^2}
  \eqdef{eq:f-c-local}$$

\subsection{Local Formulas From Prior Work}
 \figref{fig:comparison} plots the local friction coefficients from
 White~\eqref{eq:White}, Prandtl-Schlichting~\eqref{eq:c'_f},
 Mills-Hang~\eqref{eq:C_f}, and present work $\fcr$
 Formula~\eqref{eq:f-c-local}.
\unorderedlist
 \li Mills-Hang matches $\fcr$ Formula~\eqref{eq:f-c-local} within
 3.1\% over $750<{x/k_S}<2750$.
\endunorderedlist

\vbox{\settabs 2\columns
\+\hfill\figscale{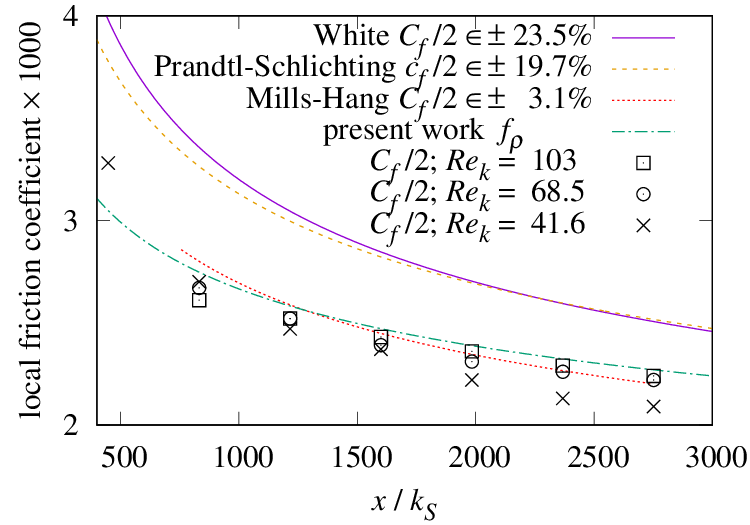}{234pt}\hfill
 &\hfill\figscale{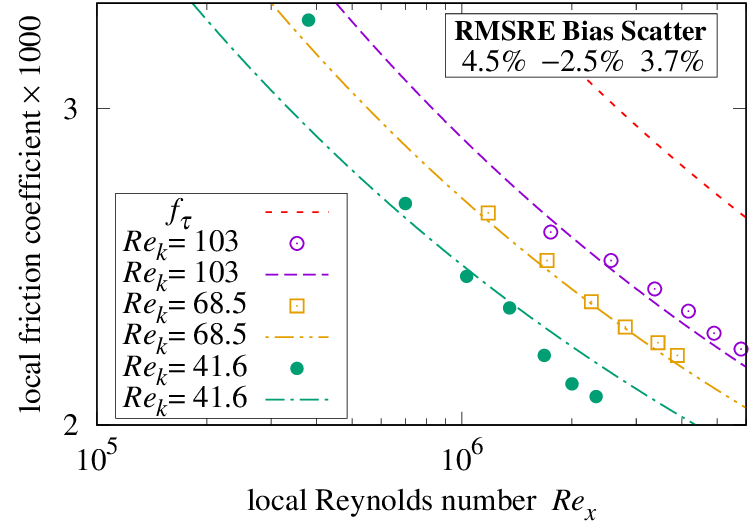}{234pt}\hfill&\cr
\+\hfill{\bf\figdef{fig:comparison}\quad Pimenta \etal versus $x/k_S$}\hfill
 &\hfill{\bf\figdef{fig:Moody-PMK}\quad Pimenta \etal versus $\Rex$}\hfill&\cr
}

\medskip
\centerline{\bf\tabdef{tab:local-performance}\quad{Local friction coefficient of sphere-roughened plate}}
\vbox{\settabs 7\columns
\toprule
\+\hfill\bf Source &&\quad~\bf Formula &\hfill\bf RMSRE &\hfill\bf Bias~~&\hfill\bf Scatter~~&\hfil\bf Used&\cr
\midrule
\+\quad Prandtl and Schlichting~\cite{prandtl1934resistance}&&\quad~\eqref{eq:c'_f}~$\cf/2$&\hfill15.6\%\quad&\hfill $-15.4\%$&\hfill 2.6\%\quad&\hfill19/19\qquad&\cr
\+\quad Mills and Hang~\cite{OSFCFRFP}&&\quad~\eqref{eq:C_f}~$\Cf/2$&\hfill3.7\%\quad&\hfill $-2.4\%$&\hfill 2.8\%\quad&\hfill19/19\qquad&\cr
\+\quad White~\cite{white2006}&&\quad~\eqref{eq:White}~$\Cf/2$&\hfill 16.6\%\quad&\hfill $-16.3\%$&\hfill 3.3\%\quad&\hfill19/19\qquad&\cr
\+\quad present work &&\quad~\eqref{eq:f-c-local}~$\fcr$&\hfill4.5\%\quad&\hfill $-3.0\%$&\hfill 3.3\%\quad&\hfill19/19\qquad&\cr
\bottomrule
}

\subsection{Comparison With Local Drag Measurements}
 The points labeled ``$C_f/2; \Rey_k =$'' in \figref{fig:comparison} show the
 local drag coefficient measurements versus~$x/k_S$ for the
 sphere-roughened plate at three rates of flow.
 \tabref{tab:local-performance} presents RMSRE of this set of 19
 measurements to each of the rough regime formulas.
 
 Formula~\eqref{eq:x/e} calculates $x/\varepsilon$ from the $\Rex$ and
 $\Rev=\Rey_k/5.333$ values supplied by Pimenta
 \etal \cite{A014219}.  \figref{fig:Moody-PMK} plots local
 $\fcr(x/\varepsilon)$ versus $\Rex$ for the
 sphere-roughened plate.

\unorderedlist
 \li The Pimenta \etal measurements have RMSRE $4.5\%$ from $\fcr$
 Formula~\eqref{eq:f-c-local}.
\endunorderedlist

\subsection{Continuous Average to Local Skin-Friction}
 For continuous boundary-layers, the transform for local friction~$\fc$
 given average friction~$\fcol$ is:

$${\fc(\Rex)\over\fcol(\Rey)}
         ={1\over\fcol(\Rey)}\,{\diff{\left([\Rex-\Rez]\,\fcol(\Rex)\right)}\over\diff\Rex}
         \eqdef{eq:f-from-f-bar}$$

 The local skin-friction coefficient for turbulent flow~$\fct$
 can be derived from~$\fctol$ equation~\eqref{eq:W-smooth},
 Formula~\eqref{eq:f-from-f-bar}, and $\Wz$
 identity~\eqref{eq:W-diff}, provided that
 $\Rex\ge\Rez\ge\sqrt{3}\,{\rm e}$ and $\Rex\gg\sqrt{3}\,{\rm e}$:
$${\diff{\Wz(\vartheta)}\over \diff\vartheta}\equiv{{\Wz(\vartheta)}\over{\vartheta\,\left[\Wz(\vartheta)+1\right]}}
   \eqdef{eq:W-diff}$$
$${\fct(\Rex)}=
  {\Wz^2\left(\Rex/\sqrt3\right)-2\,[1-\Rez/\Rex]\;\Wz\left(\Rex/\sqrt3\right)-1
   \over\left[\Wz\left(\Rex/\sqrt3\right)-1\right]^3\,\left[\Wz\left(\Rex/\sqrt3\right)+1\right]}
   \eqdef{eq:f-c-smooth}$$

\subsection{Comparison With Local Measurements}
 \tabref{tab:Churchill-local} compares measurements at
 $10^5\le\Rex\le10^{10}$ made by Smith and
 Walker~\cite{smith_walker_1959} and Spalding and
 Chi~\cite{spalding_chi_1964}
 with ``present work'' Formula~\eqref{eq:f-c-smooth} and formulas
 collected by Churchill~\cite{CHURCHILL1993231}.

\centerline{\bf\tabdef{tab:Churchill-local}\quad{Local turbulent performance}}
\moveright 0.25\hsize \vbox{\settabs 8\columns
\toprule
\+\qquad {\bf Source}&\qquad\quad {\bf RMSRE}&\hfill {\bf Bias}~~&\hfill {\bf Scatter}&\cr
\midrule
\+Schultz-Grunow&\hfill$8.5\%$&\hfill$+5.5\%$&\hfill$6.6\%$&\cr
\+White&\hfill$4.2\%$&\hfill$-2.9\%$&\hfill$3.0\%$&\cr
\+Von Karman&\hfill$3.7\%$&\hfill$+0.7\%$&\hfill$3.6\%$&\cr
\+Schlichting&\hfill$3.6\%$&\hfill$-2.4\%$&\hfill$2.7\%$&\cr
\+Spalding&\hfill$3.4\%$&\hfill$-2.5\%$&\hfill$2.3\%$&\cr
\+Churchill Eq. 16&\hfill$3.0\%$&\hfill$-2.3\%$&\hfill$2.0\%$&\cr
\+present work&\hfill$1.8\%$&\hfill$+1.1\%$&\hfill$1.4\%$&\cr
\+Churchill Eq. 18&\hfill$1.3\%$&\hfill$+0.6\%$&\hfill$1.2\%$&\cr
\bottomrule
}

 When calculating RMSRE, the error due to variation in a single sample
 tends to be larger than the error when that variation is distributed
 among multiple samples.  Thus, local measurements tend to have larger
 RMSRE than average measurements do.  Other than ``Churchill Eq. 18'',
 this is the case when comparing \tabref{tab:Churchill-local}
 with \figref{fig:Churchill-smooth}.

\unorderedlist
 \li The Churchill local data-set has~1.8\% RMSRE versus ``present work''
 Formula~\eqref{eq:f-c-smooth}.
\endunorderedlist

\subsection{Skin-Friction in Liquids}
 \figref{fig:friction} compares $\fct$ Formula~\eqref{eq:f-c-smooth}
 with skin-friction measurements in several fluids
 from \v{Z}ukauskas and \v{S}lan\v{c}iauskas~\cite{Zukauskas1987}.

 A fluid's Prandtl number~($\Pra$) is its kinetic viscosity per
 thermal diffusivity ratio.  Fluids with $\Pra\gg1$ transport heat
 primarily via fluid flow; conduction dominates heat transfer in
 fluids with $\Pra\ll1$.

\unorderedlist
 \li These data-sets have~2.5\%-5.2\% RMSRE versus ``present work''
 Formula~\eqref{eq:f-c-smooth}.

 \li No significant dependence on $\Pra$ is manifest.
\endunorderedlist

\vbox{\settabs 1\columns
\+\hfill\figscale{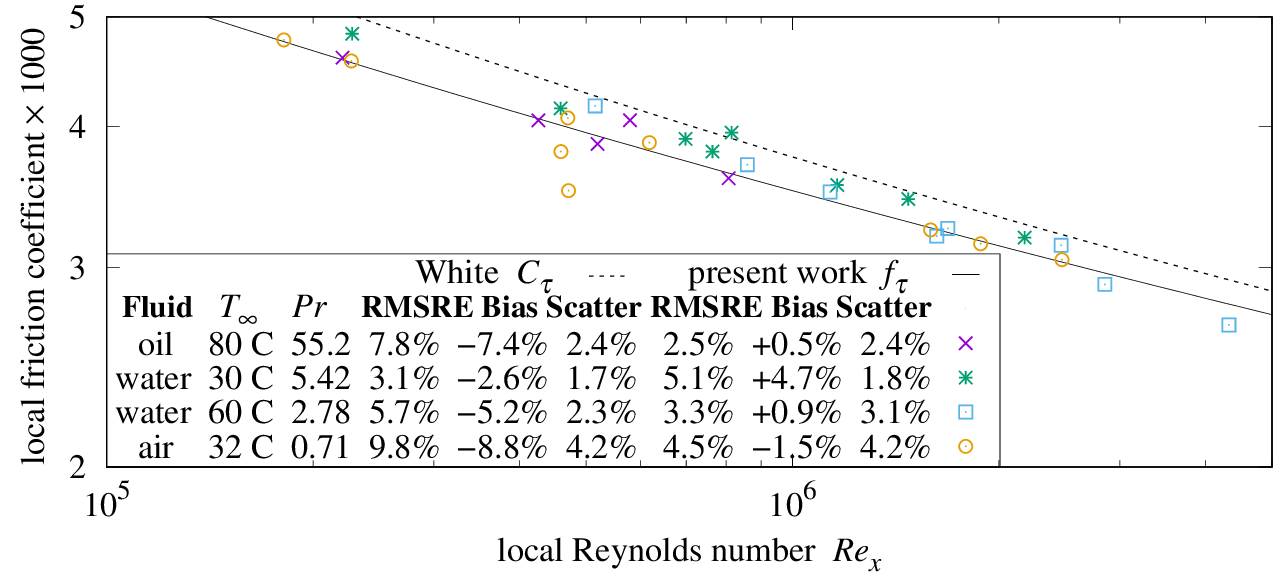}{360pt}\hfill&\cr
\+\hfill{\bf\figdef{fig:friction}\quad Local $\fct$ versus $\Rex$ of smooth plate}\hfill&\cr}

\section{Forced Convection}

 Forced convective heat transfer is expressed as the (dimensionless)
 average Nusselt number~$\Nuol\equiv{\hol\,L/k}$, where $k$ is the
 fluid's thermal conductivity (${\rm W/(m\cdot K)}$) and $\hol$ is the
 plate's average convective surface conductance in that fluid (${\rm
 W/(m^2\cdot K)}$).  The local version, $\Nu\equiv{h\,x/k}$, has the
 same units.

\subsection{Rough Plate}
 The disruption which transfers momentum can also transfer heat.
 Hence heat transfer will grow with~$\Rey\,\fcrol$.
 Jaffer~\cite{thermo3010010} finds that the
 natural convection boundary layer of an upward-facing plate is
 disrupted by collision of the flows and their detachment from the
 plate's center; the other plate orientations have continuous boundary
 layers.  Boundary layer disruption being the essence of rough
 flow, this investigation proposes that heat transfer grows
 with the same~$\root3\of\Pra$ dependence as upward-facing natural
 convection.

 Fluid heating by the leading part of the plate reduces heat transfer
 from the trailing part; hence, heat transfer is scaled by~1/2.
 Expanding $\fcrol$ from equation~\eqref{eq:f-C} yields
 Formula~\eqref{eq:forced} for rough average forced
 convection heat transfer, provided that ${L/\varepsilon}\gg1$ and
 $\Pra\ge0$:
$$\Nurol={\Rey\,\fcrol\,\Pra^{1/3}\over2}={\Rey\,\Pra^{1/3}\over6\,\ln^2{\left(L/\varepsilon\right)}}
 \eqdef{eq:forced}$$

 The present analysis for self-similar $\fcrol$ did not involve
 continuous boundary-layers; hence, it avoids the $\Pra\ge0.6$
 restriction affecting Formula~\eqref{eq:Gnielinski}.
 Formula~\eqref{eq:forced} is compared with measurements
 in \ref{Rough Heat Transfer Measurements}.

\subsection{Smooth Plate}
 For turbulent convection, Lienhard~\cite{10.1115/1.4046795}
 recommends composing the Gnielinski Formula~\eqref{eq:Gnielinski} with
 the White $\Cft$ Formula~\eqref{eq:White-smooth}, subject to
 $\Pra\ge0.6$:
$$\Nu={\Rex\,\Pra\,\Cft/2\over1+12.7\,[\Pra^{2/3}-1]\,\sqrt{\Cft/2}}
  \eqdef{eq:Gnielinski}$$

 Lienhard states that $\Nu$ Formula~\eqref{eq:gas} has similar accuracy
 for gases with $0.6\le\Pra<2$:
$$\eqalignno{\Nu&=0.0296\,\Rex^{4/5}\,\Pra^{0.6}&\eqdef{eq:gas}\cr
  \Nuol&=0.037\,\Rey^{4/5}\,\Pra^{0.6}&\eqdef{eq:gases}\cr}$$

 Smooth plate forced convection is similar to natural convection from
 a vertical plate; in both, fluid flows parallel to the plate's
 characteristic length axis, and is uniform across the plate's other
 axis.  Jaffer~\cite{thermo3010010} finds that stationary fluid
 conducts heat from the vertical plate with an effective Nusselt
 number $\Nuz={2^4/[\pi^{2}\,\root4\of2]}\approx1.363$; $\Nuz$ is a
 coefficient factor of both the static and flow-induced heat transfer
 terms.

 Flow-induced heat transfer grows with $\Nuz\,\Rey\,\fctol$.  Because
 this heat traverses boundary-layers, the~$\Pra$ dependence is more
 complex than $\Pra^{1/3}$.  In Jaffer~\cite{thermo3010010}, the
 vertical-plate natural convection dependence is
 $\root3\of{\Pra/\Xi(\Pra)}$, where $\Xi(\Pra)$ is defined using
 $\ell^p$-norm (discussed in \ref{Combining Transfer Processes})
 Formula~\eqref{eq:ell-p} with $p={\sqrt{1/3}}$:
$$\eqalignno{\left\|\varphi,\vartheta\right\|_p&\equiv\left[~|\varphi|^p+|\vartheta|^p\right]^{1/p}&\eqdef{eq:ell-p}\cr
\Xi(\Pra)&={\left\|1,~{0.5\over\Pra}\right\|_{\sqrt{1/3}}}
 &\eqdef{eq:Xi}\cr}$$

 In \figref{fig:Pr-factor}, $\root3\of{\Pra/\Xi(\Pra)}$ is asymptotically
 $\Pra^{1/3}$ at large~$\Pra$, and $\root3\of{2}\,\Pra^{2/3}$ at
 small~$\Pra$.  At small~$\Pra$, conduction does not amplify forced
 convection as it does for natural convection; the~$\Pra$ exponent
 should be 1.  An additional factor using the $\ell^3$-norm
 accomplishes this.  Formula~\eqref{eq:Pr-factors} is asymptotically
 $\root3\of{2}\,\Pra$ when~$\Pra\ll0.5$.  The ``$0.7\,\Pra^{0.6}$''
 trace shows that Formula~\eqref{eq:Pr-factors} has a slope close to
 Formulas~\eqref{eq:gas} and \eqref{eq:gases} for gases.
$$\root3\of{{\Pra\over\Xi(\Pra)}}\,\root3\of{{1\over\|1,1/\Pra\|_3}}\eqdef{eq:Pr-factors}$$

\vbox{\settabs 1\columns
\+\hfill\figscale{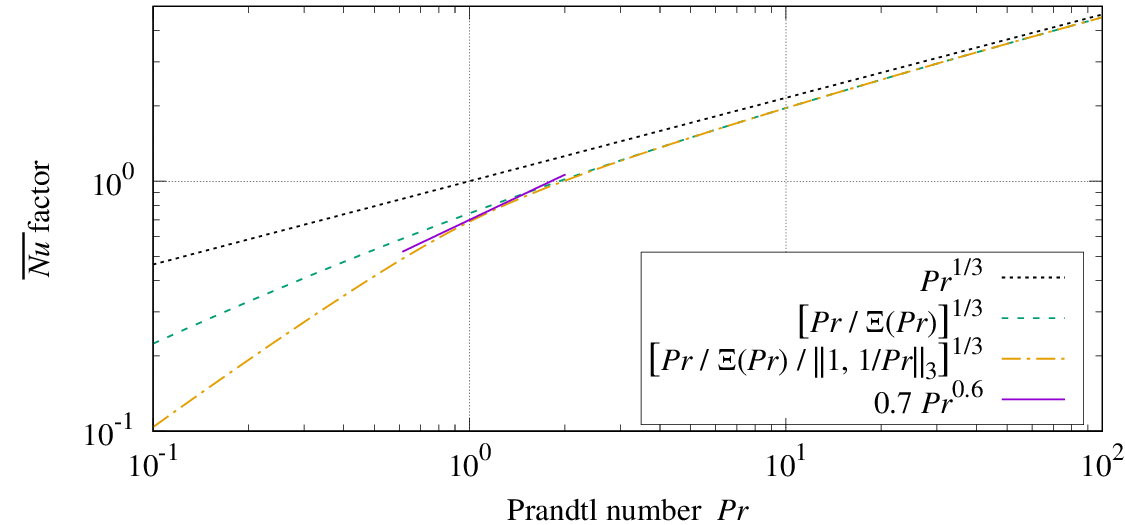}{351pt}\hfill&\cr
\+\hfill{\bf\figdef{fig:Pr-factor}\quad Smooth plate $\Nutol$ dependence on $\Pra$}\hfill&\cr
}

 The slope of Formula~\eqref{eq:Gnielinski} $\Nu(\Cft)$ decreases with
 increasing~$\Pra$; at large~$\Pra$, $\Nu$ is proportional to $\sqrt{\Cft}$
 ($\Nu\propto\sqrt{\Cft}$).  Recall
 from equations~\eqref{eq:f-C-define} and \eqref{eq:tau} that
 $\fcol\propto{u_\rho^2}$; hence $\sqrt{\fcol}\propto{u_\rho}$.  This
 indicates that transport through the boundary-layer restricts
 heat transfer at large~$\Pra$.  In order to reduce the asymptotic
 dependence from $\fctol$ to~$\sqrt{\fctol}$, the convection formula
 will include a factor which takes the square-root of an expression
 gating~$\fctol$ by~$\Pra$:
$$\sqrt{\Pra/\sqrt{162}+1\over\sqrt{162}\,\Pra\,{\fctol}+{1}}
  \qquad \sqrt{162}\equiv9\,\sqrt{2}\approx12.7 \eqdef{eq:transport-factor}$$

 Note that 12.7 is a coefficient in the Gnielinski
 Formula~\eqref{eq:Gnielinski}.

 The scaling for upstream heating was 1/2 in~$\Nurol$
 Formula~\eqref{eq:forced} for disruptive roughness; the turbulent
 boundary-layer reduces this interaction; $\sqrt{1/3}\approx0.577$
 appears correct in the smooth case.

\unorderedlist
 \li Formula~\eqref{eq:smooth-convection} is proposed for turbulent
 convection for all~$\Pra\ge0$ and $\Rey\gg\sqrt{3}\,{\rm e}$:
$$\Nutol={\Nuz\,\Rey\,{\fctol}\over\sqrt{3}}\,\sqrt{\Pra/\sqrt{162}+1\over\sqrt{162}\,\Pra\,\fctol+{1}}\,\root3\of{{\Pra/\Xi(\Pra)\over{\|1,1/\Pra\|_3}}}
 \eqdef{eq:smooth-convection}$$
\endunorderedlist

\subsection{Performance}
 \ref{Smoothness} compares~$\Nutol$
 Formula~\eqref{eq:smooth-convection} with measurements over a wide
 range of $\Pra$.

 Lienhard~\cite{10.1115/1.4046795} compares the Gnielinski-White
 convection formula with local measurements from studies of fluids
 with~$0.7\le\Pra\le257$ spanning $4000<\Rex<4.3\times10^6$.  The
 smallest turbulent $\Rex$ was $\approx10^5$.  Gas
 Formula~\eqref{eq:gas} is more accurate than Gnielinski-White for
 turbulent air ($\Pra\approx0.71$) at $\Rex<10^5$.

 \figrefs{fig:smooth-Pr} and \figrefn{fig:smooth-Re} show $\Nutol$ versus
 $\Pra$ and $\Rey$, respectively.  The ``present work'' traces are
 Formula~\eqref{eq:smooth-convection}.  The ``averaged'' traces use
 Formula~\eqref{eq:Nu-average-smooth} to numerically average the
 composition of the Gnielinski Formula~\eqref{eq:Gnielinski} with the
 White Formula~\eqref{eq:White-smooth}.
$$\Nuol(\Rey)=\int_{\Rez}^{\Rey}{\Nu(\Rex)\over\Rex}\,\diff\Rex\eqdef{eq:Nu-average-smooth}$$

\vbox{\settabs 1\columns
\+\hfill\figscale{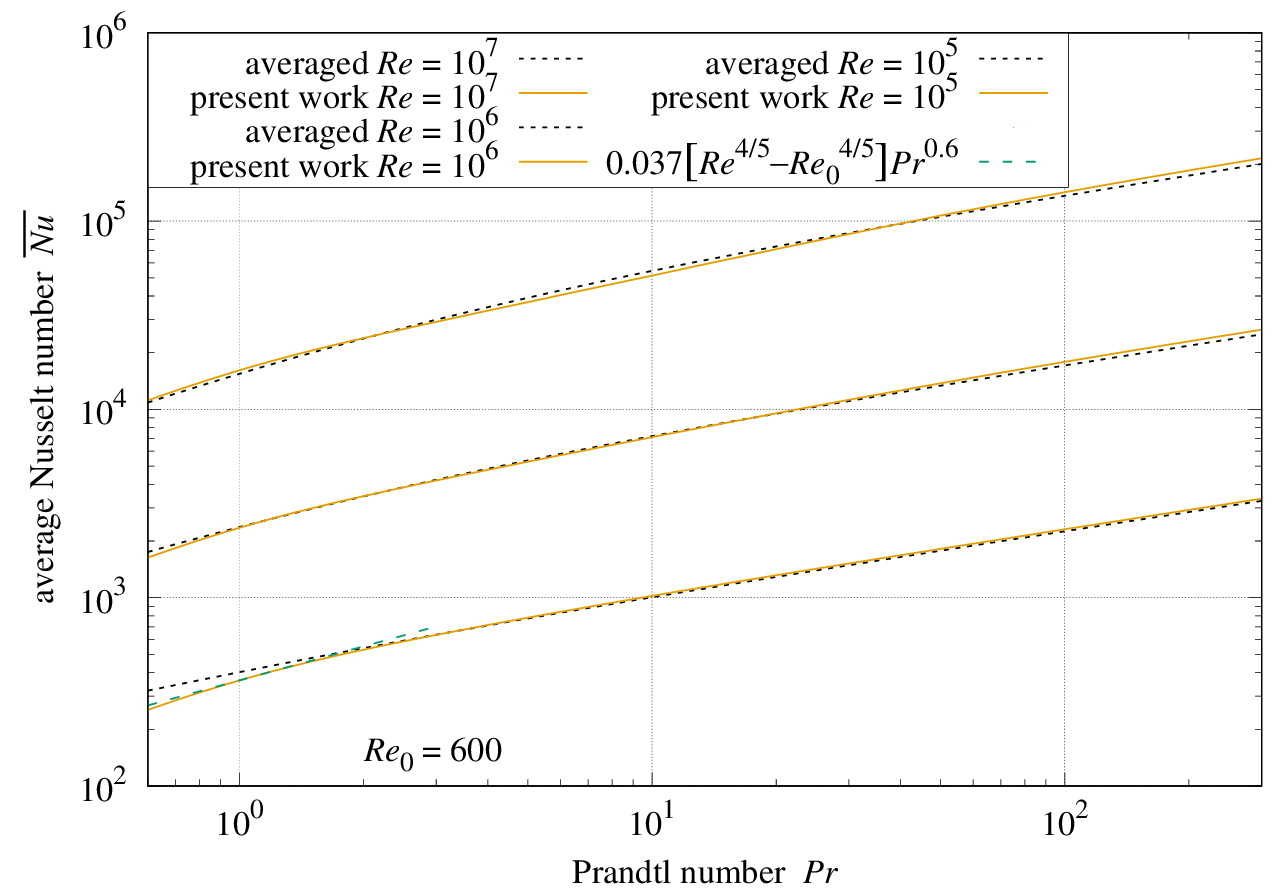}{360pt}\hfill&\cr
\+\hfill{\bf\figdef{fig:smooth-Pr}\quad Smooth plate average turbulent heat transfer versus $\Pra$ by $\Rey$}\hfill&\cr
}

\vbox{\settabs 1\columns
\+\hfill\figscale{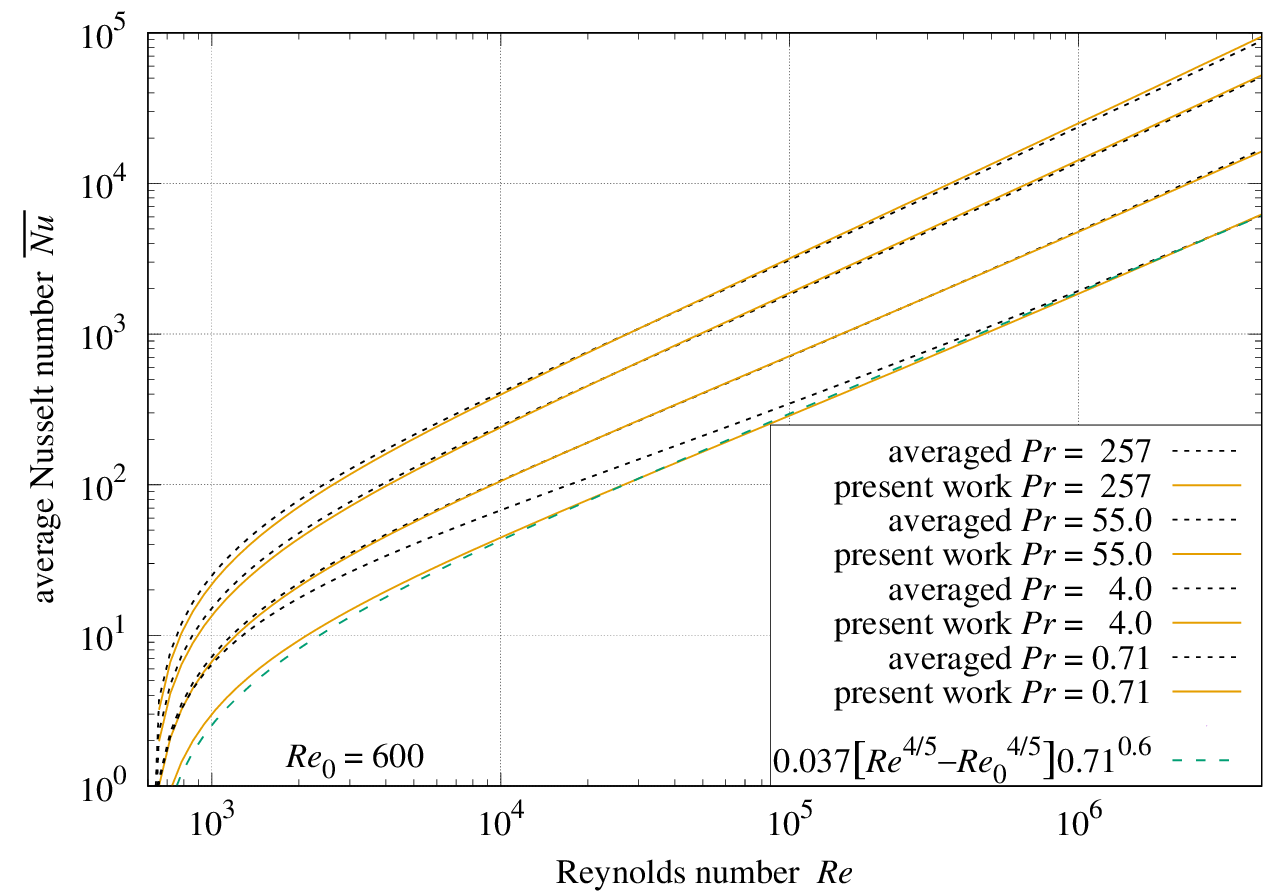}{360pt}\hfill&\cr
\+\hfill{\bf\figdef{fig:smooth-Re}\quad Smooth plate average turbulent heat transfer versus $\Rey$ by $\Pra$}\hfill&\cr
}

\vfill\eject

 \figref{fig:smooth-Re_x} shows Gnielinski
 Formula~\eqref{eq:Gnielinski} and local convection $\Nut$
 Formula~\eqref{eq:smooth-local-convection} versus~$\Rex$.
$$\Nut(\Rex)=\Rex\,{\diff\Nutol(\Rex)\over\diff\Rex}\eqdef{eq:smooth-local-convection}$$

\unorderedlist
 \li $\Nutol$ Formula~\eqref{eq:smooth-convection} matches the
 numerically averaged Gnielinski-White formula within~$\pm6.6\%$ over
 the range $10^5<\Rey<4.3\times10^6$ with~$4.0\le\Pra\le257$.

 \li At $\Pra=0.71$, $\Nutol$ matches gases Formula~\eqref{eq:gases}
 within~$\pm4\%$ over the range $10^4<\Rey<4.3\times10^6$.

 \li Over the same ranges, local convection~$\Nut$
 Formula~\eqref{eq:smooth-local-convection} matches
 Formulas~\eqref{eq:Gnielinski} and~\eqref{eq:gas} within~$\pm7.1\%$.
\endunorderedlist

\vbox{\settabs 1\columns
\+\hfill\figscale{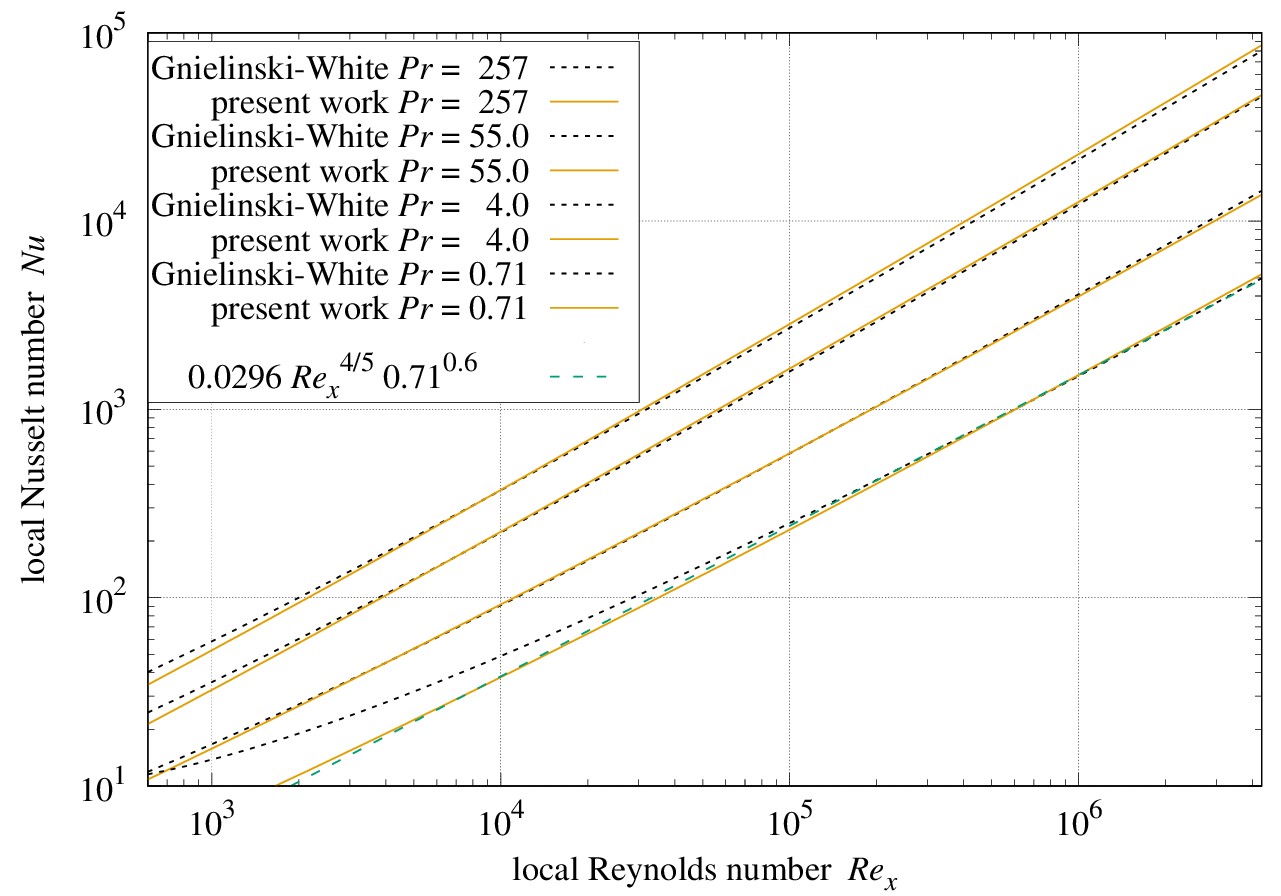}{360pt}\hfill&\cr
\+\hfill{\bf\figdef{fig:smooth-Re_x}\quad Smooth plate local turbulent heat transfer}\hfill&\cr
}

\subsection{Laminar Forced Convection}
 Formula~\eqref{eq:Colburn} is the
 Reynolds-Colburn~\cite{COLBURN19641359} analogy relating laminar
 friction to forced convective heat transfer.  Applying it to $\fclol$
 Formula~\eqref{eq:laminar-friction}, and scaling by the reduction in
 characteristic length due to an unheated starting band
 ${x_u/L}\equiv\Reu/\Rey$ , yields the laminar forced convection
 Formula~\eqref{eq:laminar-convection} given $\Rey\ge0$, $\Pra\ge0$, and
 ${0\le{x_u}\ll{L}}$.
$$\eqalignno{
  \Nuol(\Rey)&={\fcol(\Rey)}\,\Rey\,\Pra^{1/3}/2&\eqdef{eq:Colburn}\cr
  \Nulol(\Rey)&={0.664\,\Rey\,\Pra^{1/3}\over\sqrt{\Rey}+\sqrt{\Rez}}\left\{{1-\left\|1,{x_u\over L}\right\|_{-2}}\right\}&\eqdef{eq:laminar-convection}\cr
 }$$

\unorderedlist

 \li \ref{Smoothness} tests $\Nutol$
 Formula~\eqref{eq:smooth-convection} and $\Nulol$
 Formula~\eqref{eq:laminar-convection} extensively.
\endunorderedlist

\section{Onset of Rough Flow}

\unorderedlist
 \li Periodicity organizes roughness such that the approximate onset
   $\Rey$ of rough flow can be found from $\varepsilon$, $L$,
   and~$L_P$.
\endunorderedlist

 For an isotropic, periodic roughness with $0<\varepsilon<L_P\ll{L}$,
 there must be some value~$\Rel>0$ such that when $0<\Rey<\Rel$, there
 is only laminar or turbulent fluid flow along the plate.

 The boundary-layer is thinnest at the leading edge.  For isotropic,
 periodic roughness, any disruption will start within the leading band
 ($0<x<L_P$) of roughness.  This investigation considers a
 boundary-layer disrupted when $\varepsilon>\delta_2(L_P)$,
 where~$\delta_2(x)$ is the boundary-layer momentum thickness at $x$.

\subsection{Momentum Thickness}
 $\delta_2(x)$ is the thickness of bulk flow having the same momentum
 flow rate as the plate's boundary-layer at~$x$.  $\delta_2$ is not
 directly measurable.  Schlichting \cite{schlichting2014} gives
 the momentum thickness of laminar and turbulent
 boundary-layers as Formulas~\eqref{eq:Schlichting-d2l}
 and~\eqref{eq:Schlichting-d2s}, respectively:
$$\eqalignno{
  \delta_{2\lambda}(x)&={0.664\,x\over\Rex^{1/2}}=0.664\,\sqrt{\Rex}\,{L\over\Rey}&\eqdef{eq:Schlichting-d2l}\cr
  \delta_{2\tau}(x)&={0.036\,x\over\Rex^{1/5}}=0.036\,{\Rex}^{4/5}{L\over\Rey}&\eqdef{eq:Schlichting-d2s}\cr}$$
 Laminar $\delta_{2\lambda}$ derives from the Blasius
 boundary-layer model.  Turbulent $\delta_{2\tau}$ is less
 certain.

 Momentum thickness~$\delta_2(x)$ is a local property of the fluid
 flow.  In order to work locally with $\Rev$
 Formula~\eqref{eq:roughness-Re},
 change $L\to x$ and $\Rey\to\Rex$; then solve $\Rev=1$
 for~$x/\varepsilon$:
$$%
  {\Rex\over\sqrt{3}}={x\over\varepsilon}\,\ln{x\over\varepsilon}
  \qquad{x\over\varepsilon}={\exp\left(\Wz\left(\Rex\over\sqrt{3}\right)\right)}\eqdef{eq:x/e}$$

 $\Rex$ is proportional to $x$ ($\Rex\propto x$); hence, the
 turbulent momentum thickness~$\delta_2(x)$ should be proportional to
 the product of~$x$ and $u_\rho/u$ Formula~\eqref{eq:ratio-log}.
 Eliminating ${x/\varepsilon}$ using Formula~\eqref{eq:x/e}:
$$\delta_2(x)\propto{x\over\sqrt{3}\,\ln{(x/\varepsilon)}}
  ={x\over\sqrt{3}\;\Wz\left(\Rex/\sqrt{3}\right)}\eqdef{eq:prop2-d2}$$

 The proposed Formula~\eqref{eq:d2-smooth} coefficient is
 $1/3^3\approx0.0370$, which may relate to
 $\varphi^\varphi\equiv\exp(\varphi\,\ln{\varphi})$.
$$\delta_2(x)={x\over3^3\;\Wz\left({\Rex/\sqrt{3}}\right)}\eqdef{eq:d2-smooth}$$

 \figref{fig:onset_x} demonstrates that
 ``turbulent~$\delta_2$'' Formula~\eqref{eq:d2-smooth} and
 ``Schlichting $\delta_{2\tau}$'' Formula~\eqref{eq:Schlichting-d2s}
 nearly match between the origin and the intersection of the laminar
 and turbulent curves at:
$$\Rex=\left[{0.664\over0.036}\right]^{10/3}\approx16579\eqdef{eq:Re_x}$$

\unorderedlist
 \li Thus, $\delta_{2\tau}(L_P)$ is a reasonable approximation
 for~$\delta_2(L_P)$ in the leading band of roughness.
\endunorderedlist

\subsection{Flow Mode $\Rey$ Bounds}
 \figref{fig:onset} shows the leading band momentum thickness of
 laminar and turbulent flows along a 1~m long plate
 versus~$\Rey$.  The intersecting laminar~$\delta_{2\lambda}$ and
 turbulent~$\delta_2$ curves partition the graph into four
 regions labeled I, II, III, and IV.

\unorderedlist
\li When the point at coordinates $[\Rey,\varepsilon]$ is in region~I,
 the roughness is sufficient to disrupt both laminar and
 turbulent flow; hence, leading band fluid flow will be rough.

\li When $[\Rey,\varepsilon]$ is in regions~II or III, the roughness
 is not large enough to significantly disrupt laminar flow; hence,
 leading band fluid flow will be laminar, possibly transitioning to
 turbulent.

\li When $[\Rey,\varepsilon]$ is in region~IV, the roughness would
 be sufficient to disrupt laminar flow, but not large enough to
 disrupt turbulent flow; hence, leading band fluid flow
 would be turbulent.
\endunorderedlist

\medskip
\vbox{\settabs 2\columns
\+\figscale{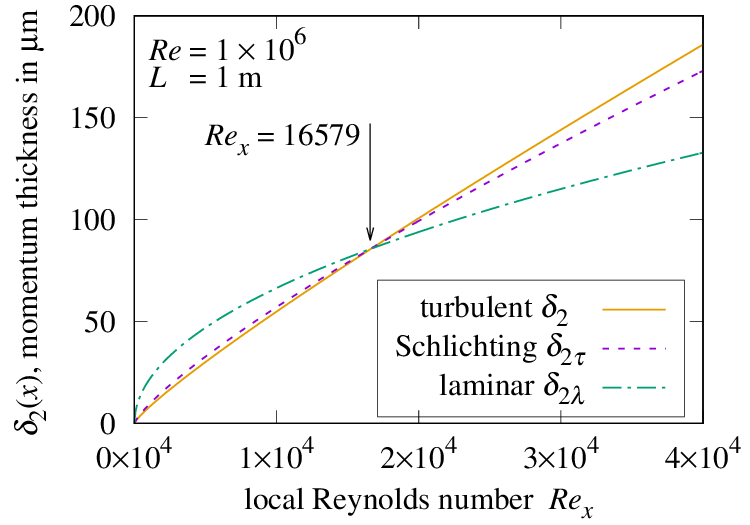}{234pt}\hfill
 &\hfill\figscale{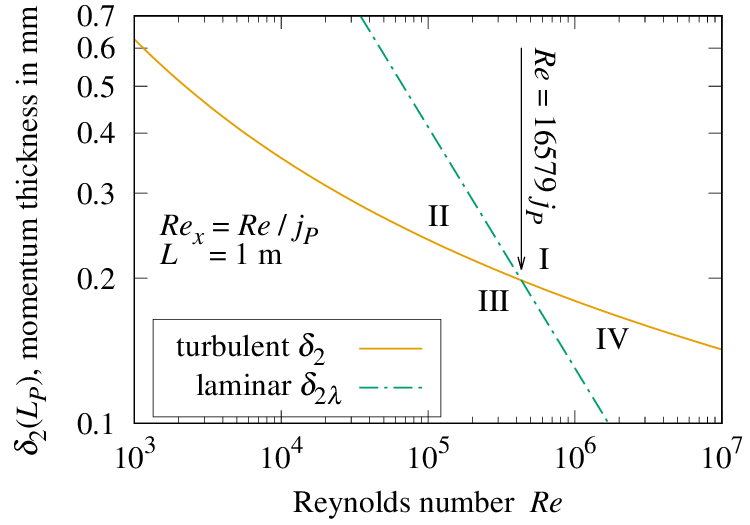}{234pt}&\cr
\+\hfill{\bf\figdef{fig:onset_x}\quad Smooth plate $\delta_2(x)$ versus $\Rex$}\hfill
 &\hfill{\bf\figdef{fig:onset}\quad Smooth plate $\delta_2(L_P)$ versus $\Rey$}\hfill&\cr
\medskip}

 With $\delta_{2\lambda}=\varepsilon$ and $x=L_P$, solve
 Formula~\eqref{eq:Schlichting-d2l} for the laminar
 upper-bound~$\Rel$:
$$\Rel=\left[{0.664\over\varepsilon}\right]^2\,{L_P\,L}\eqdef{eq:Re-laminar}$$

 With $x=L_P$ and $\delta_2(x)=\varepsilon$ in
 Formula~\eqref{eq:d2-smooth} with $\Rey\gg{\sqrt{3}\,{\rm e}\,L/L_P}$:
$$ \Wz\left({\Rey\,L_P\over\sqrt{3}\,L}\right)={L_P\over3^3\;\varepsilon}
  \eqdef{eq:Re-smooth0}$$

 The inverse of $\varphi=\Wz(\vartheta)$ is
 $\vartheta=\varphi\,\exp{\varphi}$.  Solving for the
 turbulent upper-bound $\Ret=\Rey$:
$$ \Ret={\sqrt{3}\,L\over3^3\;\varepsilon}\,\exp{L_P\over3^3\;\varepsilon}\eqdef{eq:Re-smooth}$$

 Equating $\Rel$ and $\Ret$ yields their intercept
 (found numerically) at ${L_P/\varepsilon}\approx194.3$.

 The combination of $L_P/\varepsilon>194.3$ and $L\gg L_P$ required by
 region~IV operation
 will be rare; $\Ret<\Rel$ will hold for nearly all
 isotropic, periodic roughnesses.

\section{Plateau Roughness}

 Self-similar roughness disrupts a boundary-layer at all scales.  At a
 minimum, isotropic, periodic roughness disrupts a boundary-layer only
 every period~$L_P\ll L$.  Between these disruptions the RMS height-of-roughness
 must be smaller than $\varepsilon$ because $L_P$ is the dominant
 period.  If this inter-roughness region is flat, parallel to the
 flow, and that flat is the tallest feature of each roughness cell,
 then a boundary-layer can grow along it.

\unorderedlist
 \li If an isotropic, periodic roughness lacks flats parallel to the
 flow, then its shearing stress comes from the same flow--roughness
 interaction as a self-similar roughness induces, and the
 skin-friction coefficient will be~$\fcrol$ Formula~\eqref{eq:f-C}.
\endunorderedlist

 Distinct flow mode regions can form along plates whose roughness
 peaks are all co-planar (at the same elevation) plateaus.  With
 $\Rey>\max(\Ret,\Rel)$ producing
 rough flow in the leading band of the plate, turbulent flow
 occurs downstream from where~$\delta_2(x)$ is large enough
 to bridge the gaps.  The skin-friction drag from the downstream portion of the
 surface will be proportional to~$\fctol$, not the constant $\fcrol$
 associated with rough flow.  The combined
 skin-friction drag formula is developed in~\ref{Combining Transfer Processes}.

 Informally, a ``plateau roughness'' is an isotropic, periodic
 roughness with most of its area at its peak elevation.  Of particular
 interest are plateau roughnesses where each cell contains a single
 continuous plateau area whose boundary has a convex perimeter within
 the cell.  This will either be an array of ``islands'' whose tops are
 all co-planar, or an array of ``wells'' dropping below an otherwise
 flat plane.

\subsection{Plateau Islands}
 Consider a smooth flat plate etched with a square grid of grooves
 subjected to a $\Rey>\max(\Ret,\Rel)$ flow.  When the
 boundary-layer is disrupted by a groove perpendicular to the flow,
 the turbulent boundary-layer restarts at the leading edge of
 the next island.  At the scale of the roughness period~$L_P$, the
 momentum thickness of the boundary-layer grows from 0 to nearly the
 $L$-scale~$\delta_2(x)$ value (depending on the size of the island).
 If~$\delta_2$ grows to exceed~$\varepsilon$, then the rest of the
 plate (to its trailing edge) will have a turbulent friction
 coefficient proportional to~$\fctol$, but with
 characteristic length~$L_P$.

 Along isotropic roughness, the growth of $\delta_2$ depends on
 plateau size, but not on orientation.
 An isotropic size metric is needed.
 In natural convection from an upward-facing horizontal
 plate~\cite{goldstein1973natural,lloyd1974natural}, the (isotropic)
 characteristic length metric $\Ls=A^*/{p^*}$, where $A^*$ is the
 convex region's area and ${p^*}$ is its perimeter length.  For a
 regular polygon or circle, $\Ls=r/2$, where~$r$ is the minimum radius
 of the regular polygon or circle.

 In order to find the island $\Rex$ threshold $\ReI$,
 multiply both sides of
 equation~\eqref{eq:d2-smooth} by $3^3\Rey/[\sqrt{3}\,L]$.  This
 allows~$\Rex$ to be isolated using the Lambert~$\Wz$ function
 identity $\varphi/\Wz(\varphi)=\exp\Wz(\varphi)$:
$${3^3\Rey\over\sqrt{3}\,L}\,\delta_2
 ={\exp\Wz\left({\Rex\over\sqrt{3}}\right)}\eqdef{eq:grooved2}$$

  The boundary-layer thickness needed to bridge the gap grows with
  $\varepsilon$ and shrinks with increasing $\Ls/\varepsilon$, suggesting
  $\delta_2=\varepsilon^2/\Ls$.
  The~$\Rey$ strength needed to reach $\delta_2$ thickness at $x=L_P$
  grows strongly with~$L/L_P$.  Letting~$\Rey=[L/L_P]^3$ in
  equation~\eqref{eq:grooved2}, then taking the logarithm of both
  sides:
$$\ln{3^3\,\varepsilon^2\,L^2\over\sqrt{3}\,\Ls\,L_P^3}=\Wz\left({\Rex\over\sqrt{3}}\right)\eqdef{eq:grooved3}$$

 The inverse of $\varphi=\Wz(\vartheta)$ is
 $\vartheta=\varphi\,\exp{\varphi}$.  Solving
 Formula~\eqref{eq:grooved3} for $\ReI=\Rex$:
$$\ReI={3^3\,\varepsilon^2\,L^2\over \Ls\,L_P^3}\,\ln{3^3\,\varepsilon^2\,L^2\over\sqrt{3}\,\Ls\,L_P^3}
  \qquad{[4\,\Ls]^2\over L_P^2}>{1\over2}\eqdef{eq:Re_I}$$
 With wide enough gaps, the islands are too narrow to support
 turbulent flow bridging the gaps, leading to the inequality in
 Formula~\eqref{eq:Re_I}.  Formula~\eqref{eq:Re_I} is tested against
 two square-grooved plates in \ref{Rough Heat Transfer Measurements}.

\subsection{Plateau Wells}
 Laying a perforated sheet on a flat plate turns its holes into wells.
 $\Ls=A^*/p^*$.  Fluid is forced up after diving into a well; instead
 of $\Rey=[{L/L_P}]^3$ in Formula~\eqref{eq:grooved3}, let
 $\Rey=L^3/[{2\,L_P}]^3$,
 leading to the wells $\Rex$ threshold $\ReW$:
$$\ReW={3^3\,\varepsilon^2\,L^2\over 2^3\,\Ls\,L_P^3}\,\ln{3^3\,\varepsilon^2\,L^2\over2^3\,\sqrt{3}\,\Ls\,L_P^3}
\qquad{[4\,\Ls]^2\over L_P^2}<{1\over2}\eqdef{eq:Re_W}$$
 With sufficiently wide wells, the flats between them are too narrow to
 support turbulent flow bridging the wells, leading to the
 inequality in Formula~\eqref{eq:Re_W}.  Formula~\eqref{eq:Re_W} is tested
 with perforated sheets
 in \ref{Rough Skin-Friction Measurements}.

\subsection{Openness}
 Let ``openness'' $0<\Omega<1$ be the non-plateau area per cell area
 ratio; $1-\Omega$ is thus the ``upper area fraction''
 of \ref{Periodic Roughness}.
 Let $S_{s,t}$ be a $w\times{w}$ matrix of elevations.
 The span of elevations accepted as plateau must be a length smaller
 than~${\varepsilon}$ and decrease with increasing $L_P/\varepsilon$, which
 suggests ${\varepsilon^2/L_P}$:
$$\Omega\approx{1\over{w^2}}\sum_{t=0}^{w-1}\,\sum_{s=0}^{w-1}\cases{
        1,& if $S_{s,t}<\max(S)-{\varepsilon^2/L_P}$;\cr
        0,& otherwise.\cr
        }\eqdef{eq:Omega}$$

 This allows a quantitative definition of plateau roughness:
\unorderedlist
 \li ``Plateau roughness'' is an isotropic, periodic roughness with
   $\Omega<1/2$.
\endunorderedlist

 An isotropic, periodic roughness which is not plateau roughness
 disrupts nascent boundary layers every $L_P$, but lacks the
 flat peaks necessary for turbulent layer growth.  Thus, this
 investigation proposes:

\unorderedlist
 \li When $\Omega>1/2$ and $\Rey>\max(\Ret,\Rel)$, flow along the
   entire surface will be rough.
\endunorderedlist

 Note that $\Omega<1/2$ does not replace the
 inequalities~(\eqrefn{eq:Re_I}, \eqrefn{eq:Re_W}); it is an
 additional constraint.  However, they are related by ``circularity''.

\subsection{Circularity}
 For island area ${A^*}$ with perimeter ${p^*}$, circularity
 $o^*={4\,\pi\,A^*/{p^*}^2}$ takes its maximum value, 1, in disks; it
 is $\pi\,\sqrt{3}/6\approx0.907$ in hexagons, $\pi/4\approx0.785$ in
 squares, and $\pi\,\sqrt{3}/9\approx0.605$ in equilateral triangles.
 Similarly, for square cell area $A$ with perimeter $p$, circularity
 $o={4\,\pi\,A/{p}^2}$.  Scaling by the circularity ratio derives
 ${[4\,\Ls]^2/L_P^2}$ from $[1-\Omega]$.

$$[1-\Omega]\,{o^*\over o}
 ={{A^*}^2\over{p^*}^2}\,{p^2\over A_P^2}
 ={[4\,\Ls]^2\over L_P^2}\eqdef{eq:Omega-islands}$$

 Formula~\eqref{eq:Omega-islands} is the same within a hexagonal cell,
 $A=\sqrt{3}/2\,L_P^2$ and perimeter $p=2\,\sqrt{3}\,L_P$.

 The plateau wells roughness formula replaces
 $[1-\Omega]$ with $\Omega$:

$$\Omega\,{o^*\over o}
 ={{A^*}^2\over{p^*}^2}\,{p^2\over A_P^2}
 ={[4\,\Ls]^2\over L_P^2}\eqdef{eq:Omega-wells}$$

 The slope of each trace in \figref{fig:openness} is its ${o^*/o}$
 ratio.  The ``square/square'' trace for a square array of square
 posts has slope~1.  The ``disk/hexagon'' trace for a hexagonal array
 of circular wells has slope $\sqrt{12}/\pi\approx1.1$.  The
 ``square/hexagon'' trace for a hexagonal array of square posts has
 slope $\sqrt{3}/2\approx0.87$.  Each trace spans the bounds for that
 bi-level configuration discovered in \ref{Periodic Roughness}.

\smallskip
\vbox{\settabs 1\columns
\+\hfil\figscale{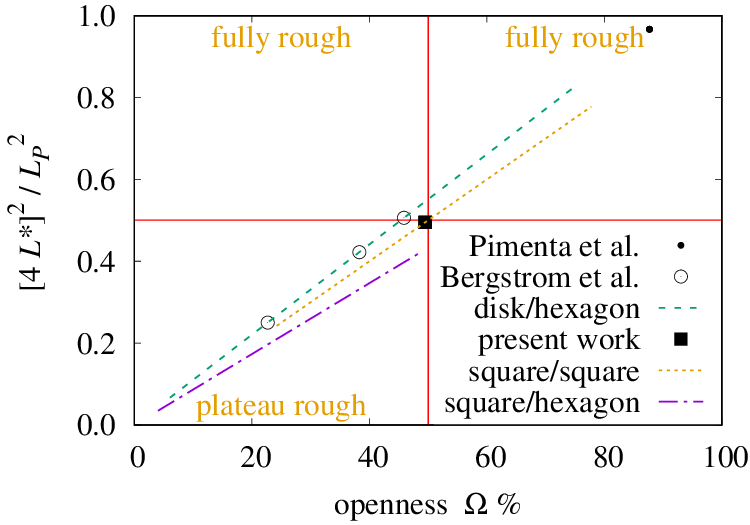}{234pt}&\cr
\+\hfil{\bf\figdef{fig:openness}\quad $[4\,\Ls]^2/L_P^2$ versus $\Omega$}&\cr
}

 Also plotted are the $\Omega$ values of experiments by
 Pimenta \etal \cite{A014219}, Bergstrom \etal \cite{Bergstrom_2005},
 and this investigation.
 The experiments of
 \refs{Local Skin-Friction Coefficients},
 \refn{Rough Skin-Friction Measurements},
 and \refn{Rough Heat Transfer Measurements}
 find friction and heat transfer consistent with the
 type of flow named in three of the quadrants.  The lower right
 quadrant
 would be reached when the convex region circularity is less than the
 cell circularity.  But \ref{Periodic Roughness} found that such
 bi-level surfaces did not qualify as isotropic, periodic roughness
 when $\Omega>1/2$.  The Pimenta \etal plate was not bi-level
 roughness.

\unorderedlist
 \li The experimental data locate the $[4\,\Ls]^2/L_P^2$ threshold
 as being between 0.495 and 0.506.
\endunorderedlist

\section{Combining Transfer Processes}

  This unnamed form appears frequently in heat transfer formulas:
$$F^p(\xi)=F_0^p(\xi)+F_\infty^p(\xi)\eqdef{eq:mixing}$$

  Churchill and Usagi~\cite{AIC:AIC690180606} stated that such
  formulas are ``remarkably successful in correlating rates of
  transfer for processes which vary uniformly between these limiting
  cases.''
 Convection and skin-friction are transfer processes.
 Convection transfers heat; skin-friction drag ($\propto\Rey\,\fc$)
 transfers momentum.

\subsection{The $\ell^p$-norm}
  Requiring $F_0(\xi)\ge0$ and $F_\infty(\xi)\ge0$, and taking
  the~$p$th root of both sides of equation~\eqref{eq:mixing} yields a
  vector-space functional form known as the
  $\ell^p$-norm, which is notated $\|F_0~,~F_\infty\|_p$~:
$$\left\|F_0~,~F_\infty\right\|_p=\left[~|F_0|^p+|F_\infty|^p\right]^{1/p}\eqdef{eq:l^p}$$

  Norms generalize the notion of distance.  Formally, a vector-space
  norm obeys the triangle inequality: $\|y~,~z\|_p\le|y|+|z|$, which
  holds only for $p\ge1$.  Both prior and present works find $p<1$
  useful as well.

 Let $y\ge0$ and $z\ge0$. When $p>1$, the processes compete and
 $\|y,z\|_p\ge\max(y,z)$; the most competitive is
 $\|y,z\|_{+\infty}\equiv\max(y,z)$.  The $\ell^2$-norm is equivalent
 to root-sum-squared; it models perpendicular competitive processes
 such as forced and natural convection from the present apparatus
 plate sides in Formula~\eqref{eq:U_S}.

 The $\ell^1$-norm models independent processes;
 $\|y,z\|_1\equiv{y+z}$.  It appears in
 Formulas~(\eqrefn{eq:staged-abrupt}, \eqrefn{eq:staged-friction}, \eqrefn{eq:staged-convection}).

 When $0<p<1$, the processes cooperate and $\|y,z\|_p\ge{y+z}$.
 Cooperation between conduction and flow-induced heat transfer uses
 the $\ell^{1/2}$-norm in natural convection
 (\cite{thermo3010010}).  Formula~\eqref{eq:Xi} uses the
 $\ell^{\sqrt{1/3}}$-norm from natural convection in forced convection.

 When $p<0$, $\|y,z\|_p\le\min(y,z)$, with
 the transition sharpness controlled by $p$; the extreme is
 $\|y,z\|_{-\infty}\equiv\min(y,z)$.  Negative $p$ can model a single
 flow through serial processes; the most restrictive process limits
 the flow.
 The $\ell^{-2}$-norm
 appears in the unheated starting length term of
 Formula~\eqref{eq:laminar-convection}, and in fan-speed
 Formula~\eqref{eq:Re-tunnel}.  The $\ell^{-4}$-norm appears in
 laminar-turbulent transition Formulas
 (\eqrefn{eq:bi-level-friction}, \eqrefn{eq:bi-level-convect}, \eqrefn{eq:wells-local}, \eqrefn{eq:staged-friction}, \eqrefn{eq:staged-convection}).

\subsection{Flow Modes}
 Isotropic, periodic surfaces with $\Ret<\Rel$ and $\Omega>1/2$
 shed the rough flow of
 Formulas~(\eqrefn{eq:f-C},~\eqrefn{eq:f-c-local},~\eqrefn{eq:forced})
 when $\Rey>\Rel$.
 Plateau roughness ($\Omega<1/2$) sheds either smooth or rough
 flow, or a combination.  \tabref{tab:flow mode} proposes the
 behaviors of plateau islands and plateau wells roughnesses.
 If the ${[4\,\Ls]^2/L_P^2}$ condition is not satisfied, then the
 $\Rex$ conditions split the plate at distance~$x$ from the leading
 edge into regions operating in different modes.
 Formula~\eqref{eq:Re_I} is the islands threshold $\ReI$;
 Formula~\eqref{eq:Re_W} is the wells threshold $\ReW$.

\centerline{\bf\tabdef{tab:flow mode}\quad{Flow modes for plateau roughness}}
\vbox{\settabs 4\columns
\toprule
\+\hfil{\bf Islands condition}&\hfil{\bf Plateau islands}&\hfil{\bf Wells condition}&\hfil{\bf Plateau wells}&\cr
\midrule
\+\hfil${[4\,\Ls]^2/L_P^2}<1/2$& rough Formulas~(\eqrefn{eq:f-C},~\eqrefn{eq:f-c-local},~\eqrefn{eq:forced})&\hfil${[4\,\Ls]^2/L_P^2}>1/2$& rough Formulas~(\eqrefn{eq:f-C},~\eqrefn{eq:f-c-local},~\eqrefn{eq:forced})&\cr
\+\hfil$\Rex<\ReI$& rough Formulas~(\eqrefn{eq:f-C},~\eqrefn{eq:f-c-local},~\eqrefn{eq:forced})&\hfil$\Rex<\ReW$& blend Formula~\eqref{eq:perf-mix-local}&\cr %
\+\hfil$\Rex>\ReI$& smooth Formulas~(\eqrefn{eq:posts},~\eqrefn{eq:posts-convect})&\hfil$\Rex>\ReW$& smooth Formula~(\eqrefn{eq:effective-smooth-area-local})&\cr %
\bottomrule
}

\subsection{Plateau Islands}
 The ``$\Rex>\ReI$'' flow mode is turbulent with
 characteristic length~$L_P$.  The island's plateau area is augmented
 by 1/2 of the non-plateau area and an area which grows with
 $\varepsilon$, combined using the $\ell^2$-norm because $\varepsilon$
 and $L_P$ are perpendicular:
$$\eqalignno{
  \fIol=&\left\{{1-\Omega}+\left\|{\Omega\over2},{2\,\varepsilon\,[4\,\Ls]\over L_P^2}\right\|_2\right\}
  {L\over L_P}\,\fctol\left({\Rey\,L_P\over L}\right)&\eqdef{eq:posts}\cr
  \NuIol=&\left\{{1-\Omega}+\left\|{\Omega\over2},{2\,\varepsilon\,[4\,\Ls]\over L_P^2}\right\|_2\right\}
  {L\over L_P}\,\Nutol\left({\Rey\,L_P\over L}\right)&\eqdef{eq:posts-convect}\cr
  }$$

 $\fcrol$ and $\Nurol$ are active at $\Rex<\ReI$; $\fIol$ and
 $\NuIol$ are active at $\Rex>\ReI$.  The ${\|\Rey, \ReI\|_{-4}}$
 term transitions between these parts gradually in
 Formulas~(\eqrefn{eq:bi-level-friction}, \eqrefn{eq:bi-level-convect}).
 The $\big[{\|\Rey, \ReI\|_{-4}/\Rey}\big]$ factor normalizes the
 characteristic length in Formula~\eqref{eq:bi-level-friction}.

$$\eqalignno{
  \fciol=&\fIol(\Rey)+\big[{\|\Rey, \ReI\|_{-4}/\Rey}\big]\big\{\fcrol-\fIol\left(\|\Rey, \ReI\|_{-4}\right)\big\}
        &\eqdef{eq:bi-level-friction}\cr
  \Nuiol=&\NuIol(\Rey)+\Nurol\left(\|\Rey, \ReI\|_{-4}\right)-\NuIol\left(\|\Rey, \ReI\|_{-4}\right)&\eqdef{eq:bi-level-convect}\cr
  }$$

 \figref{fig:islands} shows that $\Nuol$ and $\fcol$ are closely
 related in gases.  \ref{Rough Heat Transfer Measurements} compares
 convection heat transfer measurements from two bi-level plates with
 $\Nuiol$ Formula~\eqref{eq:bi-level-convect}.

\medskip
\vbox{\settabs 2\columns
\+\hfill\figscale{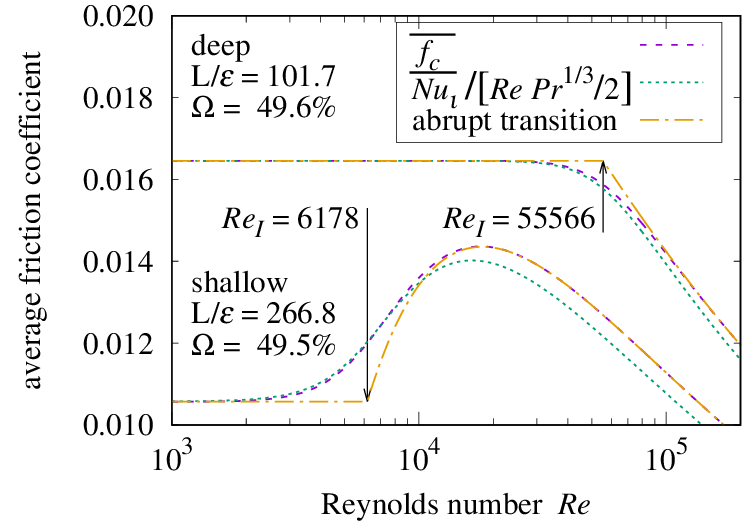}{230pt}\hfill&\hfill\figscale{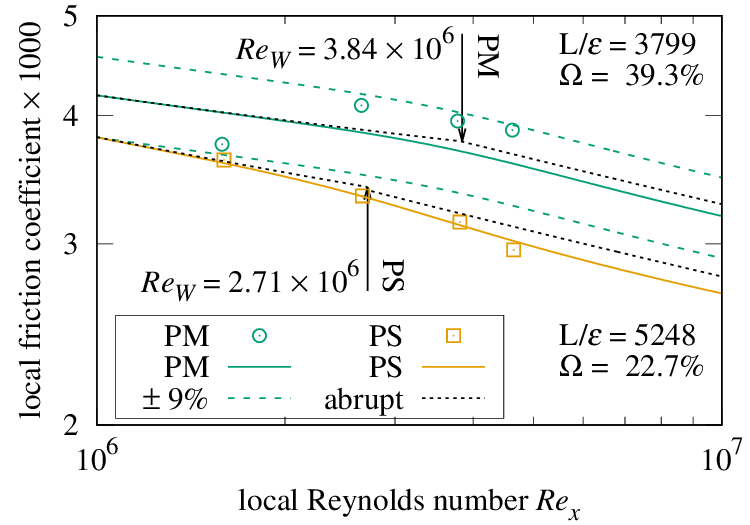}{230pt}\hfill\cr
\+\hfill{\bf\figdef{fig:islands}\quad Plateau islands friction}\hfill&\hfill{\bf\figdef{fig:wells}\quad Plateau wells friction}\hfill\cr
}

\subsection{Plateau Wells}
 In ``blend'' mode, the plateau sheds turbulent flow while its
 wells shed rough flow.  The effective friction coefficient is
 the area-proportional blend:
$$\eqalignno{
  \fcb&={\Omega\,\fcrol(L/\varepsilon)}+[1-\Omega]\,\fct(\Rey)&\eqdef{eq:perf-mix-local}\cr
  }$$

 Where $\Rex>\ReW$, the friction is turbulent, but with
 additional area $2\,\pi\,{\varepsilon\,[4\,\Ls]}$.  The well walls
 are perpendicular to the plateau, but only a portion of each well
 wall is parallel to the flow, which must divert to brush by
 both.  A strength between $\ell^{\sqrt{2}}$ and $\ell^2$ is needed;
 their geometric mean is $p=\root4\of8\approx1.682$:
$$\eqalignno{
  f_W&=\left\|1,{2\,\pi\,\varepsilon\,[4\,\Ls]/ L_P^2}\right\|_{\root4\of8}\,\fct(\Rey)&\eqdef{eq:effective-smooth-area-local}\cr
  }$$

 $\fcb$ is active at $\Rex<\ReW$; $f_W$ is active at $\Rex>\ReW$:
$$\eqalignno{
  \fcw&=f_W(\Rey)+\big[\|\Rey, \ReW\|_{-4}/\Rey\big]\big\{\fcb(\|\Rey, \ReW\|_{-4})-f_W(\|\Rey, \ReW\|_{-4})\big\}&\eqdef{eq:wells-local}\cr
  }$$

 The $\Rex=\ReW$ plane can split wells; thus the transition
 between~$\fcb$ and~$f_W$ flows must be gradual in
 Formula~\eqref{eq:wells-local}.  With the $\ell^{-4}$-norm, the
 PM$\pm9\%$ (expected uncertainty) curves in \figref{fig:wells} bound
 the PM measurements from Bergstrom \etal \cite{Bergstrom_2005}; the
 $\ell^{-2}$-norm does not.
 \ref{Rough Skin-Friction Measurements} compares
 local friction measurements of perforated sheets with
 Formulas~(\eqrefn{eq:effective-smooth-area-local}, \eqrefn{eq:wells-local}).

\subsection{Staged-Transition}
 When laminar and turbulent flow occupy distinct plate regions
 separated at $\Rey_c$, their transfers combine using the
 $\ell^1$-norm (addition).  An abrupt transition at $\Rey_c$ would
 behave as Formula~\eqref{eq:staged-abrupt}.
 In practice, staged-transitions are not abrupt, behaving as the
 $\ell^{-4}$-norm in Formulas
 (\eqrefn{eq:staged-friction}, \eqrefn{eq:staged-convection}).
 The subscript 4 is used to identify staged-transition formulas.

$$\eqalignno{
  &\quad\,\left\|{\Rey^\prime\over\Rey}\,\fclol(\Rey^\prime),~\fctol(\Rey)-{\Rey^\prime\over\Rey}\,\fctol(\Rey^\prime)\right\|_1\qquad\Rey^\prime=\min(\Rey,~\Rec)&\eqdef{eq:staged-abrupt}\cr
  \ffol&=\left\|{\Rey^\prime\over\Rey}\,\fclol(\Rey^\prime),~\fctol(\Rey)-{\Rey^\prime\over\Rey}\,\fctol(\Rey^\prime)\right\|_1\qquad\Rey^\prime=\|\Rey,~\Rec\|_{-4}&\eqdef{eq:staged-friction}\cr
  \Nufol&=\left\|\Nulol({\Rey_{4}}),~\Nutol(\Rey)-\Nutol({\Rey_{4}})\right\|_1\qquad\Rey_{4}=\|\Rey,\sqrt{2}\,\Rec\|_{-4}&\eqdef{eq:staged-convection}\cr
  }$$

 Formula~\eqref{eq:staged-convection} models staged-transition
 convection.  Convection's positive slope versus friction's negative
 slope requires scaling $\Rec$ by $\sqrt{2}$ so that the lower edge of
 the transition is at $\Rec$.  Formula~\eqref{eq:staged-convection} is
 tested in \ref{Smoothness}.

\section{Rough Skin-Friction Measurements}

 Bergstrom \etal \cite{Bergstrom_2005} has skin-friction coefficient
 measurements of sandpapers, woven wire meshes, and perforated sheets
 attached to a smooth plate, and also the
 $1.67{\rm\,m}\times1.16{\rm\,m}$ smooth plate alone.  Skin-friction
 measurements were derived from Pitot probe measurements of air
 velocity at locations which were 1.3~m downwind from the leading edge
 of the plate.  Bergstrom \etal \cite{Bergstrom_2005} estimated 5\% as
 the combined measurement uncertainty of the smooth surface friction
 coefficient, and 9\% for the rough surfaces.

 The measurement tables in~\cite{Bergstrom_2005} include a column for
 free-stream velocity, $U_e$.  In order to compute the Reynolds number
 $\Rey=U_e\,L/\nu$, the kinematic viscosity
 $\nu=16\times10^{-6}\rm~m^2/s$ was calculated for air at
 $20^\circ{\rm C}$, 25\%~RH, and 95~kPa, the mean atmospheric pressure
 at the University of Saskatchewan.

\subsection{Smooth Plate}
 Three of the four measurements labeled ``SM smooth'' are within 5\%
 of ``present work $\fct$'' in \figref{fig:Bergstrom-sand}.  The
 RMSRE versus $\fct$ Formula~\eqref{eq:f-c-smooth} is 4.9\%;
 RMSRE versus White's $\Cft$ Formula~\eqref{eq:White-smooth} is 9.6\%.

\subsection{Sandpaper}
Microscopic examination of coarse grades of sandpaper reveals glued
mounds of grits separated by canyons having depths which are several
times the mean grit diameter.  Sandpaper grit mean diameter is
standardized, but not the height-of-roughness of the mounds; it can
vary by manufacturer and lot.  The horizontal traces
in \figref{fig:Bergstrom-sand} show that skin-friction coefficients
which are independent of~$\Rex$, as in the present theory, can be
within the~9\% estimated measurement uncertainty of the data.

\subsection{Comparison With Sand-Roughness}
 The RMS height-of-roughness~$\varepsilon$ of sandpaper is much larger
 than $\varepsilon$ of sand-roughness with the same mean grit
 diameter.  For example, 40 grit sandpaper has a skin-friction
 coefficient consistent with $\varepsilon\approx503\,{\rm\mu{m}}$,
 while $k_S=425\,{\rm\mu{m}}$ sand-roughness would have
 $\varepsilon=k_S/5.333\approx80\,{\rm\mu{m}}$.

\medskip
\vbox{\settabs 1\columns
\+\hfill\figscale{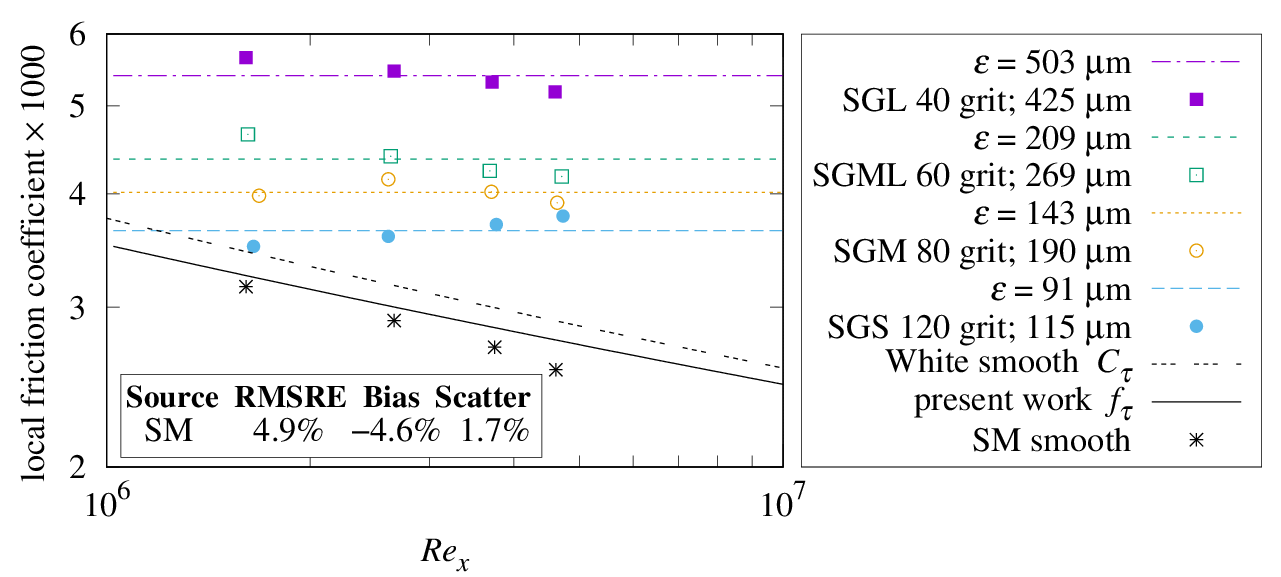}{360pt}\hfill&\cr
\+\hfill{\bf\figdef{fig:Bergstrom-sand}\quad Local $\fcr$ versus $\Rex$ of sandpaper}\hfill&\cr
}

\vfill\eject

\subsection{Woven Wire Mesh}

 Bergstrom \etal \cite{Bergstrom_2005} has local skin-friction
 coefficient measurements of woven wire meshes attached to a smooth
 plate.

\subsection{Mesh Openness}
 Woven wire meshes are specified by wire diameter~$d$ and wire center
 spacing~$s$.  Bergstrom \etal \cite{Bergstrom_2005} calculate mesh
 openness
 as $[s-\sqrt{2}\,d]^2/s^2$ instead of the $[s-d]^2/s^2$ formula used
 by manufacturers (neither metric is plateau openness $\Omega$).
 \tabref{tab:wire-mesh-dimensions}
 lists the dimensions and openness from~\cite{Bergstrom_2005} along
 with openness calculated both ways.  The WML and WMM meshes have
 $[s-\sqrt{2}\,d]^2/s^2$ values close to~\cite{Bergstrom_2005}.  The
 WMS mesh has $[s-\sqrt{2}\,d]^2/s^2\approx49\%$, versus 44\%
 from~\cite{Bergstrom_2005}.  If the 1.68~mm spacing were instead
 1.48~mm, WMS would have $[s-\sqrt{2}\,d]^2/s^2\approx44\%$, but
 significantly less friction coefficient than the WMS measurements
 in \figref{fig:Bergstrom-mesh}.  A 0.36~mm wire diameter has
 conventional openness $[s-d]^2/s^2\approx44\%$ and matches the WMS
 data and the WMM trace and data.

\centerline{\bf\tabdef{tab:wire-mesh-dimensions}\quad{Wire mesh dimensions}}
\vbox{\settabs 6\columns
\toprule
\+\hfil\bf Wire diameter $d$&\hfil\bf Spacing $s$&\hfil\bf $[s-d]^2/s^2$&\hfil\bf $[s-\sqrt{2}\,d]^2/s^2$&\hfil{\bf From} \cite{Bergstrom_2005}&\hfil{\bf Tag}&\cr
\midrule
\+\hfil1.04 mm&\hfil3.68 mm&\hfil52\%&\hfil36\%&\hfil35\%&\hfil WML&\cr
\+\hfil0.58 mm&\hfil1.77 mm&\hfil45\%&\hfil29\%&\hfil30\%&\hfil WMM&\cr
\+\hfil0.36 mm&\hfil1.68 mm&\hfil62\%&\hfil49\%&\hfil$\underline{44\%}$&\hfil WMS&\cr
\+\hfil0.36 mm&\hfil1.48 mm&\hfil58\%&\hfil$\underline{44\%}$ &\cr
\+\hfil0.56 mm&\hfil1.68 mm&\hfil$\underline{44\%}$&\hfil28\% &\cr
\bottomrule
}

\subsection{Gaps}
There are periodic gaps between the wires and the plate; so the
mesh-plate combination is not strictly a roughness.  With the gaps
filled, the RMS height-of-roughness~$\varepsilon$ would be:
$$\eqalignno{&z(x,y)=\sqrt{{d^2\over4}-x^2}+d-{d\over2}\,\cos{\pi\,y\over s}&\cr
  &\overline{z}={4\over s^2}\,\int_{d/2}^s\int_0^{d/2}z(x,y)\,\diff{x}\,\diff{y}&\cr
  &\varepsilon={4\over s^2}\int_{d/2}^s\int_0^{d/2}\left|z(x,y)-\overline{z}\right|^2\,\diff{x}\,\diff{y}+{[s-d]^2\,\overline{z}^2\over s^2}&\eqdef{eq:mesh-eps}\cr}$$

The periodic gaps between wires and the plate increase the flow's
shearing stress.  Scaling~$\varepsilon$ by the square root of the
filled-gap per empty-gap side area ratio is an increase of
about~$26\%$ for these meshes:
$$\varepsilon'=\varepsilon\,\sqrt{12\,s+\pi\,d\over8\,s}\eqdef{eq:mesh-gap}$$
The ``unscaled 1.04 mm, 3.68 mm'' trace in \figref{fig:Bergstrom-mesh} shows
the predicted WML friction without this scaling.

\vbox{\settabs 1\columns
\+\hfill\figscale{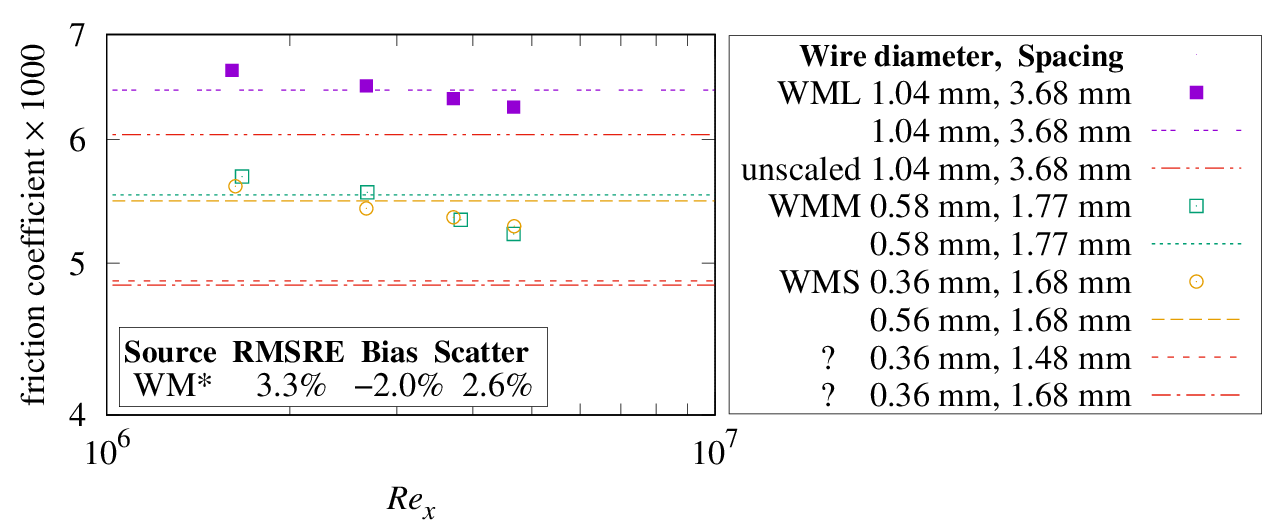}{360pt}\hfill&\cr
\+\hfill{\bf\figdef{fig:Bergstrom-mesh}\quad Local $\fcr$ versus $\Rex$ of woven wire mesh}\hfill&\cr
}

\unorderedlist
 \li Using (scaled) $\varepsilon'$, the WML and WMM measurements match
 the present theory well within the $\pm9\%$ estimated measurement
 uncertainty.  The WMS measurements do not match unless a hypothesized
 single digit misprint in~\cite{Bergstrom_2005} is corrected, changing
 the wire diameter from $0.036\rm~mm$ to $0.056\rm~mm$.
 Taken together, the (corrected) three wire meshes have 3.3\% RMSRE
 versus the present theory.
\endunorderedlist

\subsection{Perforated Sheet}

 Bergstrom \etal \cite{Bergstrom_2005} has local skin-friction coefficient
 measurements of perforated sheets attached to a smooth plate.

\subsection{Perforated Sheet Openness}
 \tabref{tab:perf-openness} checks the openness of the perforated
 sheets from Bergstrom \etal \cite{Bergstrom_2005}.  It indicates that
 the holes were hexagonally arrayed.  However, the PS sheet's
 calculated openness is 1/2 of the paper's~22\%.  There are two single
 digit changes, either of which results in hexagonal openness
 near~22\%: hole diameter $d=1.7{\rm\,mm}$ or center spacing
 $s=2.4{\rm\,mm}$.

\centerline{\bf\tabdef{tab:perf-openness}\quad{Perforated sheet openness}}
\vbox{\settabs 5\columns
\toprule
\+\quad\bf Hole diameter $d$&\quad\bf Spacing $s$&\hfil\bf Square $\Omega$&\hfil\bf Hexagonal $\Omega$&\bf From \cite{Bergstrom_2005}\hfill{\bf Tag}&\cr
\midrule
\+\quad$2.0{\rm\,mm}\approx5/64$&\quad$2.81{\rm\,mm}\approx7/64$&\hfill39.8\%\qquad\quad&\hfill45.9\%\qquad\quad&\qquad45\%\hfill PL&\cr
\+\quad$1.6{\rm\,mm}\approx4/64$&\quad$2.43{\rm\,mm}\approx6/64$&\hfill34.1\%\qquad\quad&\hfill39.3\%\qquad\quad&\qquad41\%\hfill PM&\cr
\+\quad$1.2{\rm\,mm}\approx3/64$&\quad$3.40{\rm\,mm}\approx8.6/64$&\hfill9.8\%\qquad\quad&\hfill11.3\%\qquad\quad&\qquad$\underline{22\%}$\hfill PS&\cr
\+\quad$1.7{\rm\,mm}\approx4.3/64$&\quad$3.40{\rm\,mm}\approx8.6/64$&\hfill19.6\%\qquad\quad&\hfill$\underline{22.7\%}$\qquad\quad&&\cr
\+\quad$1.2{\rm\,mm}\approx3/64$&\quad$2.40{\rm\,mm}\approx6/64$&\hfill19.6\%\qquad\quad&\hfill$\underline{22.7\%}$\qquad\quad&&\cr
\bottomrule
}

 North American suppliers of perforated sheet metal generally specify
 hole diameter and center spacing in terms of 1/64 of an
 inch.  \tabref{tab:perf-openness} provides dimensions both ways.
 $d=3/64$ with $s=6/64$ is a standard size; $d=4.3/64$ with $s=8.6/64$
 is not.  Replacing PS row $s=3.40\rm~mm$ with $s=2.40\rm~mm$:

\subsection{Plateau Wells}
 \tabref{tab:perf-parameters} shows the dimensions and metrics of the
 $\varsigma$-thick perforated sheets when laid on the flat plate.
 For PL, $[4\,\Ls]^2/L_P^2=d^2/s^2\approx0.507>1/2$; its flow will be
 rough.
 The ``$\fcr(2898)=0.0052$'' trace in \figref{fig:Bergstrom-perf} shows
 the predicted local skin-friction coefficient's close proximity to
 the PL measurements.  The ``?~2.00~mm, 2.81~mm, 0.90~mm''
 trace with transition at ``PL?'' shows the behavior predicted
 if $d^2/s^2<1/2$ had been the case.

\medskip
\centerline{\bf\tabdef{tab:perf-parameters}\quad{Perforated sheet parameters}}
\vbox{\settabs 8\columns
\toprule
\+\hfil $[4\,\Ls]=d$&\hfil $L_P=s$&\hfil\bf$\Omega$&\hfil$d^2/s^2$&\hfil$\varsigma$&\hfil$\varepsilon$&\hfil$\ReW$&\hfil{\bf Tag}&\cr
\midrule
\+\hfil2.0 mm&\hfil2.81 mm&\hfil45.9\%&\hfil0.507&\hfil0.90 mm&\hfil0.449 mm&\hfil$1.96\times10^6$&\hfil PL&\cr
\+\hfil1.6 mm&\hfil2.43 mm&\hfil39.3\%&\hfil0.434&\hfil0.90 mm&\hfil0.441 mm&\hfil$3.84\times10^6$&\hfil PM&\cr
\+\hfil1.2 mm&\hfil2.40 mm&\hfil22.7\%&\hfil0.250&\hfil0.76 mm&\hfil0.318 mm&\hfil$2.71\times10^6$&\hfil PS&\cr
\bottomrule
}

 PM and PS have $d^2/s^2<1/2$.  As $\Rex$ grows to exceed $\ReW$,
 the local drag coefficient gradually transitions from blend
 Formula~\eqref{eq:perf-mix-local} to smooth
 Formula~\eqref{eq:effective-smooth-area-local}.  The $\ReW$
 Formula~\eqref{eq:Re_W} transitions are marked by vertical
 lines.  \figref{fig:wells} details the PS and PM abrupt and smooth
 transitions.

\vbox{\settabs 1\columns
\+\hfill\figscale{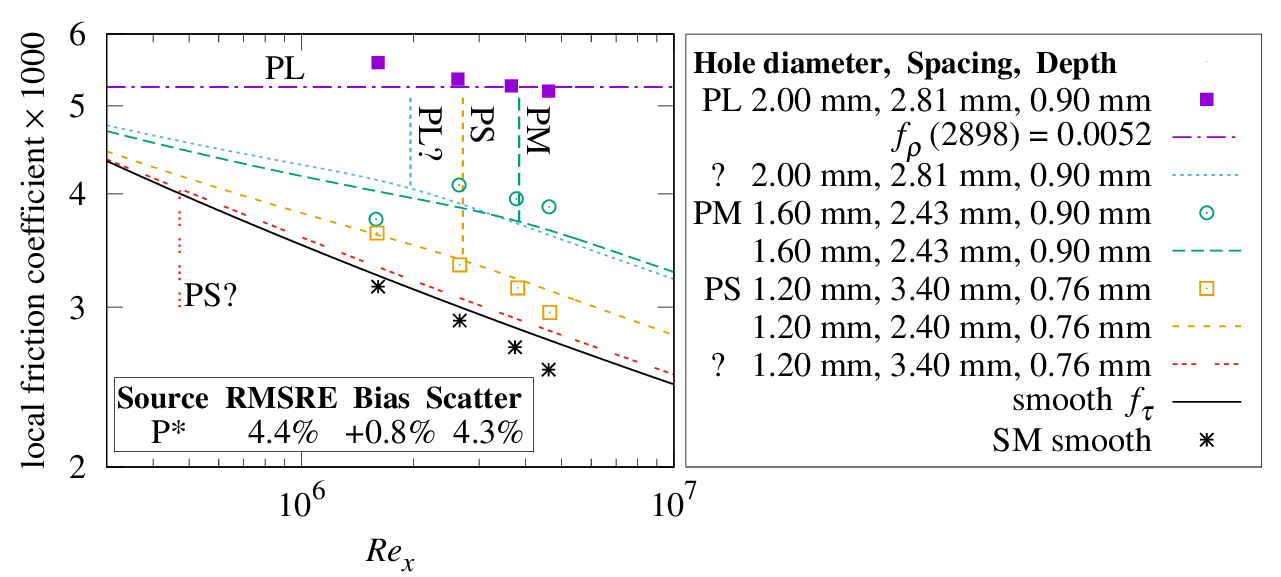}{360pt}\hfill&\cr
\+\hfill{\bf\figdef{fig:Bergstrom-perf}\quad Local $\fcr$ versus $\Rex$ of perforated sheet}\hfill&\cr
}

\unorderedlist
 \li PL and PM measurements match the present theory within the
 $\pm9\%$ measurement uncertainty.  The PS measurements do not match
 unless a hypothesized single digit misprint in~\cite{Bergstrom_2005}
 is corrected, changing the~PS hole spacing from $3.4\rm~mm$ to
 $2.4\rm~mm$.  The ``PS?'' trace shows the behavior predicted
 of the original~PS.
 Taken together, the (corrected) three perforated sheets have 4.4\%
 RMSRE versus the present theory.
\endunorderedlist

\section{Rough Heat Transfer Measurements}

 Both the Pimenta \etal \cite{A014219} and Bergstrom
 \etal \cite{Bergstrom_2005} measurements are restricted to velocity
 ranges of less than 3:1.
 For a novel theory to be persuasive, confirmations over a wider range
 of fluid velocities are needed.

 The present apparatus combined an open intake wind-tunnel, software
 phase-locked loop (PLL) fan control, and a heated aluminum plate.
 It measured average convection heat transfer in air at $2300<\Rey<93000$, a 40:1
 range.
 \ref{Appendix A: Apparatus and Measurement Methodology}
 describes the apparatus and measurement methodology.

\subsection{Bi-Level 3~mm Roughness}
 A square grid of 6~mm deep grooves in a
 $0.305{\rm~m}\times0.305{\rm~m}$ plate created the
 $\varepsilon=3.0{\rm~mm}$ bi-level roughness.

 The two peripheral $2\,\varepsilon\times L$ sides of the bi-level
 plate roughness which are parallel to the fluid flow also contribute
 to forced convection heat transfer.
 Turning to dimensional analysis, $\varepsilon$ and plate width $L_W$
 cooperate weakly, leading to an effective width of
 $\|L_W,\varepsilon\|_{\sqrt{1/2}}$, about a~5.4\% increase.

 Applying average convection Formula~\eqref{eq:forced} to the bi-level
 plate geometry, with the 5.4\% increase, yields:
$$\Nurol(\Rey)=0.00823\,\Rey\,\Pra^{1/3}\eqdef{eq:3mm-rough}$$

 Note that this correction applies only to average $\Nurol$
 measurements, not to local $\Nur$ measurements.

 \figref{fig:mixed-dn-correlation-3} shows convection measurements
 made with the plate averaging 11~K warmer than the ambient air.
 $\Nurol$ is Formula~\eqref{eq:3mm-rough}; $\Nuiol$ is
 formula \eqref{eq:bi-level-convect}.  At $\Rey<3000$, the natural
 convection component dominates the mixture; hence, measurements at
 $\Rey<3000$ are excluded from the RMSRE calculations.

\unorderedlist
 \li Measurements in the range $4000<\Rey<50000$ match~$\Nurol$
 Formula~\eqref{eq:3mm-rough} with $1.8\%$~RMSRE.
\endunorderedlist

 At $\Rey>\ReI=55566$, $\NuIol$ is Formula~\eqref{eq:posts-convect};
 its 4/5 slope shows that convection is from turbulent flow.
 Its height above the $\Nutol-\Nutol(\ReI)$ trace shows that it is
 operating with a shorter characteristic length than
 $\Nutol-\Nutol(\ReI)$.  The \figref{fig:mixed-dn-correlation-3}
 inset shows that $\Nuiol$ formula \eqref{eq:bi-level-convect} is a
 closer match to measurements at $60000<\Rey<90000$ than an abrupt
 transition.

\unorderedlist
 \li Measurements in the range $4000<\Rey<90000$ match the plateau
 islands $\Nuiol$ Formula~\eqref{eq:bi-level-convect} with
 $1.8\%$ RMSRE.
\endunorderedlist

\vbox{\settabs 1\columns
\+\hfill\figscale{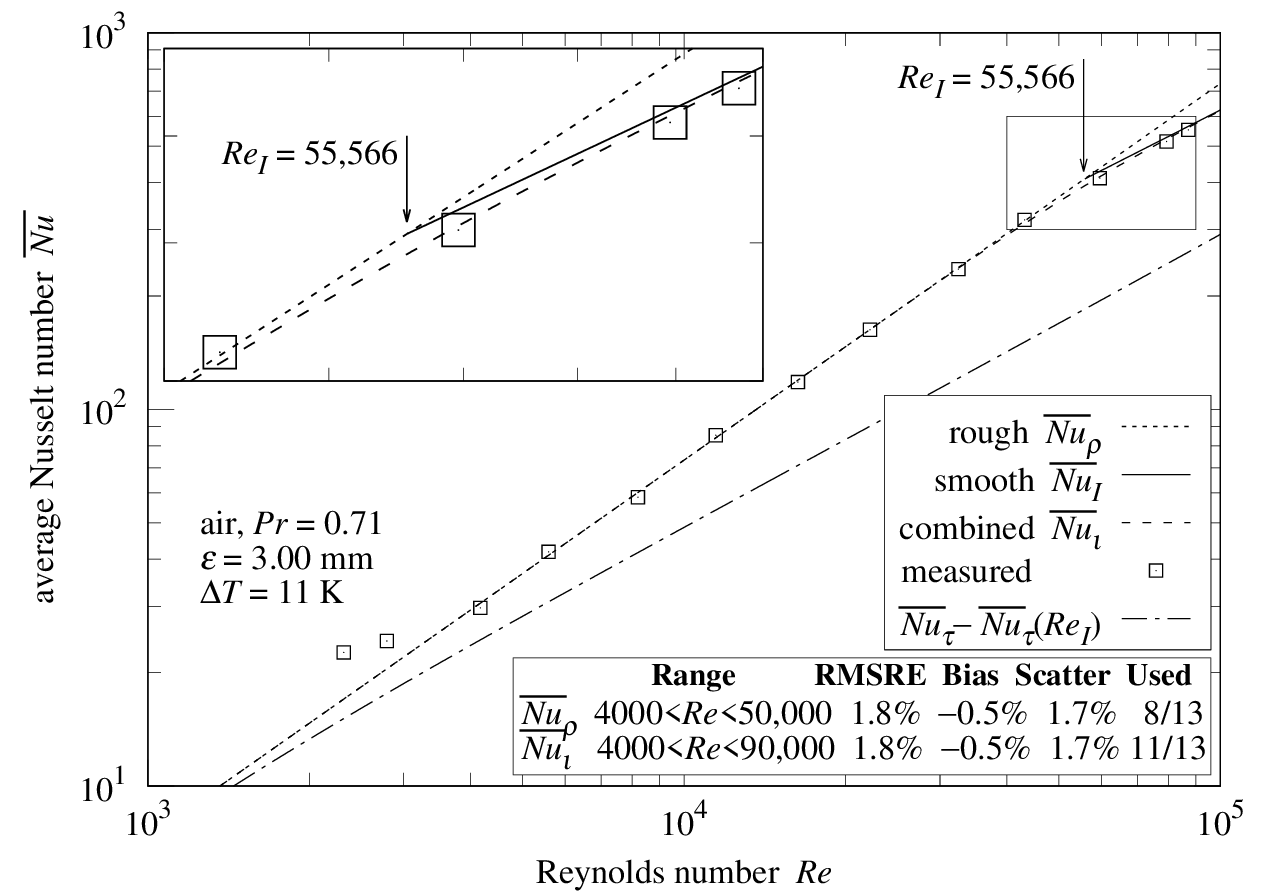}{360pt}\hfill&\cr
\+\hfill{\bf\figdef{fig:mixed-dn-correlation-3}\quad Convection from bi-level plate; $L/\varepsilon=102$}\hfill&\cr
}

\vfill\eject

\subsection{Bi-Level 1~mm Roughness}
 After making a variety of convection measurements,
 the~$\varepsilon=3~$mm plate was machined to reduce its roughness
 to $\varepsilon=1.04~$mm.

 In order to preserve the plate's wire suspension,
 the four corner posts were not shortened.  The $\ReI$ transition
 involves only the leading portion of the plate.  $\ReI=6178$ was
 calculated by Formula~\eqref{eq:Re_I} using $\varepsilon=1.14~$mm,
 the RMS height-of-roughness of the leading three rows of posts.

 The plate's effective width, $\|L_W,~\varepsilon\|_{\sqrt{1/2}}$, is
 about 2.6\% larger than~$L_W$, but affects only $\Nurol$ flow at
 $\Rey<6178$.  $\NuIol$ Formula~\eqref{eq:bi-level-convect} already
 accounts for smooth convection from the post sides.

 In \figref{fig:mixed-dn-correlation-1}, $\Nuiol$ is
 formula \eqref{eq:bi-level-convect}.  At $\Rey<5000$, the natural
 convection component dominates the mixture; hence, measurements at
 $\Rey<5000$ are excluded from the RMSRE calculations.

 The $\ell^{-4}$-norm in $\Nuiol$ Formula~\eqref{eq:bi-level-convect}
 fits very well with the measurements
 in \figref{fig:mixed-dn-correlation-1}.  Replacing it with the
 $\ell^{-2}$-norm drives some~$\Nuol$ values out of the expected
 uncertainty bounds.

\unorderedlist
 \li Convection measurements at $5000<\Rey<10^5$ match plateau-islands
 Formula~\eqref{eq:bi-level-convect} with 1.5\%~RMSRE.
\endunorderedlist

\vbox{\settabs 1\columns
\+\hfill\figscale{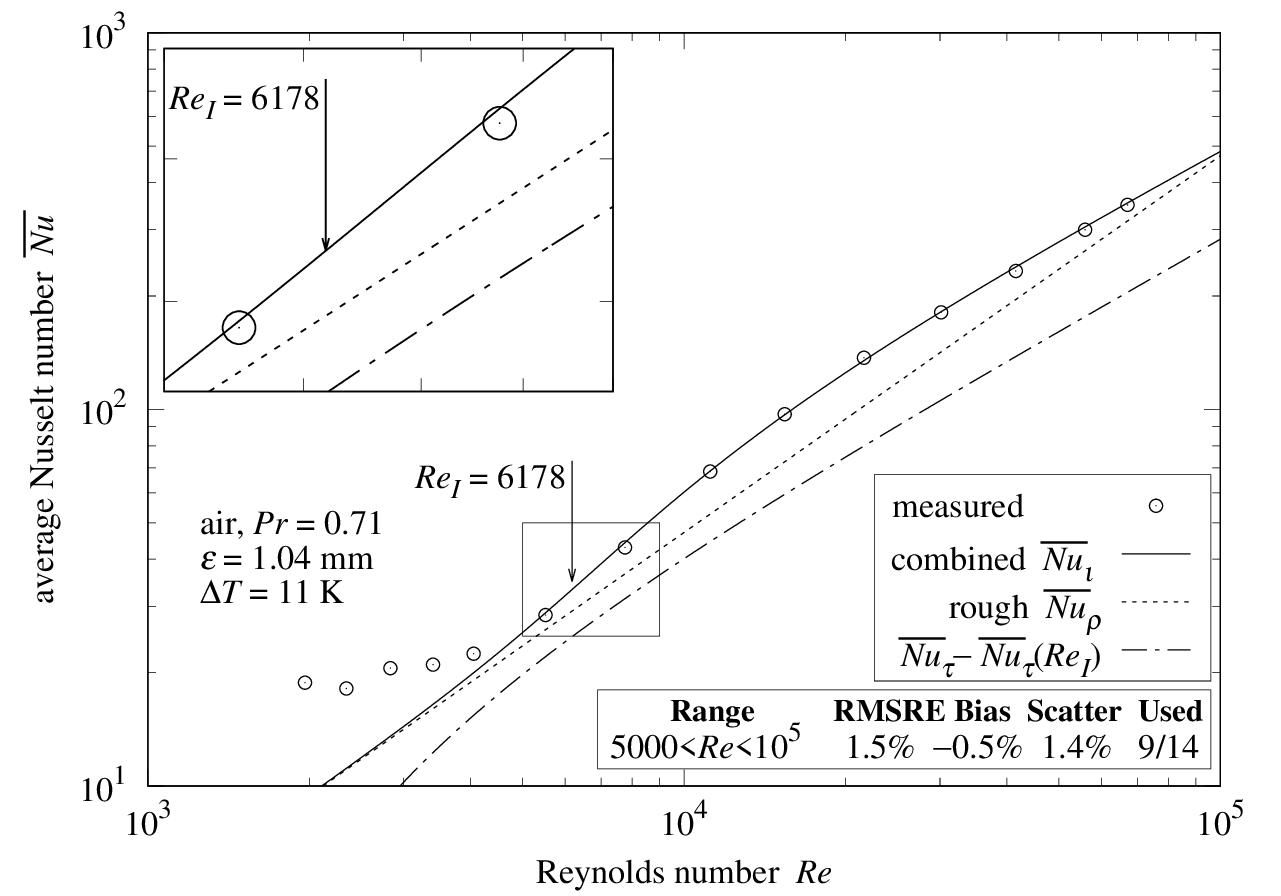}{360pt}\hfill&\cr
\+\hfill{\bf\figdef{fig:mixed-dn-correlation-1}\quad Convection from bi-level plate; $L/\varepsilon=295$}\hfill&\cr
}

\subsection{Onset of Rough Flow}
 $\Rel$ Formula~\eqref{eq:Re-laminar} predicts that the flow along the
 leading band of roughness of the 3~mm bi-level plate transitions from
 laminar to rough flow at $\Rel\approx175$, which is too slow
 to test in this apparatus.
 Formula~\eqref{eq:Re-laminar} predicts $\Rel=1473$ for the 1~mm
 bi-level plate.  The plate was found to have convection consistent
 with rough flow at $\Rey\approx2500$, which is less than~$2\,\Rel$.

\section{Smoothness}

 Thus far, this investigation has focused on rough and turbulent
 flows.  Attention now turns to laminar flows.

 Schlichting \cite{schlichting2014} describes the behavior of
 parallel flow of ``low turbulence intensity'' along a sharp-edged,
 smooth surface as a ``stable laminar flow following the leading
 edge'', transitioning to a ``fully developed
 turbulent boundary-layer'' at some $\Rex<5\times10^5$.
 Lienhard~\cite{10.1115/1.4046795} models a gradual $\Rex$ transition
 between laminar and turbulent flow along the smooth plate.

\subsection{Laminar Flow Over Roughness}
  The amount of laminar flow displaced by a periodic roughness grows
  with both~$\varepsilon$ and~$L_P$.  Dimensional analysis suggests
  that this displacement is significant when
  $\Rey>L/\sqrt{\varepsilon\,L_P}$.
  However, the laminar disturbance will be eclipsed by turbulent flow
  when $\Rey>\Ret$, leading to a proposed criterion:

\unorderedlist
 \li An isotropic, periodic roughness behaves as a ``smooth'' surface
  when $L/\sqrt{\varepsilon\,L_P}>\Ret$ and $\Rey<\Rel$.
\endunorderedlist

 A silicon wafer is an isotropic, periodic roughness with
 $L_P=543{\rm~nm}$ and $\varepsilon=31.2{\rm~nm}$.
 The ``silicon wafer'' has $L/\sqrt{\varepsilon\,L_P}>\Ret$
 in \tabref{tab:dimensionless surface parameters};
 it should behave as a smooth surface when $\Rey<\Rel$.

\centerline{\bf\tabdef{tab:dimensionless surface parameters}\quad{Dimensionless surface parameters}}
\moveright 0.125\hsize\vbox{\settabs 8\columns
\toprule
\+\qquad{\bf Surface} &&\hfill$L/\varepsilon~$\quad&\hfill$L/\sqrt{\varepsilon\,L_P}$&\hfill$\Ret$\quad&\hfill$\Rel$\quad&\cr
\midrule
\+ silicon wafer&& \hfill~$9.8\times10^6$& \hfill~$2.3\times10^6$& \hfill$>~$~$1.2\times10^6$& \hfill$\ll\,$~$7.5\times10^7$&\cr
\+ \cite{A014219} Pimenta \etal&& \hfill~$1.5\times10^4$& \hfill~$5.0\times10^3$& \hfill$>~$~$1.3\times10^3$& \hfill$\ll\,$~$5.5\times10^4$&\cr
\+ duck tape\hfill&& \hfill~$7.5\times10^3$& \hfill~$1.0\times10^3$& \hfill$<~$~$4.0\times10^3$& \hfill$\ll\,$~$1.9\times10^5$&\cr
\+ \cite{Bergstrom_2005} Bergstrom \etal PS&& \hfill~$4.6\times10^3$& \hfill~$1.5\times10^3$& \hfill$>~$~$4.3\times10^2$& \hfill$\ll\,$~$2.0\times10^4$&\cr
\+ \cite{Bergstrom_2005} Bergstrom \etal PM&& \hfill~$3.0\times10^3$& \hfill~$1.3\times10^3$& \hfill$>~$~$2.3\times10^2$& \hfill$\ll\,$~$7.2\times10^3$&\cr
\+ \cite{Bergstrom_2005} Bergstrom \etal PL&& \hfill~$2.9\times10^3$& \hfill~$1.2\times10^3$& \hfill$>~$~$2.3\times10^2$& \hfill$\ll\,$~$8.0\times10^3$&\cr
\+ \cite{Bergstrom_2005} Bergstrom \etal WMS&& \hfill~$3.0\times10^3$& \hfill~$1.5\times10^3$& \hfill$>~$~$2.2\times10^2$& \hfill$\ll\,$~$5.1\times10^3$&\cr
\+ \cite{Bergstrom_2005} Bergstrom \etal WMM&& \hfill~$2.9\times10^3$& \hfill~$1.5\times10^3$& \hfill$>~$~$2.1\times10^2$& \hfill$\ll\,$~$5.0\times10^3$&\cr
\+ \cite{Bergstrom_2005} Bergstrom \etal WML&& \hfill~$1.6\times10^3$& \hfill~$7.6\times10^2$& \hfill$>~$~$1.2\times10^2$& \hfill$\ll\,$~$3.3\times10^3$&\cr
\+ present work 1~mm bi-level&& \hfill~$2.9\times10^2$& \hfill~$8.8\times10^1$& \hfill$>~$~$2.9\times10^1$& \hfill$\ll\,$~$1.5\times10^3$&\cr
\+ present work 3~mm bi-level&& \hfill~$1.0\times10^2$& \hfill~$5.1\times10^1$& \hfill$>~$~$7.5\times10^0$& \hfill$\ll\,$~$1.8\times10^2$&\cr
\bottomrule
}

\subsection{Pierced-Laminar Flow}
 Consider ``duck tape'', the only row in
 \tabref{tab:dimensionless surface parameters}
 with $L/\sqrt{\varepsilon\,L_P}<\Ret$.

 Tuck and Kouzoubov~\cite{tuck_kouzoubov_1995} finds that slow laminar
 flow over a periodic roughness ``\dots represents a shift of the
 apparent plane boundary toward the flow domain.''  At small~$\Rey$,
 the flow from the apparent boundary plane outward is the same as
 smooth laminar flow.  Thus, the roughness has little effect on
 shear stress when $\Rey<L/\sqrt{\varepsilon\,L_P}$.

\unorderedlist
 \li When $L/\sqrt{\varepsilon\,L_P}<\Ret$,
  the purely laminar upper-bound (critical $\Rey$)
  $\Rec=L/\sqrt{\varepsilon\,L_P}$.
\endunorderedlist

 Consider the boundary-layer where $\Rec<\Rex<\Rel$.
 Between the surface and its apparent boundary plane, shearing stress
 periodically ($L_P$) exceeds that of a smooth surface, spawning
 vortexes as~$\Rey$ increases, asymptotically approaching
 turbulent flow.  This investigation terms this mixture
 ``pierced-laminar flow''; the laminar flow is pierced by vortexes.

\subsection{Smooth Laminar Flow}
 Reducing free-stream turbulence to unusually low values, Schubauer
 and Skramstad~\cite{Schubauer1948} reported that ``oscillations were
 discovered in the laminar boundary layer along a flat plate.''  These
 periodic oscillations suggest that rough and smooth laminar flows
 both spawn vortexes periodically along the plate,
 controlled by the purely laminar upper-bound~$\Rec$.

\subsection{Laminar-Turbulent Mixing}
 The hypothesized periodic vortexes pierce the laminar boundary-layer.
 The present theory holds that the laminar component is active through
 the entire $\Rey<\Rel$ range in pierced-laminar flow, resulting in
 friction Formula~\eqref{eq:pierced-friction} and convection
 Formula~\eqref{eq:pierced-convection}.

$$\eqalignno{
  \fcsol&=\left\|\fclol(\Rey),~\fctol(\Rey)-{\Rey_\gamma\over\Rey}\,\fctol\left(\Rey_\gamma\right)\,\right\|_{\gamma}\qquad\Rey_\gamma=\left\|\,\Rey,~{\Rec\over\sqrt{\gamma}}\,\right\|_\gmprm&\eqdef{eq:pierced-friction}\cr
  \Nusol&=\left\|\Nulol(\Rey),~\Nutol(\Rey)-\Nutol\left(\,\left\|\Rey,~\sqrt{\gamma}\,\Rec\,\right\|_\gmprm\right)\,\right\|_{\gamma}&\eqdef{eq:pierced-convection}\cr
 }$$

 Because laminar and turbulent flows mix in pierced-laminar flow, the
 $\ell^{1}$-norm and $\ell^{-4}$-norm of staged-transition
 Formulas~(\eqrefn{eq:staged-friction}, \eqrefn{eq:staged-convection})
 are replaced by the $\ell^{\gamma}$-norm and
 $\ell^{\gamma\prime}$-norm in
 Formulas~(\eqrefn{eq:pierced-friction}, \eqrefn{eq:pierced-convection}).
 The laminar and turbulent flows are in mild competition, so
 $1<\gamma<2$.  \figref{fig:gammas} shows $\fcsol$
 Formula~\eqref{eq:pierced-friction} curves at $1\le\gamma\le2$ and
 $-8\le\gmprm\le-4$.  Smaller $\gamma$ results in a sharper downward
 bend, while more negative $\gmprm$ results in a sharper upward bend.
 Letting $\gmprm=-8/\gamma$ links the variables to sharpen the bends
 together within their respective ranges.

 The geometric mean of 1 and 2 is~$\sqrt{2}$; for the moment, assume
 $\gamma=\sqrt{2}$ and $\gmprm=8/\gamma=4\,\sqrt{2}$.

 \figref{fig:critical-f} plots Formula~\eqref{eq:pierced-friction} at
 five~$\Rec$ values.  Note that traces with smaller $\Rec$ have larger
 friction coefficients than the turbulent $\fctol$ trace.

\medskip
\vbox{\settabs 2\columns
\+\hfil\figscale{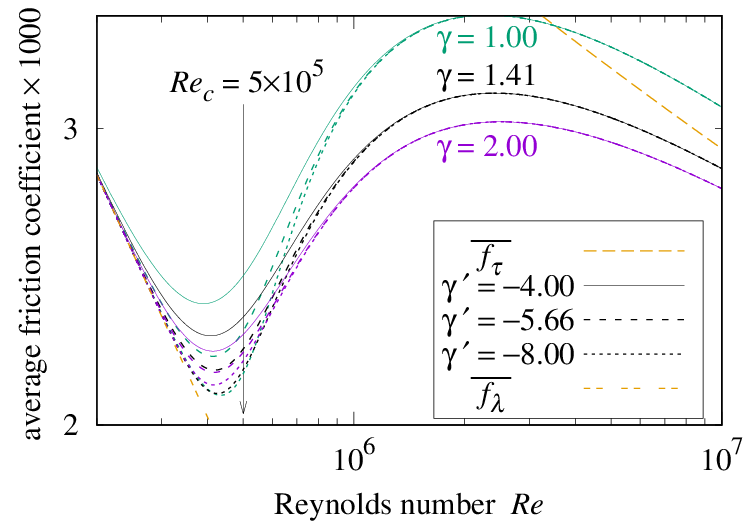}{234pt}&\hfil\figscale{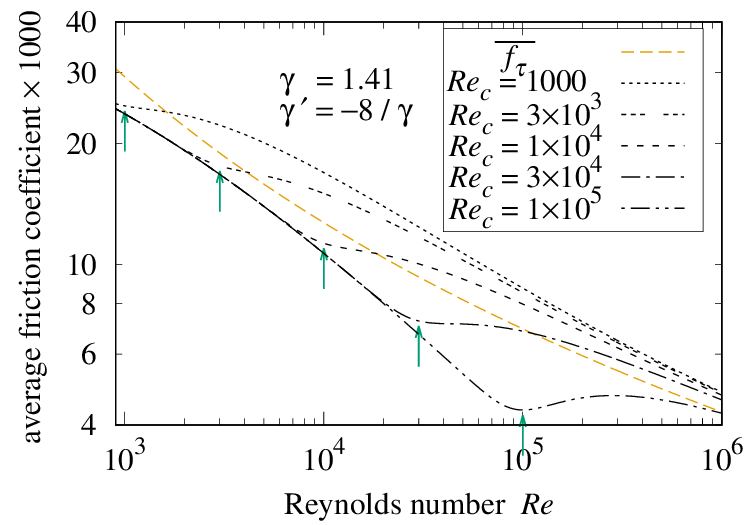}{234pt}&\cr
\+\hfil{\bf\figdef{fig:gammas}\quad Pierced-laminar friction by $\gamma, \gmprm$}&\hfil{\bf\figdef{fig:critical-f}\quad Pierced-laminar friction by $\Rec$}&\cr
}
\medskip

 \figref{fig:Gebers} compares measurements from
 Gebers~\cite{Gebers1908,Gebers1920} with staged-transition
 Formula~\eqref{eq:staged-friction}, pierced-laminar
 Formula~\eqref{eq:pierced-friction},
 and Formula~\eqref{eq:Schlichting-transition}
 from Schlichting \cite{schlichting2014} for an
 apparatus with $\Rec=5\times10^5$:
$$0.455/\log_{10}^{2.58}{\Rey}-1700/\Rey\eqdef{eq:Schlichting-transition}$$

\unorderedlist
 \li The Gebers~\cite{Gebers1908,Gebers1920} measurements have 2.9\%
 RMSRE from Formula~\eqref{eq:pierced-friction} with
 $\Rec=5\times10^5$.
\endunorderedlist

\vbox{\settabs 1\columns
\+\hfill\figscale{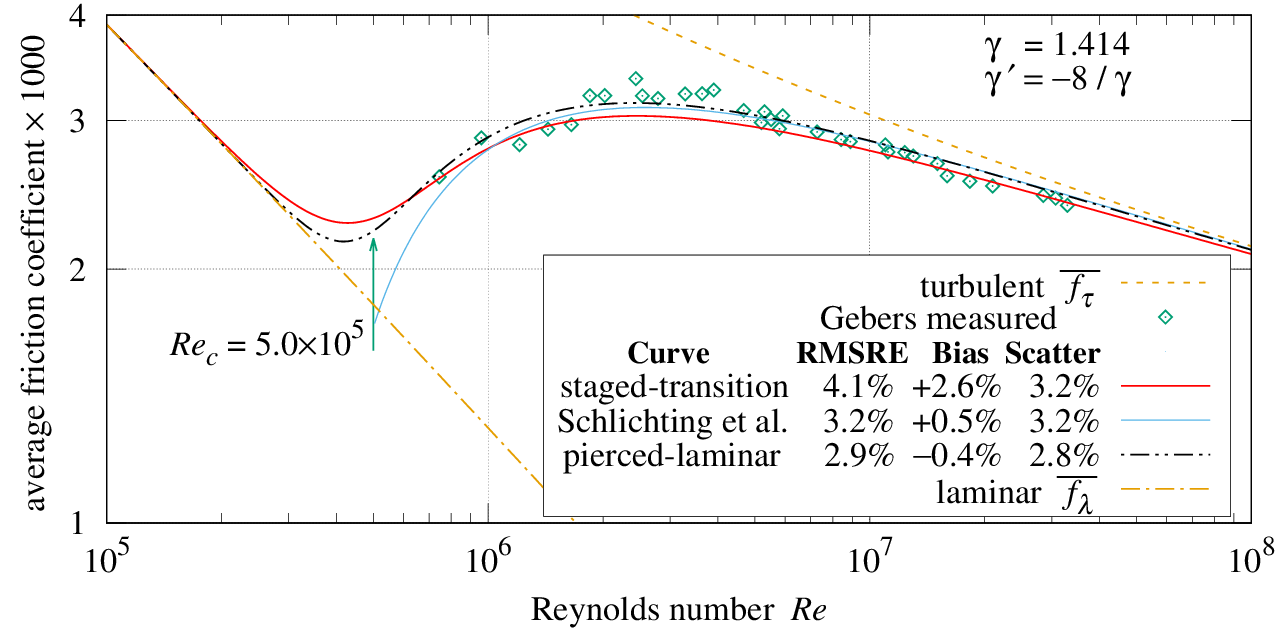}{360pt}\hfill&\cr
\+\hfill{\bf\figdef{fig:Gebers}\quad Average friction versus $\Rey$ of smooth plate}\hfill&\cr
}
\medskip

\subsection{Duck Tape}
 The bi-level test plate assembly of the present apparatus has four sides
 perpendicular to the test-surface, each a wedge of extruded
 polystyrene foam (XPS) insulation filling a 2.7~cm $45^\circ$ chamfer
 in the metal slab.  In order to isolate the convective heat flow of
 the test-surface from that of the sides, the estimated side
 convection, between 50\% at $\Rey=6000$ and 7\% at $\Rey=90000$ of
 the measured heat flow, is deducted from that measured heat flow
 (see \ref{Appendix A: Apparatus and Measurement Methodology} for
 details).

 The surface of the XPS foam board in \figref{fig:XPS-tape} was not
 smooth and not an isotropic, periodic roughness.  Without a
 theoretical basis for computing its convective heat transfer, the accuracy of
 measurements of the surface-under-test would have been limited.
 Hence, the foam was covered
 with Intertape AC6 duck tape, a $152~\mu$m-thick polyester
 cloth/polyethylene film with a pressure sensitive adhesive (shown
 in \figref{fig:XPS-tape}).
 The geometric mean of its 3.62 mm$\times$1.47 mm thread cell is
 $s\equiv{L_P}\approx2.31$~mm; the thread diameter plus the sheet
 thickness is~$d\approx0.10$~mm.  The filled-gap woven wire mesh
 Formula~\eqref{eq:mesh-eps} calculates~$\varepsilon\approx0.0403$~mm.
 These values with $L=0.305$~m yielded the ``duck tape'' row
 in \tabref{tab:dimensionless surface parameters}.

\smallskip
\vbox{\settabs 2\columns
\+\hfil\figscale{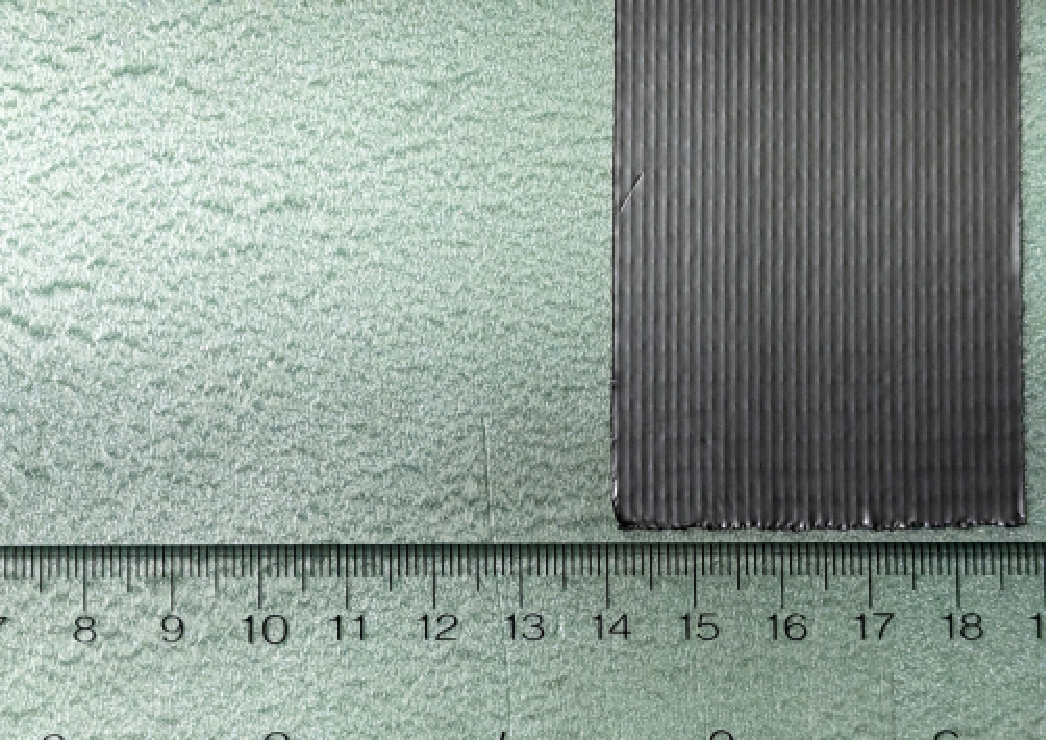}{230pt}&\hfil\figscale{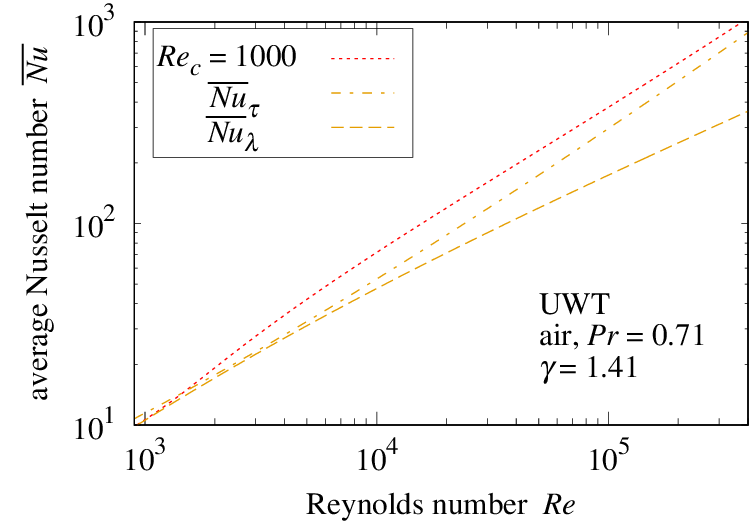}{230pt}&\cr
\+\hfil{\bf\figdef{fig:XPS-tape}\quad XPS foam board and duck tape}&\hfil{\bf\figdef{fig:critical1}\quad duck tape convective heat transfer}&\cr
}

\subsection{Pierced-Laminar Convection From Roughness}
 In the present theory,
 an $L=0.305$~m strip of duck tape fails to generate rough flow
 in the present apparatus because the apparatus's largest
 $\Rey\approx93000<\Rel\approx1.9\times10^5$.  Because
 $\Ret>L/\sqrt{\varepsilon\,L_P}$ in the duck tape row
 of \tabref{tab:dimensionless surface parameters}, the critical
 $\Rec=L/\sqrt{\varepsilon\,L_P}\approx1.0\times10^3$ in
 $\Nusol$ Formula~\eqref{eq:pierced-convection} convection.

 With duck tape applied to the foam faces, the measurements presented
 in \ref{Rough Heat Transfer Measurements} are consistent with the side
 convection modeled by pierced-laminar
 Formula~\eqref{eq:pierced-convection}.  The other curves
 in \figref{fig:critical1} are substantially less than
 Formula~\eqref{eq:pierced-convection}; they do not account for enough
 heat transfer to keep the surface-under-test measurements within the
 expected uncertainty bounds presented
 in \ref{Appendix A: Apparatus and Measurement Methodology}.

 That consistency rules out laminar $\Nulol$, turbulent
 $\Nutol$, and staged-transition $\Nufol$ as explanations of
 convection from the duck tape covered sides in combination with
 plateau islands $\Nuiol$ convection from the plate.
 This evidence is not conclusive, but supports
 pierced-laminar Formula~\eqref{eq:pierced-convection} convection from
 the duck tape surface.

\subsection{Local Skin-Friction and Convection}
Average-to-local transform~\eqref{eq:f-from-f-bar} derives
Formulas~\eqref{eq:staged-friction-local}
and~\eqref{eq:pierced-friction-local} from
Formulas~\eqref{eq:staged-friction} and~\eqref{eq:pierced-friction},
with $\Rec$ scaled by~$\sqrt{2}$.
 Formulas~\eqref{eq:staged-convection-local}
 and~\eqref{eq:pierced-convection-local} compute the local
 convections
 from the average convection
 Formulas~\eqref{eq:staged-convection}
 and~\eqref{eq:pierced-convection}, with $\Rec$ scaled
 by~$\sqrt{\gamma}$.
 These $\Rec$ scale factors are needed to make the local curve
 transitions align with~$\Rec$.
$$\eqalignno{
  f_4&={\diff[\Rex-\Rez]\,\ffol(\Rex,\sqrt{2}\,\Rec)\over\diff\Rex}&\eqdef{eq:staged-friction-local}\cr
  \Nu_4&=\Rex\,{\diff\Nufol(\Rex,\sqrt{\gamma}\,\Rec)\over\diff\Rex}&\eqdef{eq:staged-convection-local}\cr
  \fcs&={\diff[\Rex-\Rez]\,\fcsol(\Rex,\sqrt{2}\,\Rec)\over\diff\Rex}&\eqdef{eq:pierced-friction-local}\cr
  \Nus&=\Rex\,{\diff\Nusol(\Rex,\sqrt{\gamma}\,\Rec)\over\diff\Rex}&\eqdef{eq:pierced-convection-local}\cr
}$$

\subsection{Convection Transition}
 The Lienhard~\cite{10.1115/1.4046795} comprehensive smooth plate convection
 formula can be expressed using the $\ell^p$-norm:
$$\eqalignno{\Nu(\Rex)&=\left\|\Nul(\Rex),\left\{\left\|\Nul(\Rec)\left[{\Rex/\Rec}\right]^c,\Nut(\Rex)\right\|_{-10}\right\}\right\|_5&\eqdef{eq:Lienhard-transition}\cr
 \Nul(\Rex)&=0.332\,\sqrt{\Rex}\,\Pra^{1/3}\!\Bigm/\root3\of{1-[\Reu/\Rex]^{3/4}}&\eqdef{eq:Lienhard-laminar}\cr
 c&=0.9922\,\log_{10}\Rec-3.013 &\eqdef{eq:Lienhard-c}}$$

 $\Rec$ is the critical $\Rex$ upper-bound for purely laminar flow.
 Lienhard~\cite{10.1115/1.4046795} states ``The value of $\Rec$ should
 be fit to the data{set}, and the value of~$c$ may either be fitted or
 estimated from'' Formula~\eqref{eq:Lienhard-c}.  The Lienhard curves
 presented here use the $c$ estimated by
 Formula~\eqref{eq:Lienhard-c}.

 In the graphs which follow,
 the $\Rec$ and $c$ values fit by Lienhard~\cite{10.1115/1.4046795} are
 marked with a star~($*$).
 The curves are computed and labeled with the $\Rec$ value which
 minimizes the data-set RMSRE (relative to the formula) using the
 ``Golden section search'' algorithm from Kiefer~\cite{Kiefer1953}.

 \figref{fig:Kestin} plots measurements from two Kestin, Maeder, and
 Wang~\cite{KESTIN1961133} data-sets.
 \figref{fig:Reynolds} plots data from Reynolds, Kays, and
 Kline~\cite{Reynolds1958IV} at $\Pra=0.70$ with unheated $\Reu=23479$.
 Pierced-laminar Formula~\eqref{eq:pierced-convection-local} has the
 closest match (lowest RMSRE) to the measurements
 in \figrefs{fig:Kestin} and \figrefn{fig:Reynolds}.

\medskip
\vbox{\settabs 1\columns
\+\hfill\figscale{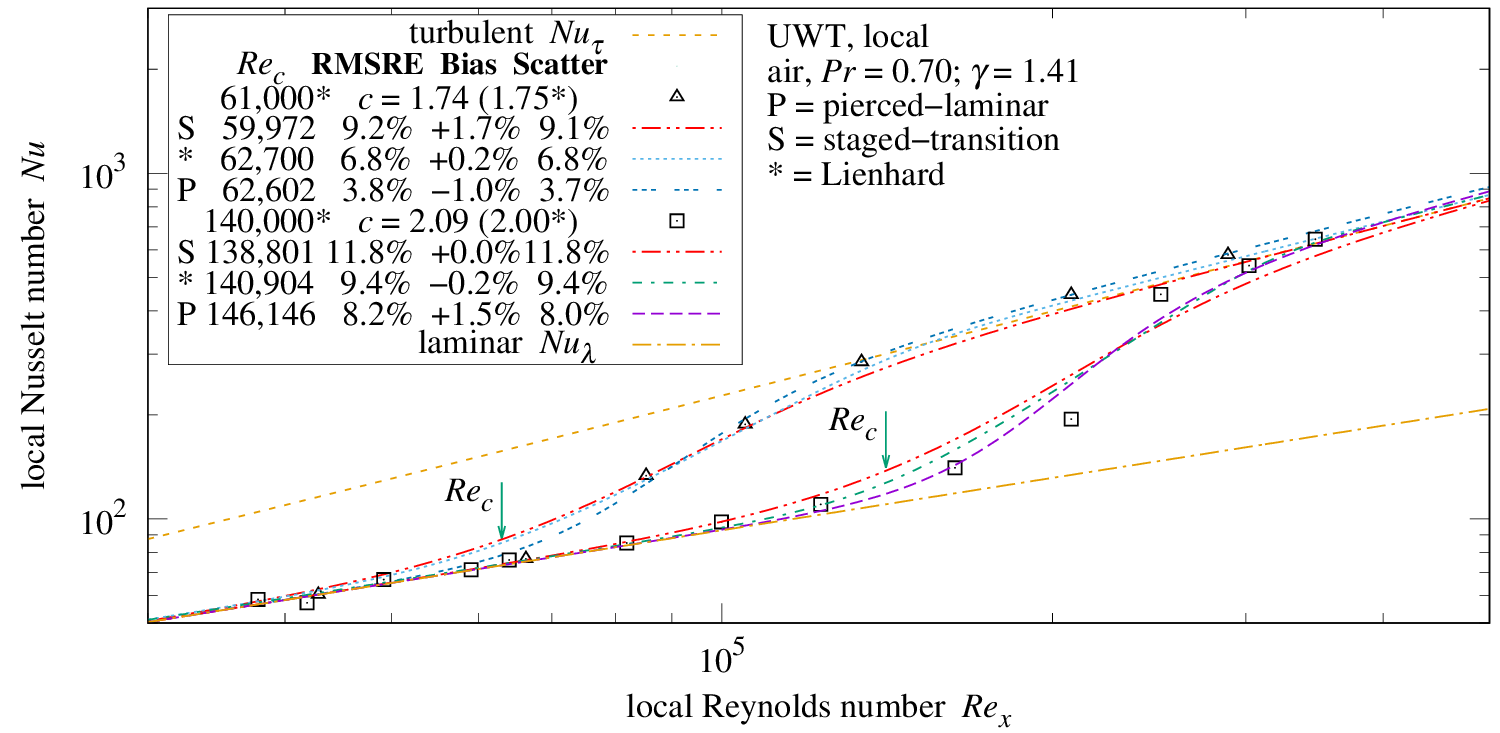}{440pt}\hfill&\cr
\+\hfill{\bf\figdef{fig:Kestin}\quad Kestin \etal \cite{KESTIN1961133} Local convection critical transition}\hfill&\cr
}
\medskip

\vbox{\settabs 1\columns
\+\hfill\figscale{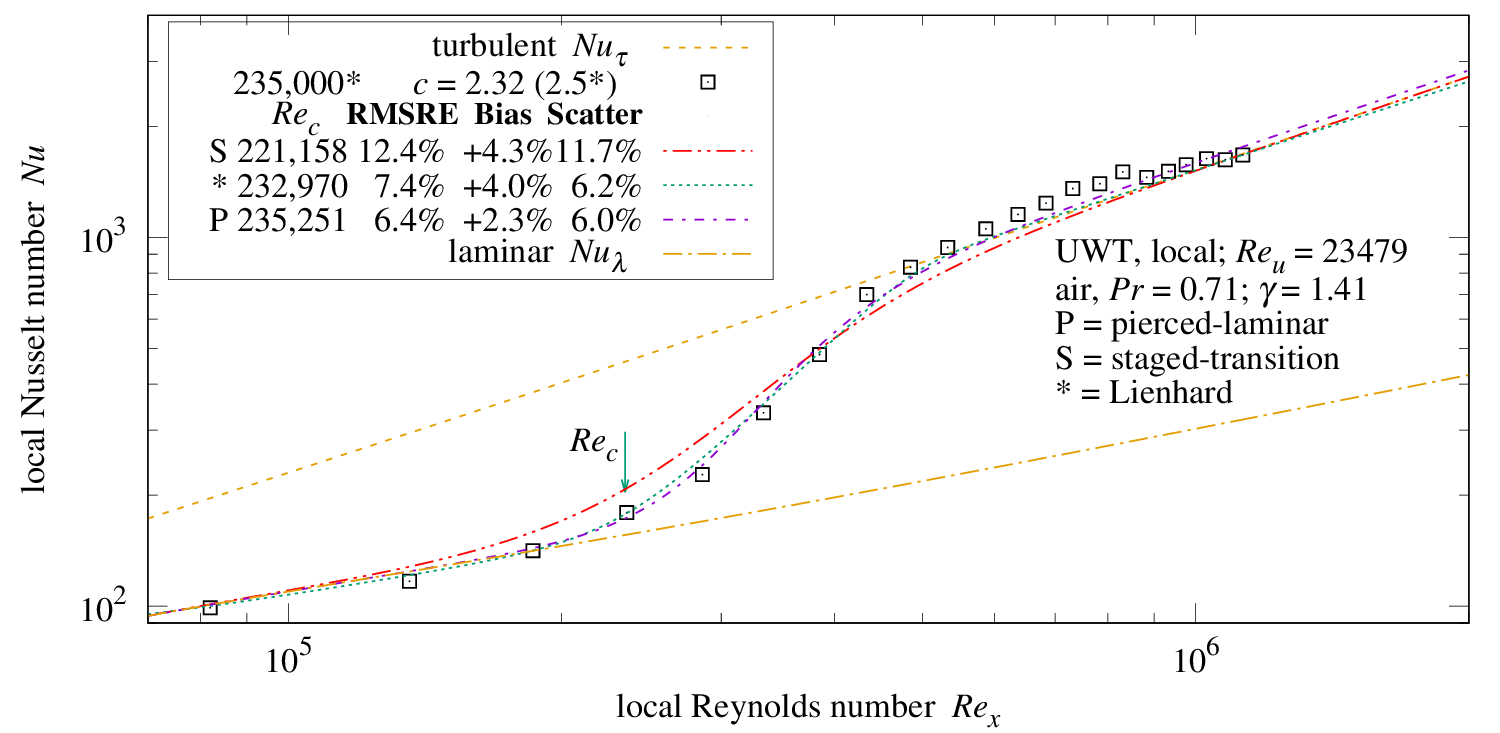}{440pt}\hfill&\cr
\+\hfill{\bf\figdef{fig:Reynolds}\quad Reynolds \etal \cite{Reynolds1958IV} Local convection critical transition}\hfill&\cr
}

\subsection{Uniform Heat Flux}
 Thus far, this investigation concerned uniform-wall-temperature (UWT,
 also termed ``isothermal'') plates.
 \v{Z}ukauskas and \v{S}lan\v{c}iauskas~\cite{Zukauskas1987} measured
 critical transitions with a uniform-heat-flux (UHF) flowing through a
 smooth surface.  In the present work, each convection graph is
 labeled UWT or UHF.

 Per Lienhard~\cite{10.1115/1.4046795}, 0.4587 replaces 0.332 in
 $\Nul$ Formula~\eqref{eq:Lienhard-laminar} for UHF plates.
 Similarly, $0.4587/0.332\approx1.382$ scales $\Nulol$
 Formula~\eqref{eq:laminar-convection} when modeling UHF plates.

 Lienhard~\cite{10.1115/1.4046795} uses the Gnielinski
 turbulent Formula~\eqref{eq:Gnielinski} for both UWT and UHF
 plates.  This investigation similarly applies its turbulent
 Formula~\eqref{eq:smooth-convection} to both UWT and UHF plates.

 When a vortex transports fluid away from the surface of a UHF plate,
 the temperature of the fluid which replaces it increases more slowly than
 it would from a UWT plate.  This reduction in local surface
 temperatures interferes with laminar heat transfer, largely
 restricting it to the $\Rex<\Rec$ region of the plate.
 Staged-transition Formula~\eqref{eq:staged-convection} models
 heat transfer from distinct laminar and turbulent areas, making it
 appropriate for UHF convection.  Note that the fluid flow is the
 same; only its heat transfer is affected.

 \figref{fig:air-transition} compares staged-transition
 Formula~\eqref{eq:staged-convection-local}, Lienhard
 Formula~\eqref{eq:Lienhard-transition}, and pierced-laminar
 Formula~\eqref{eq:pierced-convection-local} with measurements
 by \v{Z}ukauskas and \v{S}lan\v{c}iauskas~\cite{Zukauskas1987}
 of UHF plates in air.

\vbox{\settabs 1\columns
\+\hfill\figscale{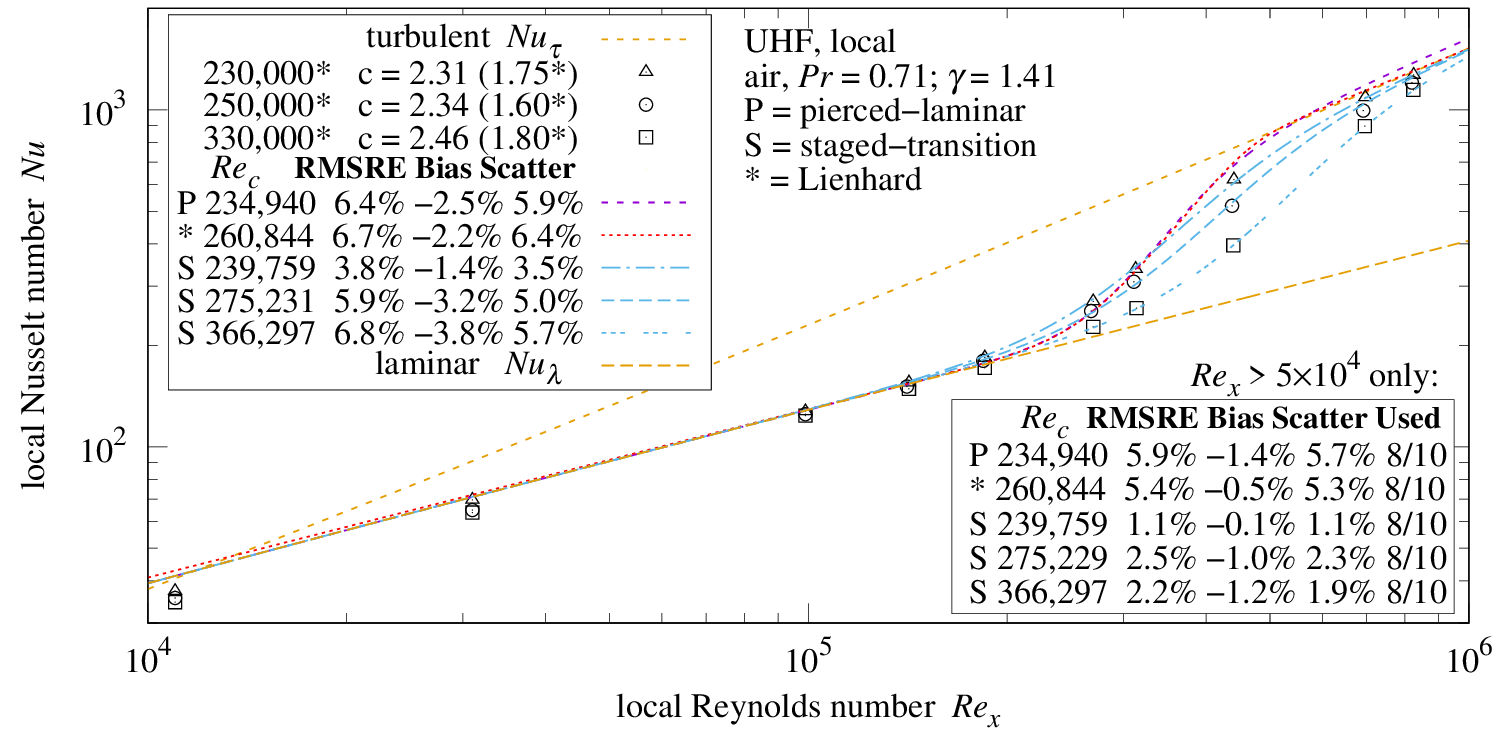}{440pt}\hfill&\cr
\+\hfill{\bf\figdef{fig:air-transition}\quad \v{Z}ukauskas and \v{S}lan\v{c}iauskas~\cite{Zukauskas1987} air critical transition}\hfill&\cr
}

 The two points of each \figref{fig:air-transition} data-set at
 $\Rex<5\times10^4$ are far from the transition range of interest.
 Disregarding them reveals that the three data-sets match (S)
 staged-transition Formula~\eqref{eq:staged-convection-local} with
 RMSRE less than 2.6\%; the other formulas have more than twice this
 RMSRE.

\unorderedlist
 \li Pierced-laminar Formula~\eqref{eq:pierced-convection-local}
 matches UWT transitions more closely than
 Formula~\eqref{eq:staged-convection-local}.

 \li Staged-transition Formula~\eqref{eq:staged-convection-local}
 matches UHF transitions more closely than
 Formula~\eqref{eq:pierced-convection-local}.

\endunorderedlist

 Thus far, all the UWT transition data-sets have been in air at
 $\Pra\approx0.71$.  \v{Z}ukauskas and \v{S}lan\v{c}iauskas had
 $\Pra>1$ transition data-sets measured from UHF plates.
 How are UWT and UHF behaviors related?

 Staged-transition and pierced-laminar heat transfer will not coincide
 at $\Rex<\Rec$ because of their different laminar heat transfer
 coefficients.
 Fluids with larger $\Pra$ will transport more heat away from the UHF
 plate's warmer regions, reducing the temperature variations across
 the plate.  Thus, UHF and UWT heat transfer formulas should converge at
 large $\Pra$ and $\Rey>\Rec$.

\unorderedlist
 \li
 At $\Rey>\root4\of2\,\Rec$ the staged-transition curve is nearly
 identical to the pierced-laminar curve with $\gamma=2$ and
 $\gmprm=-8/\gamma=-4$ in \figref{fig:gamma4}.  Therefore, convection
 $\gamma$ varies with $\Pra$, and
 $\lim_{\Pra\to\infty}\gamma(\Pra)=2$.

 \li
 Comparison of pierced-laminar curves
 with convection measurements from
 Kestin \etal \cite{KESTIN1961133} in \figref{fig:Kestin} and
 Reynolds \etal \cite{Reynolds1958IV} in \figref{fig:Reynolds}
 establishes that $\gamma(0.71)\approx1.41$.

 \li Mild competition between laminar and turbulent flows suggests
 $\lim_{\Pra\to0}\gamma(\Pra)=1$.
\endunorderedlist

\unorderedlist
 \li \figref{fig:gamma} graphs proposed Formula~\eqref{eq:gamma},
 which satisfies these three constraints.
\endunorderedlist

$$\eqalignno{&\gamma(\Pra)=1+\exp_2\left({-\Pra^{-\sqrt{1/2}}}\right)\qquad\exp_2(\varphi)\equiv2^\varphi&\eqdef{eq:gamma}}$$

 Graphing friction coefficient instead of heat transfer
 in \figref{fig:gamma4} isolates the $\ell^p$-norms from the effects
 of $\Pra$ on laminar and turbulent heat transfer.

\vbox{\settabs 2\columns
\+\hfil\figscale{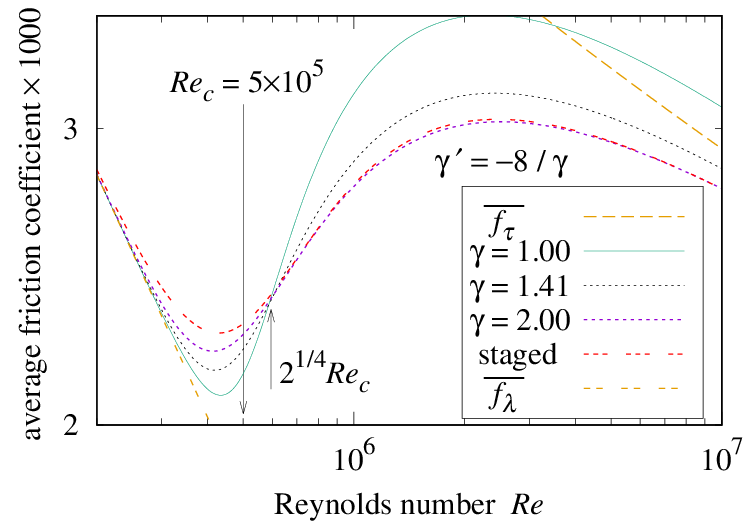}{234pt}&\hfil\figscale{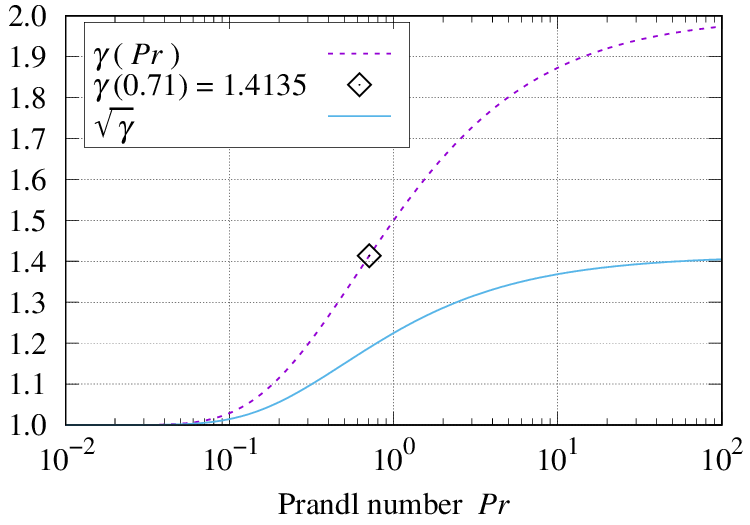}{234pt}&\cr
\+\hfil{\bf\figdef{fig:gamma4}\quad Pierced and staged friction}&\hfil{\bf\figdef{fig:gamma}\quad $\gamma(\Pra)$ and $\sqrt{\gamma}$}&\cr
}

\subsection{Viscous Liquids}
 \v{Z}ukauskas and \v{S}lan\v{c}iauskas~\cite{Zukauskas1987}
 measured critical transitions in liquid water and transformer oil.
 The $\Pra$ of those fluids is sensitive to temperature.  They
 recommended scaling the bulk fluid~$\Pra_\infty$ by
 $\root4\of{\Pra_w/\Pra_\infty}$ to yield
 $\Pra=\Pra_w^{1/4}\,\Pra_\infty^{3/4}$, where $\Pra_w$ is at plate
 temperature.
 In the presence of vortexes, the surface temperature of
 a UHF plate is not well-defined; this investigation treats UHF
 $\Pra=\Pra_\infty$.
 \v{Z}ukauskas and \v{S}lan\v{c}iauskas~\cite{Zukauskas1987}
 also scaled their UHF $\Nu$ measurements by
 $\root4\of{\Pra_\infty/\Pra_w}$; the UHF data presented here is
 descaled.

 \figrefs{fig:oil-transition} and \figrefn{fig:water-transition}
 compare Lienhard Formula~\eqref{eq:Lienhard-transition},
 pierced-laminar Formula~\eqref{eq:pierced-convection-local}, and
 staged-transition Formula~\eqref{eq:staged-convection-local} with UHF
 transition measurements in $\Pra>1$ fluids by \v{Z}ukauskas
 and \v{S}lan\v{c}iauskas~\cite{Zukauskas1987}.
 Lienhard Formula~\eqref{eq:Lienhard-transition} has 1.9\% RMSRE on
 the $\Pra_\infty=257$ data-set in \figref{fig:oil-transition}, but
 larger RMSRE than staged-transition
 Formula~\eqref{eq:staged-convection-local} on the other data-sets.

\unorderedlist
\li
 With $\Rec$ optimized to minimize RMSRE (at $\Pra=\Pra_\infty$),
 the \v{Z}ukauskas and \v{S}lan\v{c}iauskas~\cite{Zukauskas1987}
 UHF data-sets align to staged-transition
 Formula~\eqref{eq:staged-convection-local} with RMSRE less than~5\%.
\endunorderedlist

\smallskip
\vbox{\settabs 1\columns
\+\hfill\figscale{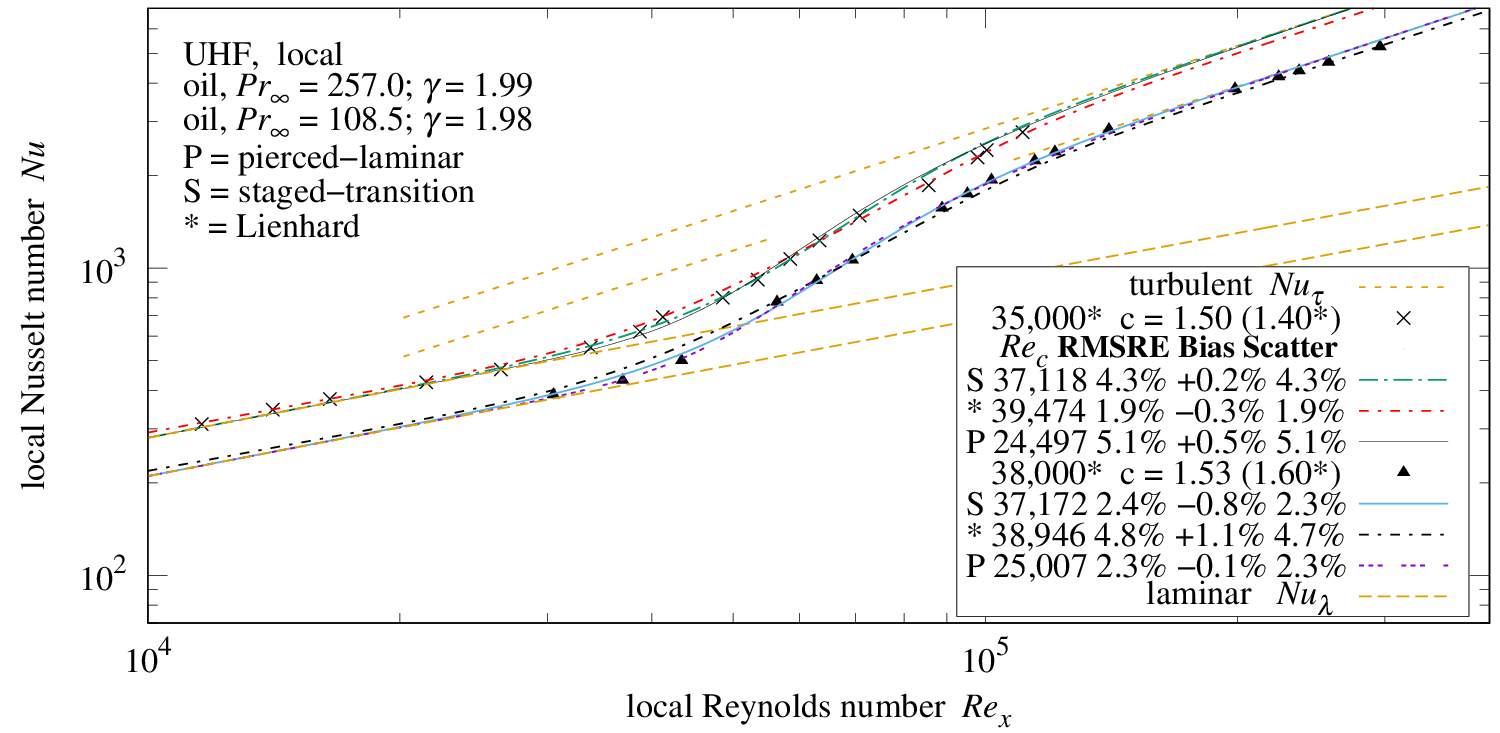}{440pt}\hfill&\cr
\+\hfill{\bf\figdef{fig:oil-transition}\quad \v{Z}ukauskas and \v{S}lan\v{c}iauskas~\cite{Zukauskas1987} oil critical transition}\hfill&\cr
}

\vbox{\settabs 1\columns
\+\hfill\figscale{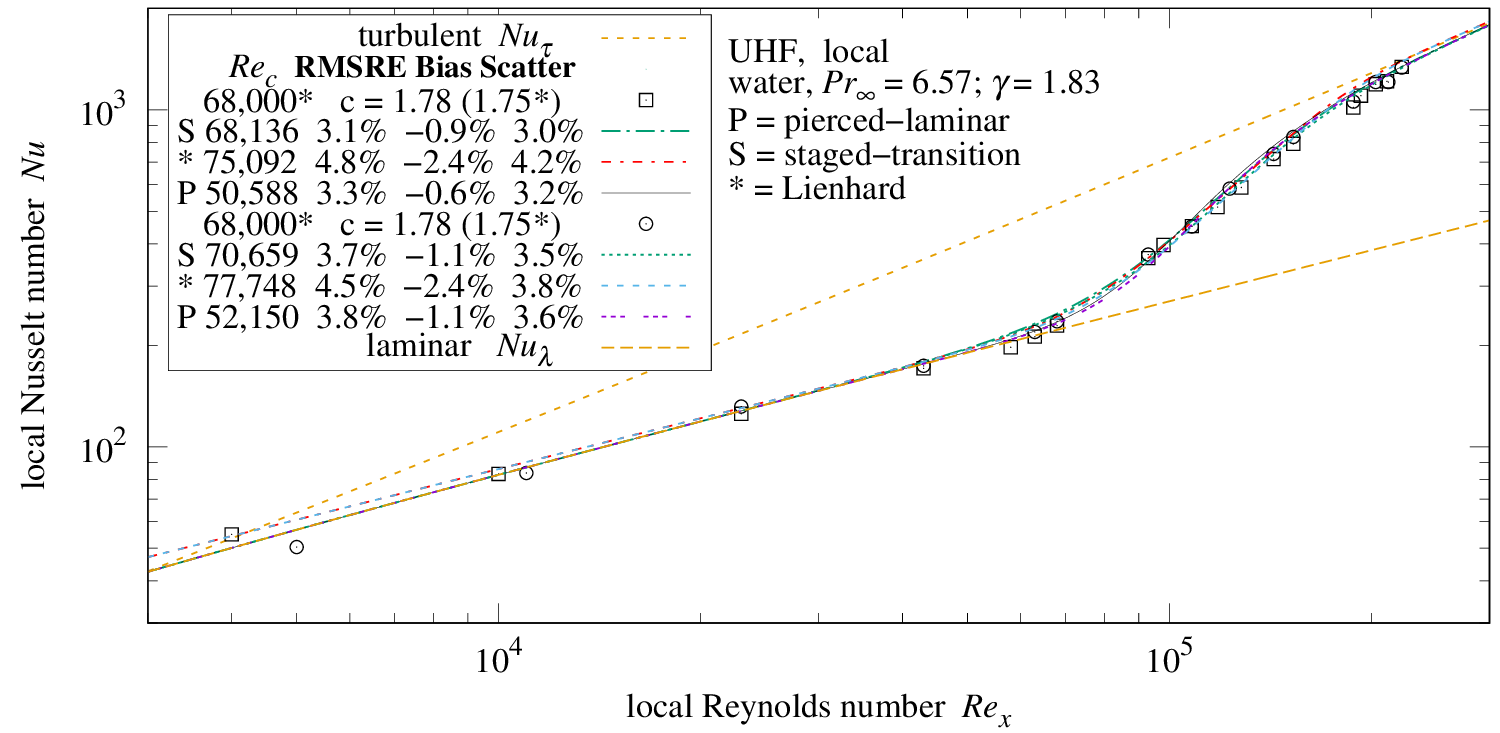}{440pt}\hfill&\cr
\+\hfill{\bf\figdef{fig:water-transition}\quad \v{Z}ukauskas and \v{S}lan\v{c}iauskas~\cite{Zukauskas1987} water critical transition}\hfill&\cr
}
\medskip

\subsection{UWT at Large Temperature Differences}
 Thus far, the heat transfer measurements have all been
 local.  \v{Z}ukauskas and \v{S}lan\v{c}iauskas~\cite{Zukauskas1987}
 also measured average UWT heat transfer in air and liquids at the
 wide variety of fluid and plate temperatures paired
 in \figref{fig:scatter}.  Testing both cooled and heated plates,
 this is a rigorous but challenging set of measurements because
 $T_\infty$, $T_w$, and $\Rey$ are independent parameters.
 Adding to that challenge, $|T_w-T_\infty|/T>10\%$ for many
 measurements in the UWT data-sets.

 While this investigation uses $\Pra=\Pra_\infty$ for UHF plates, for
 UWT plates it uses $\Pra=\Pra_w^{1/4}\,\Pra_\infty^{3/4}$
 (which \v{Z}ukauskas and \v{S}lan\v{c}iauskas~\cite{Zukauskas1987}
 recommended for both types).

 In the large temperature difference data-sets which follow, the
 numerically averaged version of Lienhard's comprehensive
 Formula~\eqref{eq:Lienhard-transition} had RMSRE values exceeding
 10\%; they are not tabulated here.

\vbox{\settabs 1\columns
\+\hfill\figscale{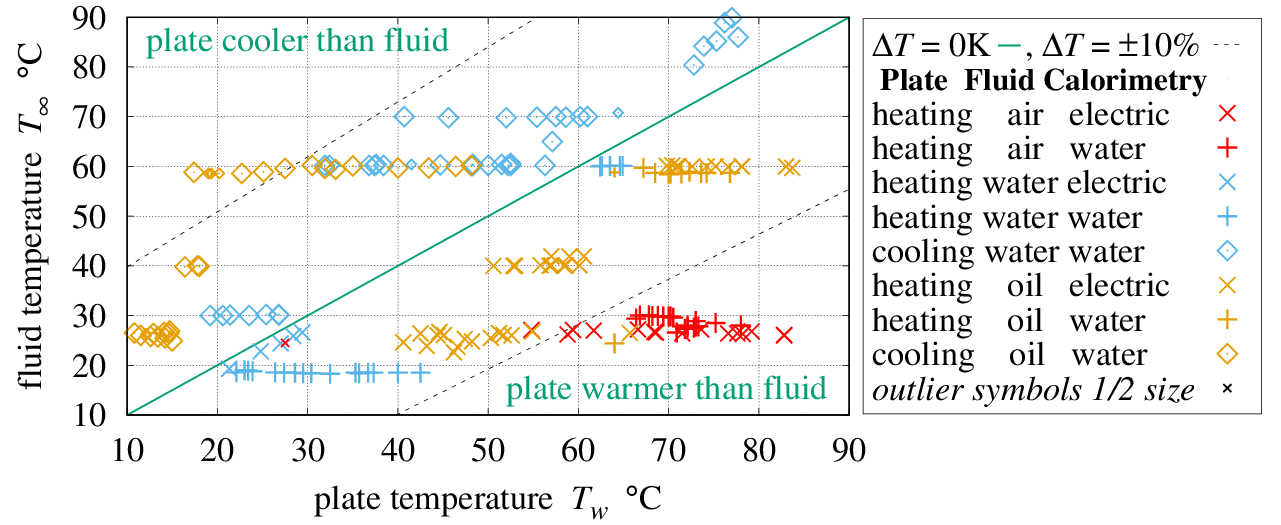}{360pt}\hfill&\cr
\+\hfill{\bf\figdef{fig:scatter}\quad UWT plate-fluid temperature pairs}\hfill&\cr
}

\subsection{Air}
 \v{Z}ukauskas and \v{S}lan\v{c}iauskas~\cite{Zukauskas1987} contains
 a table of fluid properties by temperature.  However the air
 properties were for dry air, as shown in \figref{fig:air-fits1}.  The
 monthly average relative humidity (RH) in Kaunas, Lithuania, where
 the experiments were likely performed, varies between 70\% and
 90\%.  \ref{Appendix B: Thermal and Transport Properties of Humid
 Air} details the humid-air model assembled from formulas in Kadoya,
 Matsunaga and Nagashima~\cite{doi:10.1063/1.555744},
 Wexler~\cite{Wexler1976}, Tsilingiris~\cite{TSILINGIRIS20081098}, and
 Morvay and Gvozdenac~\cite{Toolbox6}.
 The large relative errors in \figref{fig:air-fits80} shows that
 the \v{Z}ukauskas and \v{S}lan\v{c}iauskas~\cite{Zukauskas1987}
 fluid table is inconsistent with 80\%~RH at air pressure
 $P=100.725\rm~kPa$.

 Using the humid air model at $80\%$ relative
 humidity, \figref{fig:air-convection} shows the air data-sets versus
 turbulent $\Nutol$ and pierced-laminar $\Nusol$.  A single
 measurement excluded as an outlier is represented as a half-sized
 symbol.  \figref{fig:scatter} also represents outliers with
 half-sized symbols.

 These measurements were obtained from plates incorporating either an
 electric heater or a water-based calorimeter in a separate apparatus
 for each type of fluid.
 \v{Z}ukauskas and \v{S}lan\v{c}iauskas~\cite{Zukauskas1987}
 intended to measure turbulent convection, not critical transitions
 in these UWT convection data-sets; most of the $\Rey>\Rec$.  Hence,
 $\Rec$ is the data-set fitted value or 600, whichever is
 greater.

\smallskip
\vbox{\settabs 2\columns
\+\hfil\figscale{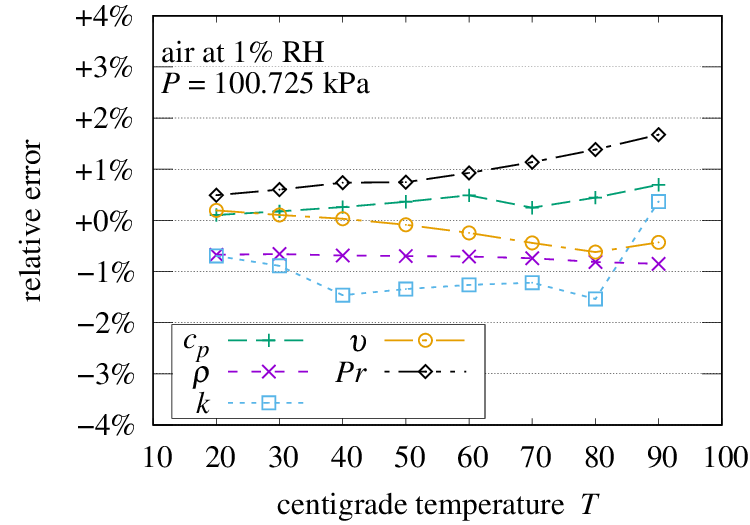}{234pt}&\hfil\figscale{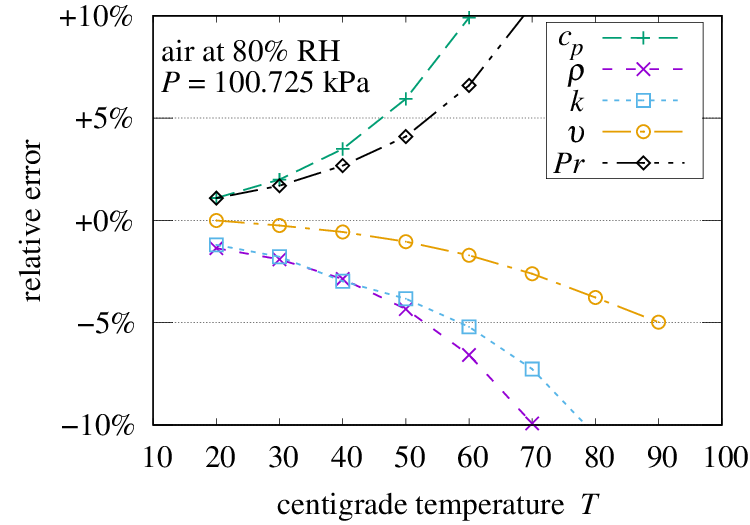}{234pt}&\cr
\+\hfil{\bf\figdef{fig:air-fits1}\quad Air property fits at 1\% RH}&\hfil{\bf\figdef{fig:air-fits80}\quad Air property fits at 80\% RH}&\cr
}

\subsection{Triggered Turbulence}
 In these tests, \v{Z}ukauskas
 and \v{S}lan\v{c}iauskas~\cite{Zukauskas1987} placed a roughness
 strip across the leading plate edge, asserting that the downstream
 flow was all turbulent.  But \figref{fig:air-convection}
 shows that $\Nutol$ is significantly less than $\Nusol$.  At
 $\Rey>\Rec$, (staged-transition) $\Nufol<\Nutol$.  This rules out
 $\Nutol$ and $\Nufol$ as explanations of \figref{fig:air-convection}.
 While these measurements exceed both laminar and turbulent
 convective heat transfer, pierced-laminar $\Nusol$ has RMSRE less than 4.3\%.

\smallskip
\vbox{\settabs 1\columns
\+\hfill\figscale{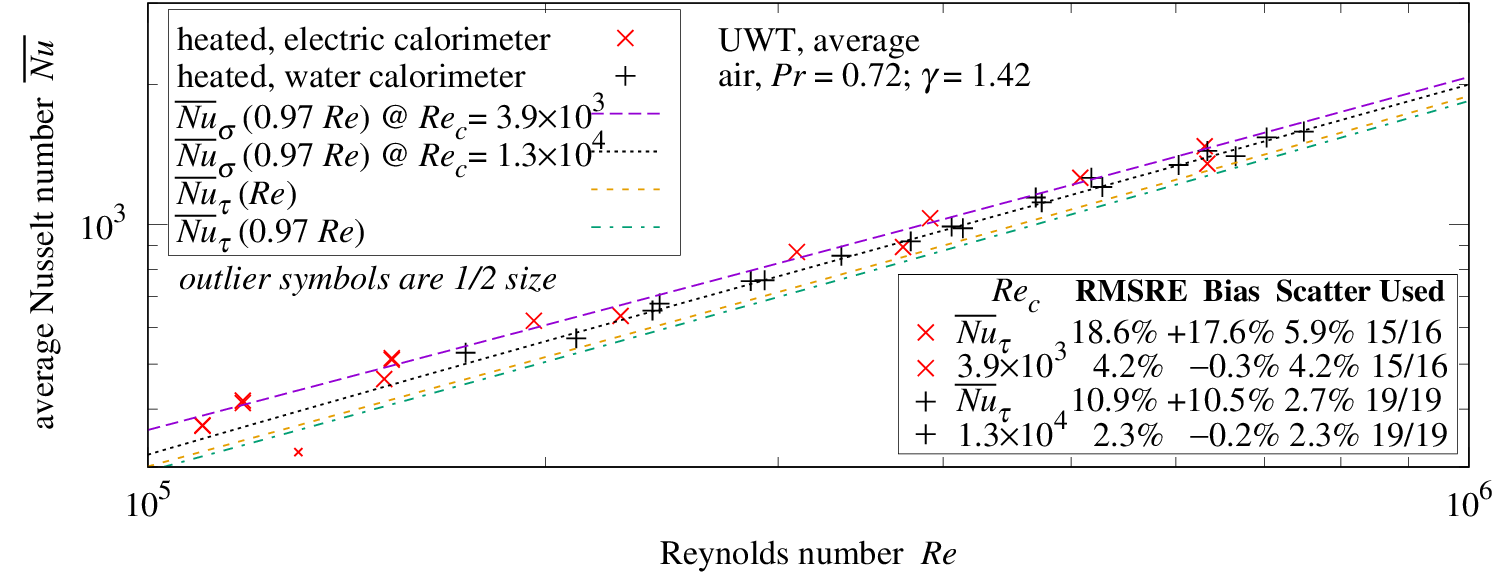}{440pt}\hfill&\cr
\+\hfill{\bf\figdef{fig:air-convection}\quad UWT average convection in air}\hfill&\cr
}

\subsection{Water}
 Unlike air, water's viscosity and thermal properties vary
 significantly with temperature.  Grouping by $\Pra$ is insufficient to
 characterize these data sets having three independent variables
 ($T_\infty$, $T_w$, and $\Rey$).  Most are between the $\Nutol$
 traces for heated and cooled plates
 in \figref{fig:water-convection}.  The measurements below the
 ``$\Nutol(0.98\,\Rey)~@~\Pra=2.05$'' trace correspond to the group of
 measurements near the top of \figref{fig:scatter}.

 The initial RMSRE calculations exceeded
 10\%.  \figref{fig:water-fits} compares the \v{Z}ukauskas
 and \v{S}lan\v{c}iauskas water property values with formulas from
 Pramuditya~\cite{Pramuditya2011} based on Wagner and
 Pru\ss~\cite{IAPWS1995}.  Correcting per these formulas did not
 significantly reduce the RMSRE values.

 This investigation hypothesizes that the $\Nuol$ measurements had
 been calculated with a constant $k=0.54\rm~W/(m\cdot{K})$ instead of
 temperature dependent values.  Correcting per this hypotheses and
 water property values, the RMSRE values for pierced-laminar $\Nusol$
 are 5.1\% or less.

\unorderedlist
 \li UHF convection is staged-transition $\Nufol$ with
 $\Pra=\Pra_\infty$.

 \li UWT convection is pierced-laminar $\Nusol$ with
 $\Pra=\Pra_w^{1/4}\,\Pra_\infty^{3/4}$.

 \li Triggered turbulence leading a smooth plate can be modeled
 as a smooth plate with a small $\Rec$.

\endunorderedlist

\vbox{\settabs 1\columns
\+\hfill\figscale{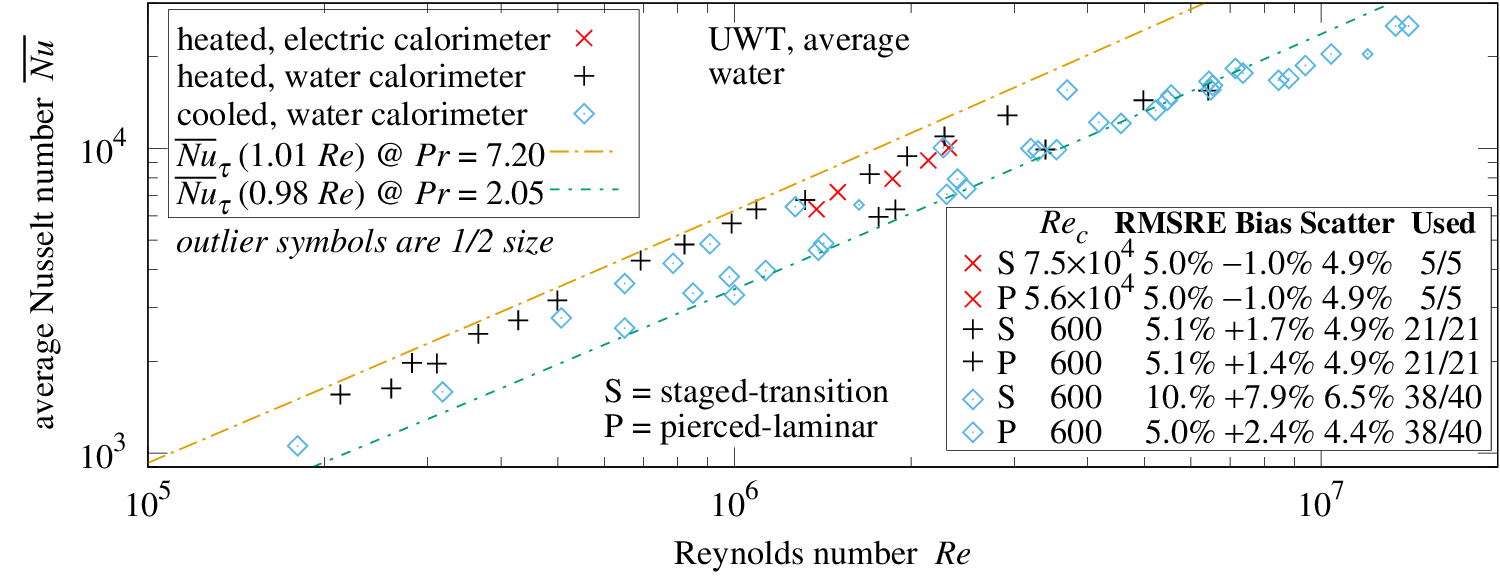}{440pt}\hfill&\cr
\+\hfill{\bf\figdef{fig:water-convection}\quad UWT average convection in water}\hfill&\cr
}

\subsection{Transformer Oil}
 There are multiple sources for the viscosity and thermal properties
 of air and water; the only source for the transformer oil used in
 these experiments
 is a 10-row table ($30^\circ{\rm C}-120^\circ{\rm C}$) in the
 appendix of \v{Z}ukauskas
 and \v{S}lan\v{c}iauskas~\cite{Zukauskas1987}.  The data-set
 temperatures span $18^\circ{\rm C}-90^\circ{\rm C}$.  Thus, there is
 no information about the oil's behavior between $18^\circ{\rm C}$ and
 $30^\circ{\rm C}$, where the slopes of the $\nu$ and $\Pra$ curves
 are changing most rapidly.  Dynamic viscosity was fit by
 $\mu=91.877\times10^{-6}\,\exp(587/[T+86.45])$.
 Having less variation, $\rho$, $k$, and specific heat (at constant
 pressure) $c_p$ were modeled by linear ramps.  \figref{fig:oil-fits}
 shows the curves hewing to \v{Z}ukauskas
 and \v{S}lan\v{c}iauskas~\cite{Zukauskas1987} within
 $\pm0.7\%$.  \figrefs{fig:oil-Pr} and \figrefn{fig:oil-nu} plot the
 table values and this investigation's hypothesized $\Pra(T)$
 and~$\nu(T)$ curves at $10^\circ{\rm C}-120^\circ{\rm C}$.

\smallskip
\vbox{\settabs 2\columns
\+\hfil\figscale{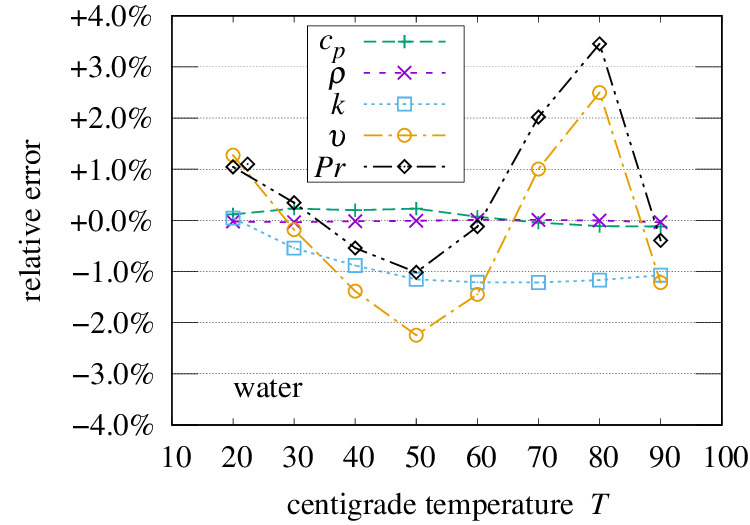}{234pt}&\hfil\figscale{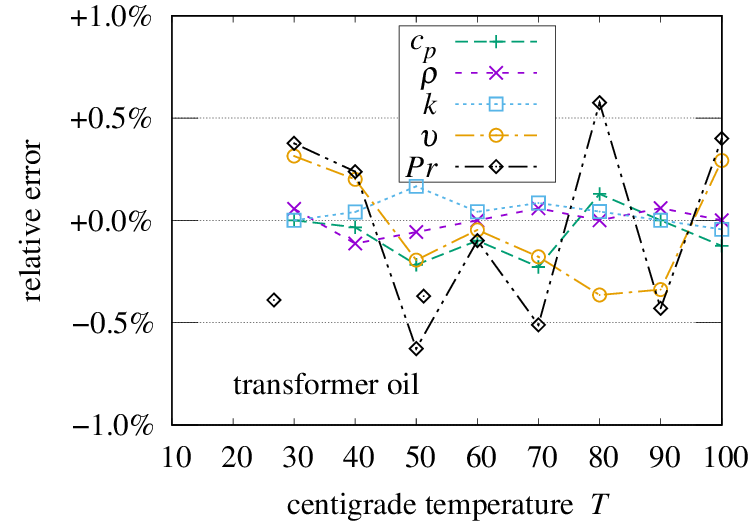}{234pt}&\cr
\+\hfil{\bf\figdef{fig:water-fits}\quad Water property fits}&\hfil{\bf\figdef{fig:oil-fits}\quad Transformer oil property fits}&\cr
}
\smallskip

\vbox{\settabs 2\columns
\+\hfil\figscale{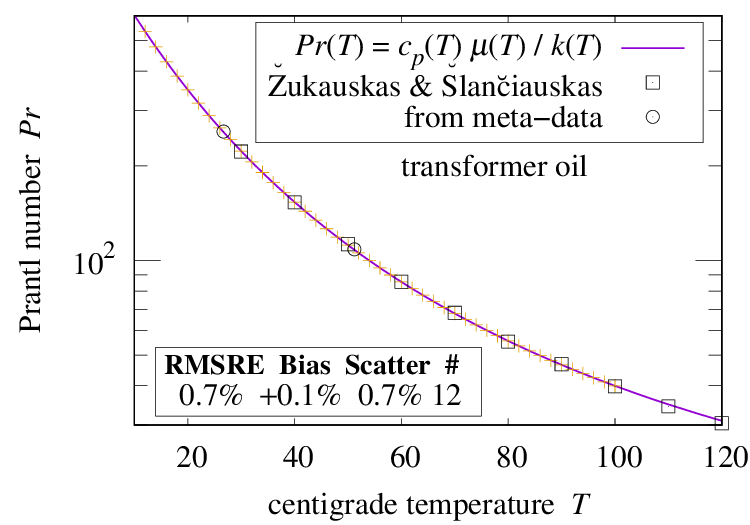}{234pt}&\hfil\figscale{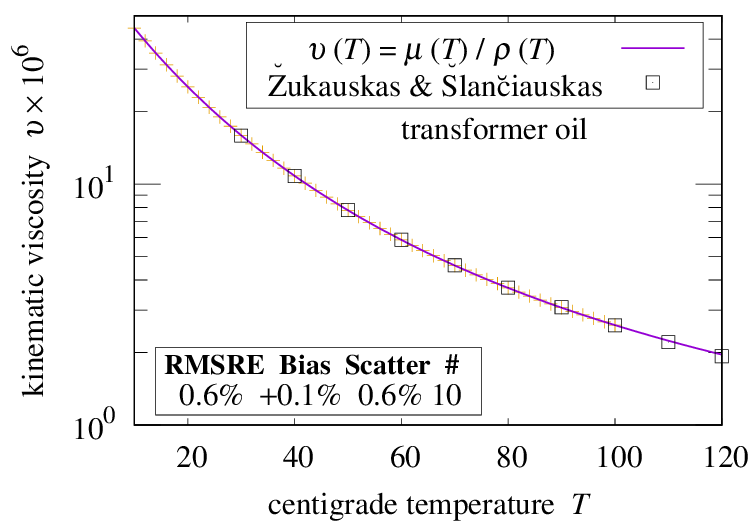}{234pt}&\cr
\+\hfil{\bf\figdef{fig:oil-Pr}\quad Transformer oil $\Pra$}&\hfil{\bf\figdef{fig:oil-nu}\quad Transformer oil $\nu$}&\cr
}

 \v{Z}ukauskas and \v{S}lan\v{c}iauskas~\cite{Zukauskas1987}
 apparently calculated transformer oil $\Nuol$ with a constant
 $k(10^\circ{\rm C})=0.126\rm~W/(m\cdot{K})$ instead of temperature
 dependent values.  As with the water measurements, this is corrected
 in this investigation's plots and RMSRE calculations.
 \figref{fig:oil-convection} plots the transformer oil measurements,
 which are nearly bounded by pierced-laminar $\Nusol$ traces for
 heated and cooled plates.

\smallskip
\vbox{\settabs 1\columns
\+\hfill\figscale{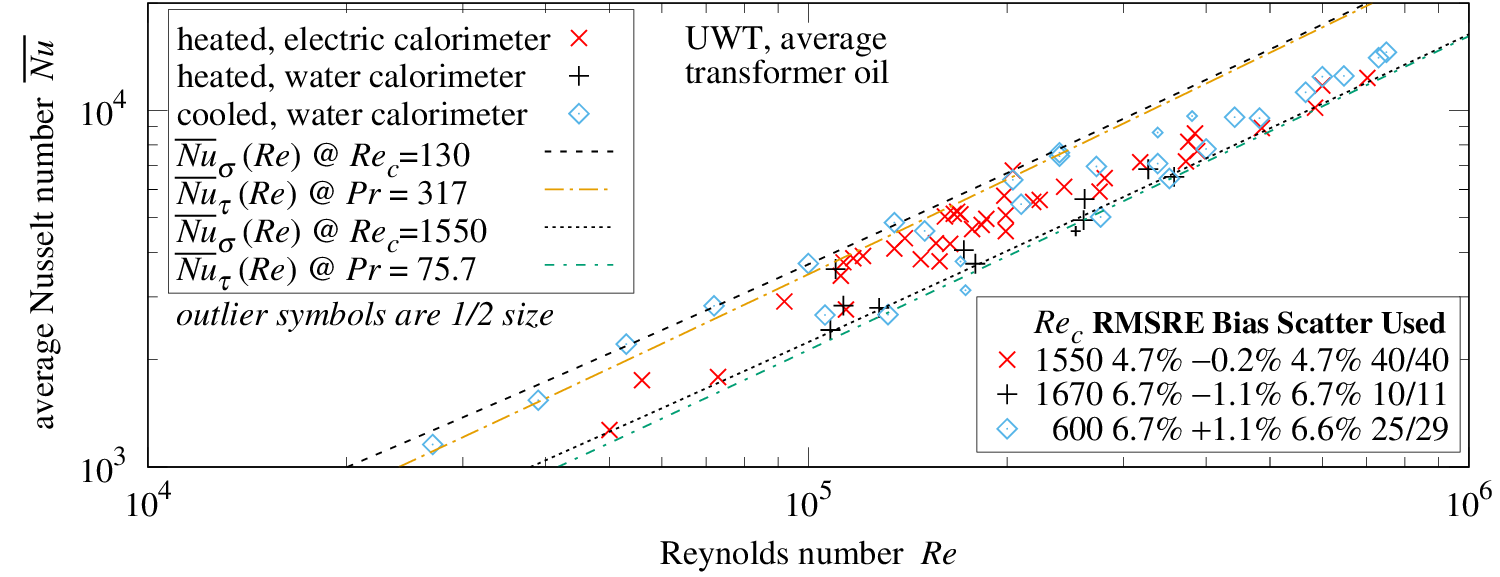}{440pt}\hfill&\cr
\+\hfill{\bf\figdef{fig:oil-convection}\quad UWT average convection in oil}\hfill&\cr
}
\smallskip

 Relative to pierced-laminar $\Nusol$
 Formula~\eqref{eq:pierced-convection}, the heating electric
 calorimeter data-set of 40 measurements had 4.7\% RMSRE.  Excluding
 one negative outlier from 11 points, the heating water calorimeter
 set had 6.7\% RMSRE.
 Excluding two positive and two negative outliers from 29 points, the
 cooling water calorimeter set also had 6.7\% RMSRE.

 Having only an incomplete source of transformer oil properties
 creates additional uncertainty for the transformer oil convection
 measurements.

 \unorderedlist \li Relative to pierced-laminar $\Nusol$
 Formula~\eqref{eq:pierced-convection}, the \v{Z}ukauskas
 and \v{S}lan\v{c}iauskas~\cite{Zukauskas1987} UWT data sets for
 air, water, and transformer oil have RMSRE values between $2.3\%$ and
 $6.7\%$.

 \endunorderedlist

\section{Results}

 All of the present work formulas which were tested are listed here
 with their prerequisites.  \tabref{tab:coverage} lists the top-level
 formulas tested by one or more data-sets.

\smallskip
\centerline{\bf\tabdef{tab:coverage}\quad Top-level formula coverage}
\vbox{\settabs 5\columns
\toprule
\+\bf Measured&\hfil\bf Plate&\bf Rough $\Pra<1$&\bf Smooth $\Pra<1$&\bf Smooth $\Pra>1$&\cr
\midrule
\+average friction&& & \eqref{eq:f-C-smooth}~$\fctol$~,~\eqref{eq:pierced-friction}~$\fcsol$&&\cr
\+local friction&& \eqref{eq:f-c-local}~$\fcr$~,~\eqref{eq:wells-local}~$\fcw$& \eqref{eq:f-c-smooth}~$\fct$& \eqref{eq:f-c-smooth}~$\fct$&\cr
\midrule
\+average convection&\hfil UWT& \eqref{eq:forced}~$\Nurol$~,~\eqref{eq:bi-level-convect}~$\Nuiol$& \eqref{eq:pierced-convection}~$\Nusol$& \eqref{eq:pierced-convection}~$\Nusol$&\cr
\+local convection&\hfil UWT& & \eqref{eq:pierced-convection-local}~$\Nus$& &\cr
\+local convection&\hfil UHF& & \eqref{eq:staged-convection-local}~$\Nu_4$& \eqref{eq:staged-convection-local}~$\Nu_4$&\cr
\bottomrule
}
\smallskip

 The local and average skin-friction coefficients from a smooth, flat,
 isothermal surface are:
$$\eqalign{
  \fcs=&{\diff[\Rex-\Rez]\,\fcsol(\Rex,\gamma\,\Rec)\over\diff\Rex}\cr
  \fcsol=&\left\|\fclol(\Rey),~\fctol(\Rey)-{\Rey_\gamma\over\Rey}\,\fctol\left(\Rey_\gamma\right)\,\right\|_{\gamma}\qquad\Rey_\gamma=\left\|\,\Rey,~{\Rec\over\sqrt{\gamma}}\,\right\|_{-8/\gamma}\cr
  &{\fclol(\Rey)}={1.328\over\sqrt{\Rey}+\sqrt{\Rez}}\qquad~\,
  \fctol(\Rey)={{2^{-5/4}}\over\left[{\Wz\left(\Rey/\sqrt{3}\right)}-1\right]^2}\cr
 }$$

 The local and average Nusselt numbers from a smooth, flat, isothermal
 plate are:

$$\eqalign{
 \Nus&=\Rex\,{\diff\Nusol(\Rex,\sqrt{\gamma}\,\Rec)\over\diff\Rex}\cr
 \Nusol&=\left\|\,\Nulol(\Rey), \Nutol(\Rey)-\Nutol\left(\|\Rey,\sqrt\gamma\,\Rec\|_{-8/\gamma}\right)\,\right\|_{\gamma}\cr
  \gamma&=1+\exp_2\left({-\Pra^{-\sqrt{1/2}}}\right)\qquad\exp_2(\varphi)\equiv2^\varphi\cr
 \Nutol(\Rey)&={\Nuz\,\Rey\,\fctol(\Rey)\over\sqrt{3}}\,\sqrt{\Pra/\sqrt{162}+1\over\sqrt{162}\,\Pra\,\fctol(\Rey)+{1}}\,\root3\of{{\Pra/\Xi\over{\|1,1/\Pra\|_3}}}\cr
 \Nulol(\Rey)&={0.664\,\Rey\,\Pra^{1/3}\over\sqrt{\Rey}+\sqrt{\Rez}}
 \qquad\Nuz={2^4\over\pi^2\,\root4\of2}
 \qquad\Xi={\left\|1,~{0.5\over\Pra}\right\|_{\sqrt{1/3}}}\cr
 }$$

  $\Pra=\Pra_w^{1/4}\,\Pra_\infty^{3/4}$; $\Pra_w$ is at
  plate temperature; $\Pra_\infty$ is at free stream temperature.

 For smooth, flat, uniform-heat-flux plates, $\Pra=\Pra_\infty$ and:
$$\eqalign{\Nu_4&=\Rex\,{\diff\Nufol(\Rex,\sqrt{\gamma}\,\Rec)\over\diff\Rex}\cr
  \Nufol&=\left\|\Nulol({\Rey_{4}}),~\Nutol(\Rey)-\Nutol({\Rey_{4}})\right\|_1\qquad\Rey_{4}=\|\Rey,\sqrt{2}\,\Rec\|_{-4}}$$

 Algorithms were presented to find~$\varepsilon$, $L_P$, and~$\Omega$
 from an elevation grid of a square portion of the rough surface.
 A plate surface is isotropic, periodic roughness when its
 $L/L_P\gg1$.

Turbulent momentum thickness $\delta_{2\tau}=\varepsilon$ at $\Ret$;
laminar momentum thickness $\delta_{2\lambda}=\varepsilon$ at $\Rel$.

$$\Ret={\sqrt{3}\,L\over3^3\;\varepsilon}\,\exp{L_P\over3^3\;\varepsilon}\qquad
  \Rel=\left[{0.664\over\varepsilon}\right]^2{L_P\,L}$$

\unorderedlist
 \li An isotropic, periodic roughness with
  $\Ret>L/\sqrt{\varepsilon\,L_P}$ behaves as a smooth surface with
  $\Rec=L/\sqrt{\varepsilon\,L_P}$ when $\Rey<\Rel$.

 \li When $\Rey>\Rel$ and $\Omega>1/2$, flow along the entire surface will
   be rough.
\endunorderedlist

 The average skin-friction coefficient~$\fcrol$ and average
 Nusselt number~$\Nurol$ of rough flow are:
$$\fcrol={1\over3\,\ln^2{\left(L/\varepsilon\right)}}\qquad
 \Nurol={\Rey\,\Pra_\infty^{1/3}\over6\,\ln^2{\left(L/\varepsilon\right)}}\qquad
 {L\over\varepsilon}\gg1$$

 Plateau islands roughness:

$$\eqalign{
\Nuiol&=\NuIol(\Rey)+\Nurol\left(\|\Rey, \ReI\|_{-4}\right)-\NuIol\left(\|\Rey, \ReI\|_{-4}\right)\cr
\NuIol&=\left\{{1-\Omega}+\left\|{\Omega\over2},{2\,\varepsilon\,[4\,\Ls]\over L_P^2}\right\|_2\right\}
  {L\over L_P}\,\Nutol\left({\Rey\,L_P\over L}\right)\cr
\ReI&={3^3\,\varepsilon^2\,L^2\over \Ls\,L_P^3}\,\ln{3^3\,\varepsilon^2\,L^2\over\sqrt{3}\,\Ls\,L_P^3}
  \qquad{[4\,\Ls]^2\over L_P^2}>{1\over2}\cr
  }$$

 Plateau wells roughness:

$$\eqalign{
 \fcw&=f_W(\Rey)+\big[\Rey_\omega/\Rey\big]\big\{\fcb(\Rey_\omega)-f_W(\Rey_\omega)\big\}
 \qquad\Rey_\omega=\|\Rey, \ReW\|_{-4}\cr
 f_W&=\left\|1,{2\,\pi\,\varepsilon\,[4\,\Ls]/ L_P^2}\right\|_{\root4\of8}\,\fct(\Rey)\cr
 \fcb&={\Omega\,\fcrol(L/\varepsilon)}+[1-\Omega]\,\fct(\Rey)\cr
 \ReW&={3^3\,\varepsilon^2\,L^2\over 2^3\,\Ls\,L_P^3}\,\ln{3^3\,\varepsilon^2\,L^2\over2^3\,\sqrt{3}\,\Ls\,L_P^3}
 \qquad{[4\,\Ls]^2\over L_P^2}<{1\over2}
 }$$

\vfill\eject

\subsection{Conformance}
 \tabrefs{tab:friction-conformance}
 and \tabrefn{tab:convection-conformance} summarize the present
 theory's conformance with 456 measurements in 32 data-sets from
 one book,
 six peer-reviewed studies, and the present apparatus.

 $L/\varepsilon=\infty$ signifies a smooth plate;
 in \tabref{tab:convection-conformance} it is followed by ``UHF'',
 ``UWT'' or ``UWT$-$'' indicating a heated UHF plate, a heated UWT
 plate, or a cooled UWT plate, respectively.

\unorderedlist
 \li Relative to the present work formulas, the 32 data-set RMSRE
 values span 0.75\% through 8.2\%.

 \li Only four of the 32 data-sets have RMSRE exceeding 6\%.

 \li Prior work formulas have smaller RMSRE on only four of the data-sets.
\endunorderedlist

\centerline{\bf\tabdef{tab:friction-conformance}\quad{Friction measurements versus present theory}}
\vbox{\settabs 9\columns
\toprule
\+\bf\hfil Source&& &$L/\varepsilon$&\bf Formula&\hfil $\Pra_\infty$&\hfil\bf RMSRE&\bf~~Bias\quad Scatter&\hfill\bf Used&\cr
\midrule
\+ \cite{CHURCHILL1993231,smith_walker_1959,spalding_chi_1964} Churchill && & $\infty$&\eqref{eq:f-C-smooth}~$\fctol$& \hfil 0.71&\hfill .74\%~\quad&\hfill$+.02\%$\qquad& .74\%\hfill 9/9~&\cr
\+ \cite{CHURCHILL1993231,smith_walker_1959,spalding_chi_1964} Churchill && & $\infty$&\eqref{eq:f-c-smooth}~$\fct$& \hfil 0.71&\hfill 1.8\%~\quad&\hfill$+1.1\%$\qquad& 1.4\%\hfill 11/11&\cr
\+ \cite{Zukauskas1987} \v{Z}ukauskas \& \v{S}lan\v{c}iauskas && & $\infty$&\eqref{eq:f-c-smooth}~$\fct$& \hfil 55.2&\hfill 2.5\%~\quad&\hfill$+0.6\%$\qquad& 2.4\%\hfill 5/5~&\cr
\+ \cite{Zukauskas1987} \v{Z}ukauskas \& \v{S}lan\v{c}iauskas && & $\infty$&\eqref{eq:f-c-smooth}~$\fct$& \hfil 5.42&\hfill 5.2\%~\quad&\hfill$+4.9\%$\qquad& 1.8\%\hfill 8/8~&\cr
\+ \cite{Zukauskas1987} \v{Z}ukauskas \& \v{S}lan\v{c}iauskas && & $\infty$&\eqref{eq:f-c-smooth}~$\fct$& \hfil 2.78&\hfill 3.3\%~\quad&\hfill$+1.0\%$\qquad& 3.1\%\hfill 8/8~&\cr
\+ \cite{Zukauskas1987} \v{Z}ukauskas \& \v{S}lan\v{c}iauskas && & $\infty$&\eqref{eq:f-c-smooth}~$\fct$& \hfil 0.71&\hfill 4.4\%~\quad&\hfill$-1.4\%$\qquad& 4.2\%\hfill 9/9~&\cr
\+ \cite{Gebers1908,Gebers1920} Gebers                           && & $\infty$&\eqref{eq:pierced-friction}~$\fcsol$& \hfil &\hfill 2.9\%~\quad&\hfill$-0.4\%$\qquad& 2.8\%\hfill 33/33&\cr
\+ \cite{A014219} Pimenta \etal && & $1.5\times10^3$&\eqref{eq:f-c-local}~$\fcr$& \hfil 0.71& \hfill4.5\%~\quad&\hfill$-3.0\%$\qquad&3.3\%\hfill 19/19&\cr
\+ \cite{Bergstrom_2005} Bergstrom \etal && & $\infty$&\eqref{eq:f-c-smooth}~$\fct$& \hfil 0.71& \hfill4.9\%~\quad&\hfill$-4.6\%$\qquad&1.7\%\hfill 4/4~&\cr
\+ \cite{Bergstrom_2005} Bergstrom \etal -- mesh&& &1600-3000&\eqref{eq:f-c-local}~$\fcr$& \hfil 0.71& \hfill3.3\%~\quad&\hfill$-2.0\%$\qquad&2.6\%\hfill 12/12&\cr
\+ \cite{Bergstrom_2005} Bergstrom \etal -- wells&& &1200-1500&\eqref{eq:wells-local}~$\fcw$& \hfil 0.71& \hfill4.4\%~\quad&\hfill$+0.8\%$\qquad&4.3\%\hfill 12/12&\cr
\bottomrule
}
\noindent Note: Churchill~\cite{CHURCHILL1993231} extracted its
measurements from Smith and Walker~\cite{smith_walker_1959} and
Spalding and Chi~\cite{spalding_chi_1964}.

\medskip
\centerline{\bf\tabdef{tab:convection-conformance}\quad{Convection measurements versus present theory}}
\vbox{\settabs 9\columns
\toprule
\+\bf\hfil Source&& &$L/\varepsilon$&\bf Formula&\hfil $\Pra_\infty$&\hfil\bf RMSRE&\bf~~Bias\quad Scatter&\hfill\bf Used&\cr
\midrule
\+ \cite{KESTIN1961133} Kestin \etal                             && & $\infty$ UWT&\eqref{eq:pierced-convection-local}~$\Nus$& \hfil 0.7&\hfill 3.8\%~\quad&\hfill$-1.0\%$\qquad&3.7\%\hfill  7/7~&\cr
\+ \cite{KESTIN1961133} Kestin \etal                             && & $\infty$ UWT&\eqref{eq:pierced-convection-local}~$\Nus$& \hfil 0.7&\hfill 8.2\%~\quad&\hfill$+1.5\%$\qquad&8.0\%\hfill 13/13&\cr
\+ \cite{Reynolds1958IV} Reynolds \etal                          && & $\infty$ UWT&\eqref{eq:pierced-convection-local}~$\Nus$& \hfil 0.71&\hfill6.4\%~\quad&\hfill$+2.3\%$\qquad&6.0\%\hfill 22/22&\cr
\+ \cite{Zukauskas1987} \v{Z}ukauskas \& \v{S}lan\v{c}iauskas && & $\infty$ UHF&\eqref{eq:staged-convection-local}~$\Nu_4$& \hfil 0.71&\hfill1.1\%~\quad&\hfill$-0.1\%$\qquad&1.1\%\hfill  8/10&\cr
\+ \cite{Zukauskas1987} \v{Z}ukauskas \& \v{S}lan\v{c}iauskas && & $\infty$ UHF&\eqref{eq:staged-convection-local}~$\Nu_4$& \hfil 0.71&\hfill2.5\%~\quad&\hfill$-1.0\%$\qquad&2.3\%\hfill  8/10&\cr
\+ \cite{Zukauskas1987} \v{Z}ukauskas \& \v{S}lan\v{c}iauskas && & $\infty$ UHF&\eqref{eq:staged-convection-local}~$\Nu_4$& \hfil 0.71&\hfill2.2\%~\quad&\hfill$-1.2\%$\qquad&1.9\%\hfill  8/10&\cr
\+ \cite{Zukauskas1987} \v{Z}ukauskas \& \v{S}lan\v{c}iauskas && & $\infty$ UHF&\eqref{eq:staged-convection-local}~$\Nu_4$& \hfil 6.57&\hfill3.7\%~\quad&\hfill$-1.1\%$\qquad&3.5\%\hfill 19/19&\cr
\+ \cite{Zukauskas1987} \v{Z}ukauskas \& \v{S}lan\v{c}iauskas && & $\infty$ UHF&\eqref{eq:staged-convection-local}~$\Nu_4$& \hfil 6.57&\hfill3.1\%~\quad&\hfill$-0.9\%$\qquad&3.0\%\hfill 15/15&\cr
\+ \cite{Zukauskas1987} \v{Z}ukauskas \& \v{S}lan\v{c}iauskas && & $\infty$ UHF&\eqref{eq:staged-convection-local}~$\Nu_4$& \hfil 108.&\hfill2.4\%~\quad&\hfill$-0.8\%$\qquad&2.3\%\hfill 17/17&\cr
\+ \cite{Zukauskas1987} \v{Z}ukauskas \& \v{S}lan\v{c}iauskas && & $\infty$ UHF&\eqref{eq:staged-convection-local}~$\Nu_4$& \hfil 257.&\hfill4.3\%~\quad&\hfill$+0.2\%$\qquad&4.3\%\hfill 17/17&\cr
\+ \cite{Zukauskas1987} \v{Z}ukauskas \& \v{S}lan\v{c}iauskas && & $\infty$ UWT&\eqref{eq:pierced-convection}~$\Nusol$& \hfil 0.71&   \hfill 4.2\%~\quad&\hfill$-0.3\%$\qquad&2.3\%\hfill 15/16&\cr
\+ \cite{Zukauskas1987} \v{Z}ukauskas \& \v{S}lan\v{c}iauskas && & $\infty$ UWT&\eqref{eq:pierced-convection}~$\Nusol$& \hfil 0.71&   \hfill 2.3\%~\quad&\hfill$-0.2\%$\qquad&2.3\%\hfill 19/19&\cr
\+ \cite{Zukauskas1987} \v{Z}ukauskas \& \v{S}lan\v{c}iauskas && & $\infty$ UWT&\eqref{eq:pierced-convection}~$\Nusol$& \hfil 5.8-7.1&\hfill 5.0\%~\quad&\hfill$-1.0\%$\qquad&4.9\%\hfill  5/5~&\cr
\+ \cite{Zukauskas1987} \v{Z}ukauskas \& \v{S}lan\v{c}iauskas && & $\infty$ UWT&\eqref{eq:pierced-convection}~$\Nusol$& \hfil 2.9-7.2&\hfill 5.1\%~\quad&\hfill$+1.4\%$\qquad&4.9\%\hfill 21/21&\cr
\+ \cite{Zukauskas1987} \v{Z}ukauskas \& \v{S}lan\v{c}iauskas && & $\infty$ UWT$-$&\eqref{eq:pierced-convection}~$\Nusol$& \hfil 2.0-5.8&\hfill 5.0\%~\quad&\hfill$+2.4\%$\qquad&4.4\%\hfill 38/40&\cr
\+ \cite{Zukauskas1987} \v{Z}ukauskas \& \v{S}lan\v{c}iauskas && & $\infty$ UWT&\eqref{eq:pierced-convection}~$\Nusol$& \hfil 75-246& \hfill 4.7\%~\quad&\hfill$-0.2\%$\qquad&4.7\%\hfill 40/40&\cr
\+ \cite{Zukauskas1987} \v{Z}ukauskas \& \v{S}lan\v{c}iauskas && & $\infty$ UWT&\eqref{eq:pierced-convection}~$\Nusol$& \hfil 80-205& \hfill 6.7\%~\quad&\hfill$-1.1\%$\qquad&6.7\%\hfill 10/11&\cr
\+ \cite{Zukauskas1987} \v{Z}ukauskas \& \v{S}lan\v{c}iauskas && & $\infty$ UWT$-$&\eqref{eq:pierced-convection}~$\Nusol$& \hfil 92-317& \hfill 6.7\%~\quad&\hfill$+1.1\%$\qquad&6.6\%\hfill 25/29&\cr
\+ present apparatus -- 3~mm bi-level&& & 102 UWT&\eqref{eq:forced}~$\Nurol$&\hfil 0.71 &\hfill1.8\%~\quad&\hfill$-0.5\%$\qquad&1.7\%\hfill  8/13&\cr
\+ present apparatus -- 3~mm bi-level&& & 102 UWT&\eqref{eq:bi-level-convect}~$\Nuiol$&\hfil 0.71 &\hfill1.8\%~\quad&\hfill$-0.5\%$\qquad&1.7\%\hfill 11/13&\cr
\+ present apparatus -- 1~mm bi-level&& & 295 UWT&\eqref{eq:posts-convect}~$\NuIol$&\hfil 0.71 &\hfill1.5\%~\quad&\hfill$-0.5\%$\qquad&1.4\%\hfill  9/14&\cr
\bottomrule
}

\section{Discussion}

 Rather than trying to tease rough flow from a nearly smooth
 surface, this investigation started with an analysis of roughness
 deep enough to disrupt boundary layer flow.

\subsection{Skin-Friction}
 With boundary layers disrupted by self-similar roughness, the flow's
 roughness velocity~$u_\rho$ was used to derive the average
 skin-friction coefficient~$\fcrol$.
 Deriving the roughness Reynolds number~$\Rev$ led to a formula for
 average turbulent skin-friction~$\fctol$ with unprecedented
 0.75\% RMSRE fidelity to the Smith and
 Walker~\cite{smith_walker_1959}, and Spalding and
 Chi~\cite{spalding_chi_1964} measurements (via
 Churchill~\cite{CHURCHILL1993231}).

 \ref{Local Skin-Friction Coefficients} and subsequent comparisons with
 measurements established that transforms between local and average
 friction differ for continuous versus disrupted boundary layers.
 Thus, these transforms are not valid for combined flows over plateau
 roughnesses which shed rough and turbulent flow simultaneously.

\subsection{Laminar Friction}
 $\fclol$ Formula~\eqref{eq:laminar-friction} differs from the
 traditional $1.328/\sqrt{\Rey}$ formula because
 Formula~\eqref{eq:laminar-friction} is valid for all $\Rey\ge0$, with
 $\fclol(0)=1.328/\sqrt{\Rez}\approx0.0542$.
 This removes the need to treat the leading edge differently from
 the rest of the plate in many cases.

 Measuring smooth plate $\fclol$ at $\Rey=1000$ in a viscous liquid,
 then solving Formula~\eqref{eq:laminar-friction} for $\Rez$
 would refine the $\Rez$ value.

\subsection{Forced Convection}
 Combining Formula~\eqref{eq:Colburn} with rough friction $\fcrol$
 Formula~\eqref{eq:f-C} results in the formula
 $\Nurol={\Rey\,\Pra^{1/3}/[6\,\ln^{2}{\left(L/\varepsilon\right)}]}$,
 which is identical to rough convection
 Formula~\eqref{eq:forced}.

 Formula~\eqref{eq:Colburn} is the original (1933) form of the
 Reynolds-Colburn analogy.  Lienhard~\cite{10.1115/1.4046795}
 demonstrates that the analogy can fail for turbulent flows.
 In particular, the $\Pra$ exponent should be $0.6$ in gasses.

 Formula~\eqref{eq:forced} using $\Pra^{1/3}$ matches rough
 convection from the $\varepsilon=3$~mm apparatus spanning
 $4000<\Rey<50000$ within its estimated measurement uncertainties
 (plotted in \ref{Appendix A: Apparatus and Measurement Methodology}).
 With $\Pra=0.71$ (air), $\Pra^{0.6}\approx0.814$ is 9\% smaller than
 $\Pra^{1/3}\approx0.892$; this mismatch exceeds some of those
 estimated measurement uncertainties.

 Thus, the $\Pra$ dependence of rough and turbulent convection
 differ.

 None of the present or cited experiments measured fluids with
 $0<\Pra<0.7$.  Forced convection heat transfer measurements in this
 range are needed.

\subsection{Onset of Rough Flow}
 The present theory predicts that an isotropic, periodic rough plate
 with $\Omega>1/2$ switches from all laminar to all rough flow
 as $\Rey>\Rel$.  The prior and present work measurements from rough
 plates were taken at $\Rey>\Rel$; most were at $\Rey\gg\Rel$.
 Measurements at $\Rey$ values closely bracketing $\Rel$ are needed to
 test $\Rel$ Formula~\eqref{eq:Re-laminar}.

\subsection{Pierced-Laminar}
 \ref{Rough Heat Transfer Measurements} presented indirect evidence of
 duck tape generating pierced-laminar convection exceeding both
 laminar and turbulent convection.

 To definitively test pierced-laminar flow from roughness, convection
 or skin-friction measurements at $\Rey$ values near $\Rec$ are needed
 from a surface having $\Rec=L/\sqrt{\varepsilon\,L_P}<\Ret<\Rel$.

\subsection{Expected Measurement Uncertainty}
 The present work made several claims based on measurements being
 within the present apparatus's expected measurement uncertainties.
 Attributing discrepancies to parameters was robust because
 each parameter affected different $\Rey$ ranges:
\unorderedlist
 \li The pierced-laminar convection of duck tape-covered sides affected
 both plates at~$\Rey>1000$.

 \li Blending plateau islands convection using the $\ell^{-4}$-norm
 affected plates only near their $\ReI$ values of 6178 and 55566,
 respectively.

 \li The $\Pra^{1/3}$ factor in rough convection affected
 the $\varepsilon=3$~mm plate at $\Rel<\Rey<\ReI$.
\endunorderedlist

\subsection{Plate-Parallel Forced Flow}
 This investigation treats forced flow which is parallel to the plate.
 If the slope of flow relative to the plane of a rough surface exceeds
 $\varepsilon/L$, then they are not effectively parallel.
 Flow along the $\varepsilon/L=3/305$ plate should be aligned within
 $0.56^\circ$; the $\varepsilon/L=1/305$ plate, within $0.188^\circ$.

 If the flow slope relative to a smooth surface exceeds the ratio of
 momentum thickness per characteristic length, then they are not
 effectively parallel.  For laminar and turbulent flows:
$${\delta_{2\lambda}\over{L}}={0.664\over\sqrt{\Rey}}\qquad
  {\delta_2\over{L}}={1/27\over\Wz\left({\Rey/\sqrt{3}}\right)}$$

\subsection{Chung, Hutchins, Schultz, and Flack}
 Of the studies cited by
 Chung \etal \cite{doi:10.1146/annurev-fluid-062520-115127}, some
 plates would qualify as isotropic, periodic roughness; testing them
 against the present theory requires isobaric flow, roughness height
 maps (or enough dimensional information to create them), free-stream
 $\Rey$, and skin-friction coefficient or convection heat transfer
 measurements.  Although some of the studies are likely to have this
 information, few, if any, report all of these details in the published
 articles.

\section{Conclusions}

\unorderedlist
 \li The pipe-plate analogy fails for roughness because rough
   skin-friction coefficients can be less than smooth regime
   coefficients for external plates, but not inside pipes.

 \li While understanding the nature of the flow shed by
   roughness is of theoretical interest, it is not needed for
   determining the skin-friction coefficient from a rough surface in
   an isobaric flow.  The present theory is independent of turbulence
   theory.

 \li Modeling the flow along a rough plate as repeated boundary-layer
  disruptions leads to exact formulas calculating the skin-friction
  coefficient and Nusselt number of a flat plate given its characteristic length,
  RMS height-of-roughness, isotropic spatial
  period, openness, Reynolds number, and the
  fluid's Prandtl numbers at free-stream and plate temperatures.

 \li These new equations
 offer improved accuracy or improved range relative to prior works.
 They were tested with 456 heat transfer and friction measurements in 32
 data-sets from one book, six peer-reviewed studies, and the present
 apparatus.

\unorderedlist
 \li Relative to the present work formulas, the 32 data-set's RMSRE
 values span 0.75\% through 8.2\%.

 \li Only four of the 32 data-sets have RMSRE exceeding 6\%.

 \li Prior work formulas have smaller RMSRE on only four of the data-sets.
 \endunorderedlist

\endunorderedlist

\beginsection{Acknowledgments}

 Thanks to John Cox (1957-2022) and Doug Ruuska for machining the
 bi-level plate.
 Thanks to Martin Jaffer for critiques and insights.
 Thanks to anonymous reviewers for their useful suggestions.

\beginsection{Abbreviations}

 The following abbreviations are used in this manuscript:

\abbreviation{ARM}{computer processor architecture}
\abbreviation{LM35C}{temperature sensor integrated circuit}
\abbreviation{MIC-6 Al}{an aluminum alloy}
\abbreviation{MPXH6115A6U}{air pressure sensor integrated circuit}
\abbreviation{PIR}{polyisocyanurate foam}
\abbreviation{PLL}{phase-locked loop, synchronization control method}
\abbreviation{PL}{Bergstrom \etal \cite{Bergstrom_2005} perforated sheet large}
\abbreviation{PM}{Bergstrom \etal \cite{Bergstrom_2005} perforated sheet medium}
\abbreviation{PS}{Bergstrom \etal \cite{Bergstrom_2005} perforated sheet small}
\abbreviation{RAM}{random access memory}
\abbreviation{RH}{relative humidity (\%)}
\abbreviation{RMS}{root-mean-squared}
\abbreviation{RMSRE}{root-mean-squared relative error (\%)}
\abbreviation{RSS}{root-sum-squared}
\abbreviation{SGL}{Bergstrom \etal \cite{Bergstrom_2005} sandpaper grit large}
\abbreviation{SGML}{Bergstrom \etal \cite{Bergstrom_2005} sandpaper grit medium-large}
\abbreviation{SGM}{Bergstrom \etal \cite{Bergstrom_2005} sandpaper grit medium}
\abbreviation{SGS}{Bergstrom \etal \cite{Bergstrom_2005} sandpaper grit small}
\abbreviation{SM}{Bergstrom \etal \cite{Bergstrom_2005} smooth plate}
\abbreviation{STM}{STMicroelectronics, integrated circuit manufacturer}
\abbreviation{UHF}{uniform heat flux (W/m$^2$)}
\abbreviation{USB}{Universal Serial Bus}
\abbreviation{UWT}{uniform wall temperature, isothermal (K)}
\abbreviation{WML}{Bergstrom \etal \cite{Bergstrom_2005} wire mesh large}
\abbreviation{WMM}{Bergstrom \etal \cite{Bergstrom_2005} wire mesh medium}
\abbreviation{WMS}{Bergstrom \etal \cite{Bergstrom_2005} wire mesh small}
\abbreviation{XPS}{extruded polystyrene foam}

\section{Nomenclature}

\nomenclature[A]{$A$}{surface area ($\rm m^2$)}
\nomenclature[A]{$\Cf/2, \Cfol/2$}{local, average skin-friction coefficient Mills-Hang~\cite{OSFCFRFP} and Pimenta \etal \cite{A014219}}
\nomenclature[A]{$c_p$}{fluid specific heat at constant pressure (${\rm J/(kg\cdot K)}$)}
\nomenclature[A]{$\cf, \cfol$}{local, average skin-friction coefficient Prandtl-Schlichting~\cite{prandtl1934resistance}}
\nomenclature[A]{$\fc, \fcol$}{local, average skin-friction coefficient present work}
\nomenclature[A]{$G(t,w)$}{Gray-code self-similar ramp-permutation}
\nomenclature[A]{$h,\overline{h}$}{local, average convective surface conductance (${\rm W/(m^2\cdot K)}$)}
\nomenclature[A]{$j_P$}{period index, the index of largest $X_j$ or $X_{j,k}$}
\nomenclature[A]{$k$}{fluid thermal conductivity (${\rm W/(m\cdot K)}$)a}
\nomenclature[A]{$k_S$}{sand-roughness (m)}
\nomenclature[A]{$L$}{plate characteristic length (m)}
\nomenclature[A]{$L_P$}{roughness spatial period (m)}
\nomenclature[A]{$\Ls$}{ratio of plateau convex region area to its perimeter (m)}
\nomenclature[A]{$\Nu,\Nuol$}{local, average Nusselt number (convection)}
\nomenclature[A]{$\Pra$}{Prandtl number of fluid}
\nomenclature[A]{$q$}{positive integer $=\log_2w$}
\nomenclature[A]{$\Rey$}{Reynolds number of flow parallel to the plate}
\nomenclature[A]{$\Rec$}{purely laminar upper-bound}
\nomenclature[A]{$\ReI, \ReW$}{$\Rex$ rough-to-turbulent flow threshold}
\nomenclature[A]{$\Rel, \Ret$}{laminar, turbulent $\Rey$ upper-bound}
\nomenclature[A]{$\Rev, \Rey_k$}{roughness, sand-roughness Reynolds number}
\nomenclature[A]{$\Rex$}{local Reynolds number $={x\,\Rey/L}$}
\nomenclature[A]{$\Rez$}{$\Rex$ integration lower bound}
\nomenclature[A]{$\Reu$}{$\Rex$ of leading unheated band}
\nomenclature[A]{$S_{j,k}$}{matrix of elevations}
\nomenclature[A]{$t$}{integer}
\nomenclature[A]{$u, u_\rho$}{bulk fluid, friction velocity ($\rm m/s$)}
\nomenclature[A]{$W(t,w)$}{wiggliest integer self-similar ramp-permutation}
\nomenclature[A]{$\Wz$}{principal branch of the Lambert $\W$ function}
\nomenclature[A]{$w$}{integer power of two $=2^q$}
\nomenclature[A]{$X_j, X_{j,k}$}{discrete Fourier transform coefficient}
\nomenclature[A]{$x, x_u$}{distance, unheated distance from leading edge of plate (m)}
\nomenclature[A]{$Y(t,w)$}{integer self-similar ramp-permutation}
\nomenclature[A]{$Z$}{roughness random variable (m)}
\nomenclature[A]{$z(x),z(x,y)$}{roughness elevation function (m)}
\nomenclature[A]{$\overline{z}$}{mean elevation of roughness function (m)}

\subsection{Greek Symbols}

\nomenclature[G]{$\gamma,\gmprm$}{exponent $p$ of the $\ell^p$-norm}
\nomenclature[G]{$\delta_2$}{momentum thickness of boundary-layer flow (m)}
\nomenclature[G]{$\delta_{2\lambda},\delta_{2\tau}$}{laminar, turbulent flow momentum thickness (m)}
\nomenclature[G]{$\epsilon,\varepsilon$}{profile, surface RMS height-of-roughness (m)}
\nomenclature[G]{$\nu$}{fluid kinematic viscosity (${\rm m^2/s}$)}
\nomenclature[G]{$\Omega$}{ratio of non-plateau area to cell area (${\rm m^2/m^2}$)}
\nomenclature[G]{$\rho$}{fluid density (${\rm kg/m^3}$)}
\nomenclature[G]{$\tau, \tau_2$}{fluid shearing stress (${\rm N/m^2}$)}
\nomenclature[G]{$\varsigma$}{peak elevation of roughness (m)}
\nomenclature[G]{$\varphi, \vartheta$}{mathematical scalar variables}

\subsection{Subscripts}

\nomenclature[U-a]{$~_\lambda$}{Laminar flow}
\nomenclature[U-b]{$~_\tau$}{Turbulent flow}
\nomenclature[U-c]{$~_\rho$}{Rough flow}
\nomenclature[U-d]{$~_\sigma$}{Pierced-laminar flow (smooth UWT plate)}
\nomenclature[U-e]{$~_4$}{Staged-transition (smooth UHF plate)}
\nomenclature[U-f]{$~_I$}{Platform islands}
\nomenclature[U-g]{$~_\iota$}{Platform islands}
\nomenclature[U-h]{$~_\beta$}{Platform wells}
\nomenclature[U-i]{$~_W$}{Platform wells}
\nomenclature[U-j]{$~_\omega$}{Platform wells}

\section{References}

\unskip

\bibliographystyle{unsrtDOI}
\bibliography{citations}

\section{Appendix A: Apparatus and Measurement Methodology}

 The goal of the present apparatus was to measure forced convection
 heat transfer from a precisely rough plate over the widest practical
 span of airflow velocities.

 Although more complicated to analyze, the plate was suspended, not
 embedded, in the wind-tunnel.  The measurements from prior
 investigations which embedded the plate in a wind-tunnel wall were
 largely incompatible with the present theory because their flows were
 not isobaric.

\subsection{The Plate}
 \figref{fig:front} shows the rough surface of the test plate; it was
 milled from a slab of MIC-6 aluminum (Al) to have (676 of) square
 $8.33{\rm\,mm}\times8.33{\rm\,mm}\times6{\rm\,mm}$ posts spaced on
 11.7~mm centers over the $30.5~{\rm cm} \times 30.5~{\rm cm}$ plate.
 The area of the top of each post was $0.694{\rm\,cm}^2$, which was
 50.4\% of its $1.38{\rm\,cm}^2$ cell.  The RMS height-of-roughness
 $\varepsilon=3.00{\rm~mm}$.  Openness $\Omega\approx49.6\%$.
 Embedded in the plate are 9 electronic resistors as heating elements
 and a Texas Instruments LM35 Precision Centigrade Temperature Sensor.
 2.54~cm of thermal insulating foam separates the back of the plate
 from a 0.32~mm thick sheet of aluminum with an LM35 at its center.
 \figref{fig:crosect} is a cross-section illustration of the plate assembly.

\medskip
\vbox{\settabs 2\columns
\+\hfil\figscale{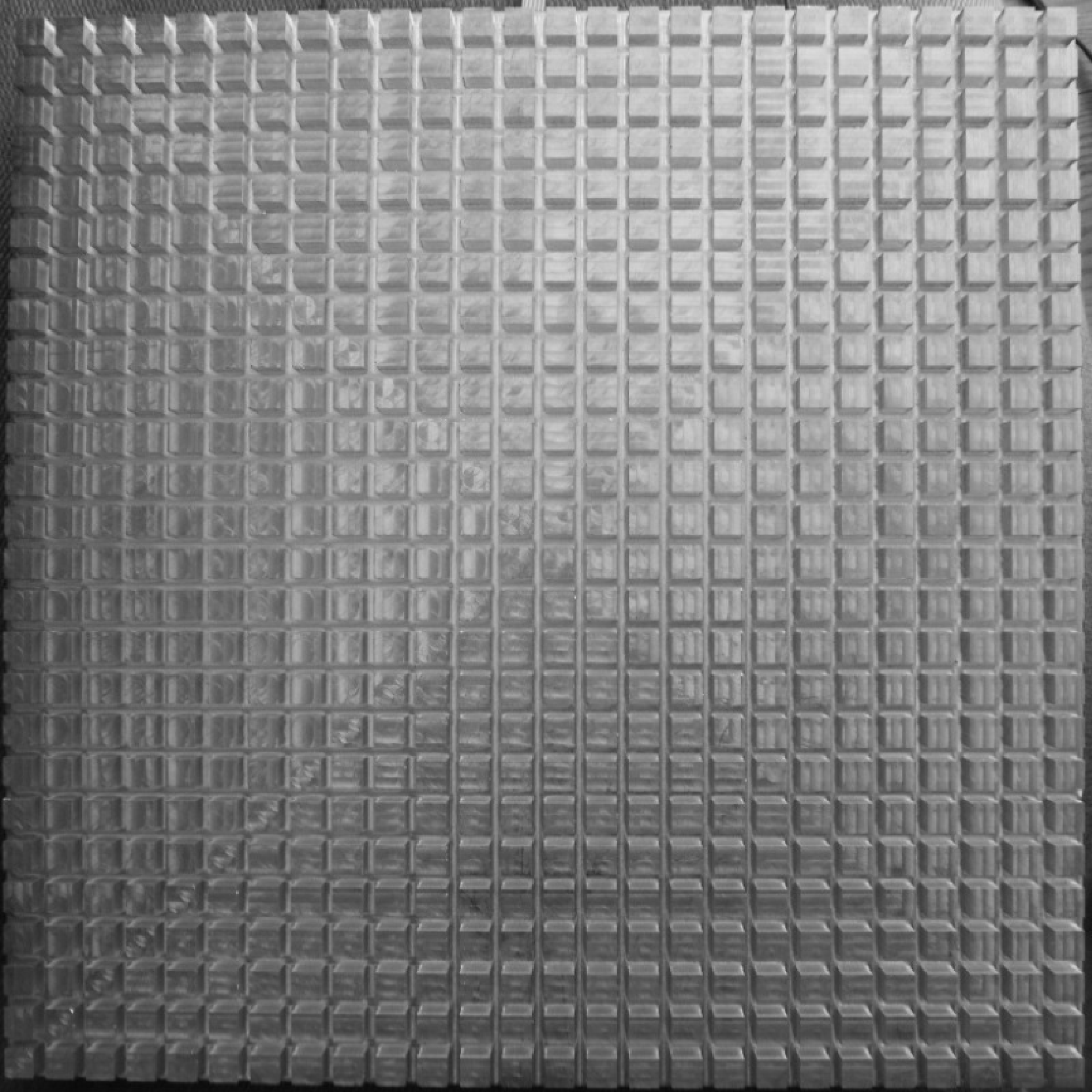}{200pt}&
  \hfil\figscale{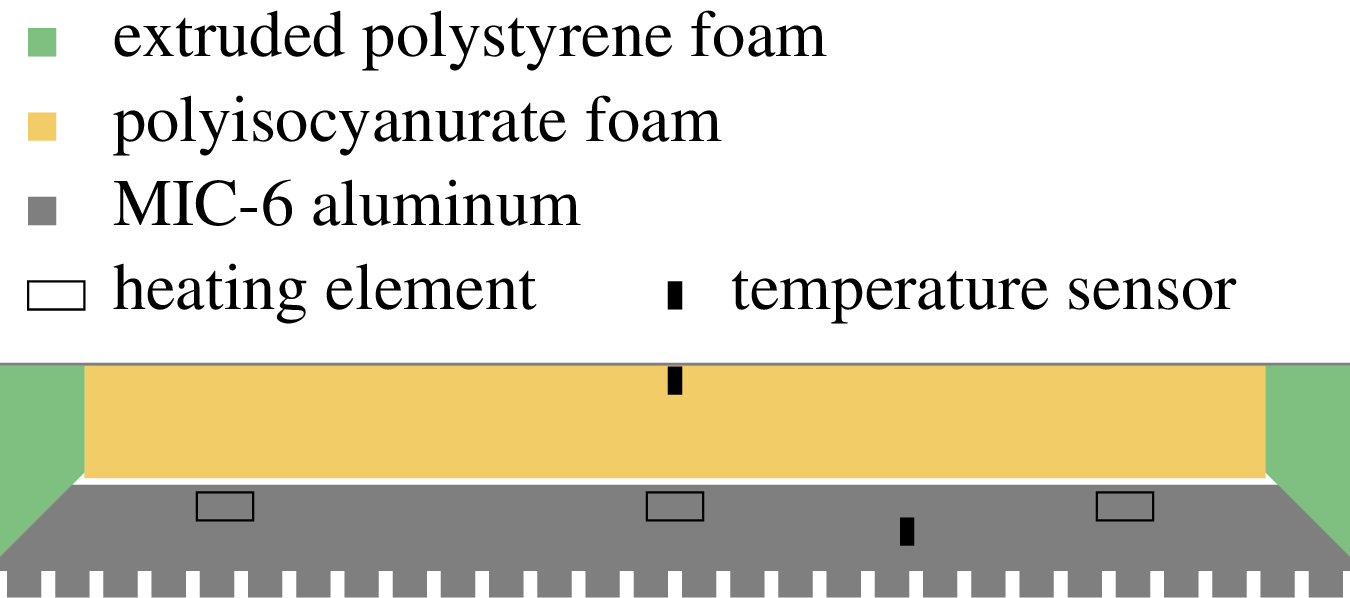}{200pt}&\cr
\+\hfil{\bf\figdef{fig:front}\quad Rough surface of plate}&
  \hfil{\bf\figdef{fig:crosect}\quad Plate assembly cross-section}&\cr
}

\subsection{Wind Tunnel}
 The fan pulls air from the test chamber's open intake through the
 test chamber.  The fan blows directly into a diffuser made of folded
 plastic mesh to disrupt vortexes generated by the fan.  In a
 sufficiently large room, the disrupted vortexes dissipate before
 being drawn into the open intake.

 To guarantee isobaric (no pressure drop) flow, the wind-tunnel must
 be sufficiently large that its test chamber and plate assembly
 boundary-layers do not interact at fan-capable airspeeds.

\medskip
\vbox{\settabs 2\columns
\+\hfil\figscale{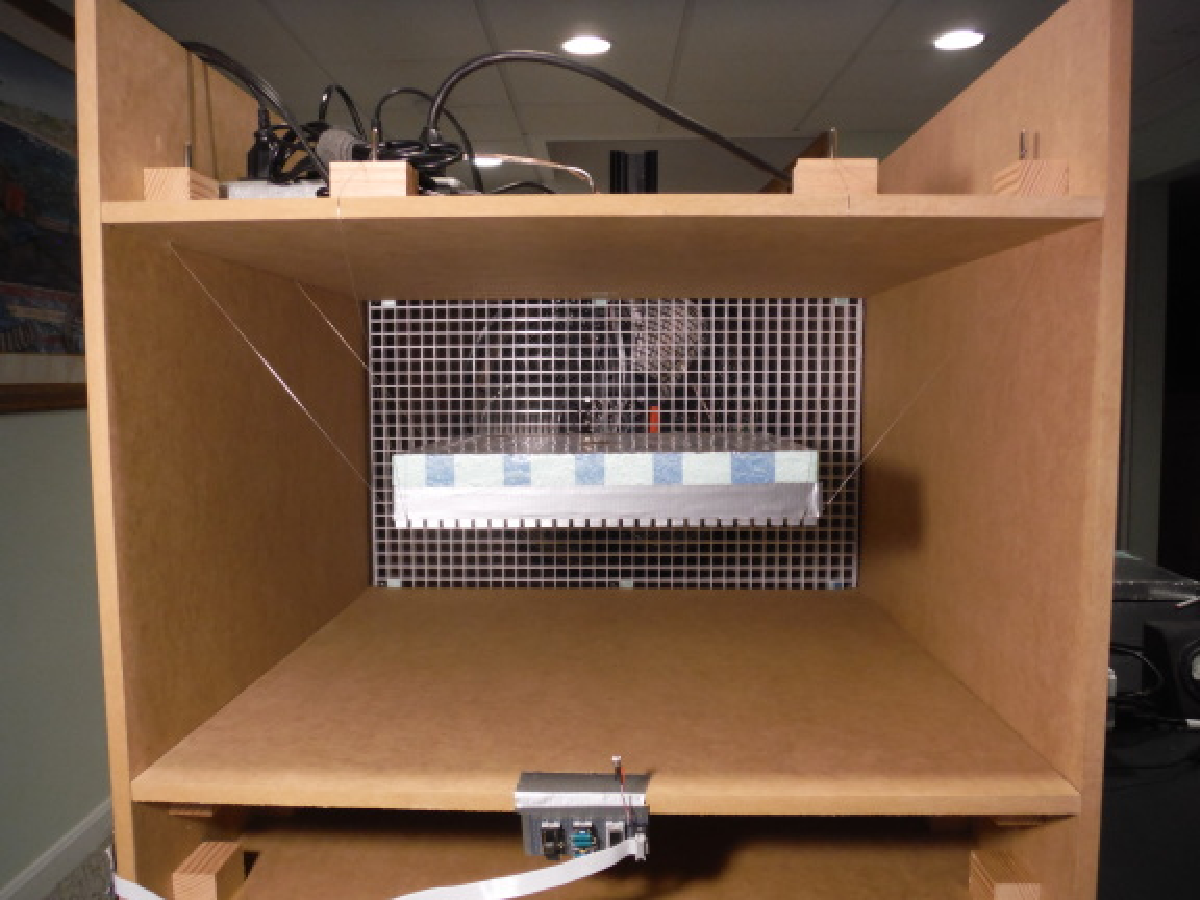}{217pt}&
  \hfil\figscale{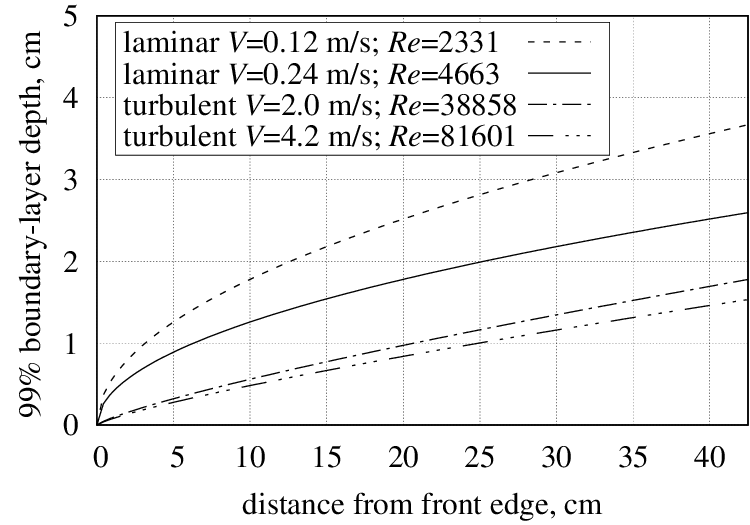}{234pt}\cr
\+\hfil{\bf\figdef{fig:suspend}\quad $\varepsilon=3\rm~mm$ plate in wind-tunnel}&
  \hfil{\bf\figdef{fig:WTBL}\quad Wind-tunnel boundary-layers}\cr
}

The wind-tunnel test chamber in \figref{fig:suspend} has a
$61{\rm~cm}\times35.6{\rm~cm}$ cross-section and a 61~cm depth.  This
allows the plate assembly to be centered in the wind-tunnel with 15~cm
of space on all sides.
The fan pulling air through the test chamber produces a maximum
airspeed of 4.65~m/s ($\Rey\approx9.2\times10^4$ along the 30.5~cm
square plate).  Its minimum nonzero airspeed is 0.12~m/s
($\Rey\approx2300$).

 Test chamber laminar and turbulent 99\% boundary-layer
 thicknesses (Schlichting \cite{schlichting2014}) are:
$$\delta_\lambda=4.92\,\sqrt{x\nu\over u}\qquad
  \delta_\tau=0.37\,x^{4/5}\left[\nu\over u\right]^{1/5}\eqdef{eq:tunnel-WTBL}$$

\figref{fig:WTBL} shows that the 15~cm clearance between the plate and
the test chamber walls is sufficient to prevent their boundary-layers
from interacting at airspeeds within the fan's capabilities.

 The plate assembly is suspended from six lengths of 0.38 mm-diameter
 steel piano wire terminated at twelve zither tuning pins in wooden
 blocks fastened to the exterior of the test chamber.
The plate is suspended face down to minimize the natural convection
from the test-surface.
With the plate assembly in the test chamber, the airspeed increases in
proportion to the reduction of test chamber aperture $A_e$ by the
plate's cross-sectional area $A_\times$:
$${u_\times\over u}={A_e\over A_e-A_\times}\approx107.6\%\eqdef{eq:u-aperture}$$

\subsection{Automation}
Data capture and control of convection experiments are performed by an
``STM32F3 Discovery 32-Bit ARM M4 72MHz'' development board.
The program written for the STM32F3 captures readings and writes them
to the microprocessor's non-volatile RAM, controls the plate heating,
servos the fan speed, and later uploads its data to a computer through
a USB cable.

Once per second during an experiment, the program calibrates and reads
each on-chip 12~bit analog-to-digital converter 16 times, summing the
sixteen 12~bit readings to create a 16~bit reading per converter.

Rotations of the fan are sensed when a fan blade interrupts
an infrared beam.  The microprocessor controls a solid-state relay
(supplying power to the fan) to maintain a fan rotation rate,
$\omega$, which is dialed into switches.  At~$\omega\le210$~r/min, the
microprocessor pulses power to the fan to phase-lock the beam
interruption signal to an internal clock.  At~$\omega>210$~r/min, the
microprocessor servos the duty cycle of a 7.5~Hz square-wave gating
power to the fan.  This system operates at
$32{\rm~r/min}<\omega<1400{\rm~r/min}$.

\medskip
\vbox{\settabs 2\columns
\+\hfill\figscale{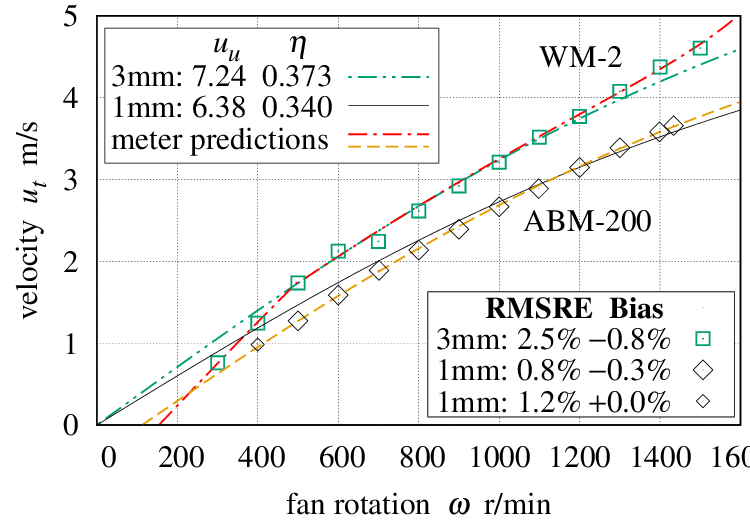}{234pt}\hfill
 &\hfill\figscale{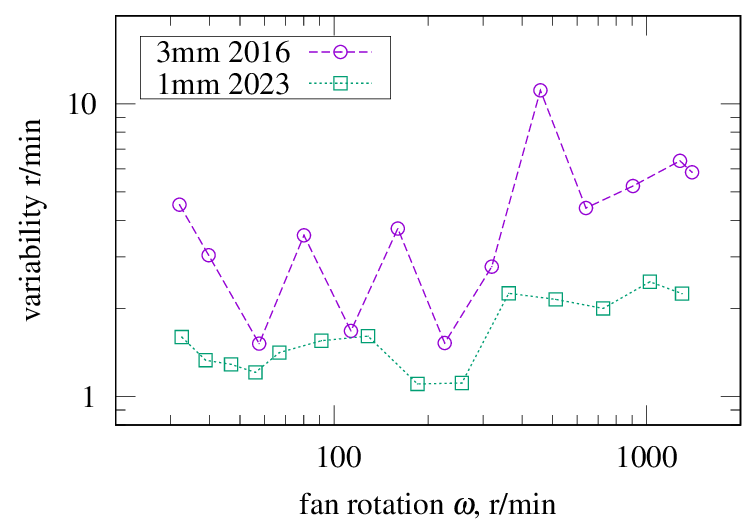}{234pt}\hfill&\cr
\+\hfill{\bf\figdef{fig:tunnel-fit}\quad Airspeed versus fan speed}\hfill
 &\hfill{\bf\figdef{fig:rpm-dn}\quad Fan PLL variability}\hfill&\cr
}

\subsection{Calibration}
 The correspondence between fan rotation rate~$\omega$ and test
 chamber airspeed~$u$ was determined using an ``Ambient
 Weather~WM-2'', which specifies an accuracy of $\pm3\%$ of reading.
 After 2017 an ``ABM-200 Airflow \& Environmental Meter'' specifying
 an accuracy of $\pm0.5\%$ of reading between 2.2~m/s and 62.5~m/s,
 was used.

 The ``UtiliTech 20 inch 3-Speed High Velocity Floor Fan'' has three
 blades with maximum radius $r=0.254$~m.  Its characteristic length is
 its hydraulic-diameter, $D_H=0.550$~m.  The velocity of the blade tips
 is $2\,\pi\,r\,\omega/60$, where $\omega$ is the number of rotations
 per minute.  The Reynolds number of the fan is:
$$\Rey_f={2\,\pi\,r\,D_H\,\omega/60\over3\,\nu}\eqdef{eq:fan-Re}$$

 The 3 blade tips trace the whole circumference in only 1/3 of a
 rotation, hence the 3 in the denominator.

 Faster fan rotation $\omega$ yields diminishing increases of
 test-chamber airspeed $u_t$, suggesting Formula~\eqref{eq:Re-tunnel},
 where $u_u$ is the limiting velocity for arbitrarily fast rotation,
 and coefficient $\eta$ converts fan $\Rey_f$ to test-chamber $\Rey_t$.
 \figref{fig:tunnel-fit} gives the parameters and
 measurements at $300{\rm~r/min}\le\omega\le1500{\rm~r/min}$.  The
 ``3mm'' points are the WM-2 measurements of the 3~mm plate in the
 original wind-tunnel; The ``1mm'' points are the ABM-200 measurements
 of the 1~mm plate in the tunnel with a new diffuser and fan cowling.
$$\Rey_t=\|\eta\,\Rey_f,~D_H\,u_u/\nu\|_{-2}\qquad
  u_t=\|\pi\,\eta\,r\,\omega/90,~u_u\|_{-2}\eqdef{eq:Re-tunnel}$$

 Airspeeds slower than 2~m/s should be nearly proportional to
 $\omega$.  Both anemometers show evidence of dry (bearing) friction
 in \figref{fig:tunnel-fit}.  The ABM-200 ``meter predictions'' trace
 plots $1.125\,u_t-0.381$; the WM-2 ``meter predictions'' trace plots
 $1.477\,u_t-0.81$ when $u_t<1.725$ and $u_t$ otherwise.
 A mistake in the 2016 measurement software under-counted fan
 rotations at $\omega>1200$~r/min.  It is compensated by replacing
 $\omega$ in Formulas (\eqrefn{eq:fan-Re}, \eqrefn{eq:Re-tunnel}) with
 $[\,\omega^{-6}-1750^{-6}]^{-1/6}$ in the WM-2 ``meter predictions''.
  The RMSRE and Bias are relative to the ``meter predictions''.  The
  second ``1mm'' row includes the point at 400~r/min.

 \figref{fig:rpm-dn} shows the fan speed variability for each
 downward-facing experiment; these are used by the measurement
 uncertainty calculations.
 The differences reflect improvements in the fan-control software
 written by the author (28 September 2023).

\subsection{Ambient Sensing}
 \figref{fig:ambient} shows the ambient sensor board which was at the
 lower edge of the test chamber in \figref{fig:suspend}.  It measures
 the pressure, relative humidity (RH), and air temperature at the
 wind-tunnel intake.  Wrapped in aluminum tape to minimize radiative
 heat transfer, the LM35 temperature sensor projects into the tunnel.
 To minimize self-heating, the LM35 is powered only while being
 sampled.

\smallskip
\vbox{\settabs 2\columns
\+\hfill\figscale{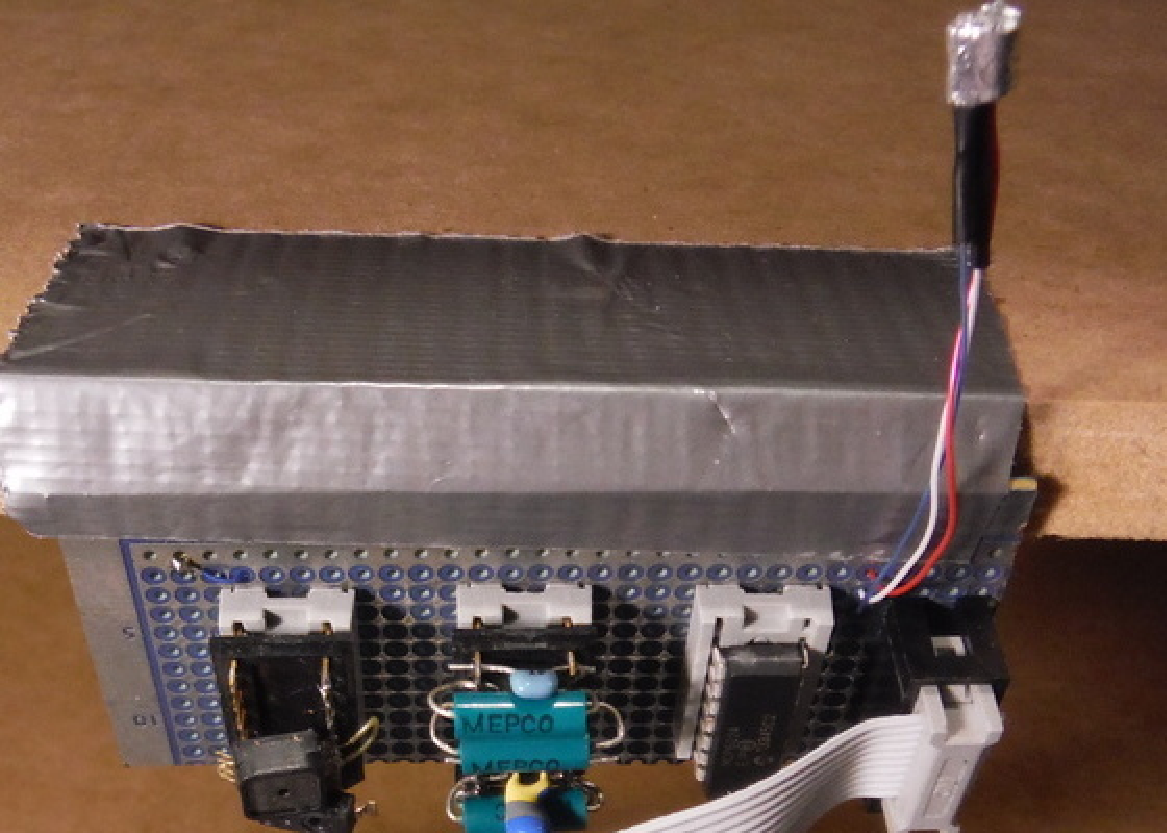}{182pt}\hfill&\hfill\figscale{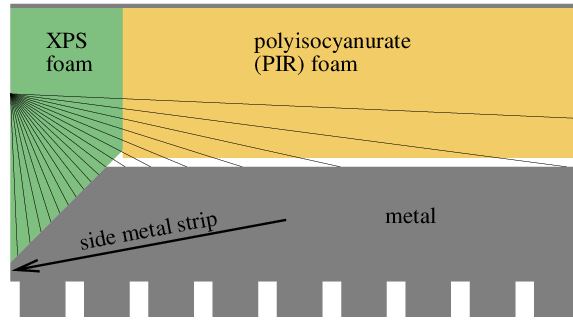}{234pt}&\cr
\+\hfill{\bf\figdef{fig:ambient}\quad Ambient sensors}\hfill&\hfill{\bf\figdef{fig:wedge}\quad XPS wedge conduction}\hfill&\cr
}

\subsection{Physical Parameters}
 \tabref{tab:physical parameters} lists the static parameters
 from measurements and specifications.

\centerline{\bf\tabdef{tab:physical parameters}\quad{Physical parameters}}
\moveright 0.0625\hsize\vbox{\settabs 8\columns
\toprule
\+{\bf Symbol}&{\bf Values}&&{\bf Description}&&&&\cr
\midrule
\+$L$&0.305 m&&length of flow along test-surface&&&&\cr
\+$A$&0.093 $\rm m^2$&&area of test-surface&&&&\cr
\+$\varepsilon$&3.00 mm & 1.04 mm & RMS height-of-roughness&&&&\cr
\+$C_{pt}$&4691 J/K & 4274 J/K & plate thermal capacity&&&&\cr
\+$D_{\rm{Al}}$&19.4 $\rm mm$&&metal slab thickness&&&&\cr
\+$D_{\rm{PIR}}$&25.4 $\rm mm$&&polyisocyanurate (PIR) foam thickness&&&&\cr
\+$D_w$&19.05 $\rm mm$&&XPS foam wedge height&&&&\cr
\+$k_{\rm PIR}$&$0.0222~\rm{W/(m\cdot{K})}$&&PIR foam thermal conductivity&&&&\cr
\+$k_{\rm XPS}$&$0.0285~\rm{W/(m\cdot{K})}$&&XPS foam thermal conductivity&&&&\cr
\+$U_I$&0.075 W/K&&front-to-back insulation thermal conductance&&&&\cr
\+$\epsilon_{\rm{Al}}$&0.04&&test-surface (MIC-6 Al) emissivity&&&&\cr
\+$\epsilon_{\rm{XPS}}$&0.515&&XPS foam emissivity (see text)&&&&\cr
\+$\epsilon_{dt}$&0.89&&duck tape emissivity&&&&\cr
\+$\epsilon_{wt}$&0.90&&test chamber interior emissivity&&&&\cr
\bottomrule
}

 The effective $\epsilon_{wt}$ may differ from the
 medium-density-fiberboard emissivity given by
 Rice~\cite{rice2004emittance} because the temperatures of the test
 chamber surfaces may not be uniform.  Through the open intake, the
 plate also exchanges thermal radiation with objects in the room having
 different temperatures.

\subsection{Modeling of Parasitic Heat Flows}
 At low airflow velocities, the sides of the insulation behind the test
 plate can leak more heat than the test-surface transfers, shrinking
 to 6\% at {1300\rm~r/min}.  To compensate, expected side heat
 transfers will be subtracted from the (combined) measured heat flow.

 \figref{fig:suspend} shows duck tape applied to the lower 54\% of the
 plate's side, which corresponds to 50\% coverage of the XPS foam
 wedge.  For this partial tape coverage, $\epsilon_W$
 Formula~\eqref{eq:emissivity} is the area proportional mean of the
 duck tape emissivity and XPS emissivity.  Barreira, Almeida, and
 Sim\~oes~\cite{s21061961} measured $\epsilon_{dt}$ emissivities
 of 0.86 and 0.89 from two brands of ``duck tape''; the larger value
 is used for the aged tape on the plate sides.  As of this writing,
 published emissivity measurements of XPS foam have not been located.

$$\epsilon_W=50\%\,\epsilon_{dt}+50\%\,\epsilon_{\rm{XPS}}\eqdef{eq:emissivity}$$

 Relative to the theory in Jaffer~\cite{thermo3010010}, natural
 convection measurements ($u=0$) from the plate assembly over the span
 of inclinations have less than 3.0\% RMSRE when calculated with
 $\epsilon_{\rm{XPS}}=0.515$; the RMSRE increases to either side of
 0.515.  This value is consistent with natural convection measurements
 of the plate assembly without tape.

The four sides are not isothermal; a 3.5~mm metal strip (see
cross-section \figref{fig:wedge}) runs the length of the side; and a
$D_w$-tall wedge of extruded polystyrene foam (XPS) insulation fills
the metal slab's $27{\rm~mm}~(=\sqrt{2}\,D_{\rm{Al}})$ $45^\circ$
chamfer.  The local surface conductance $h_W(z)$ at elevation $z$
(from the wedge point) is found by averaging the reciprocal distance
to slab metal with respect to angle~$\theta$:
$$\eqalign{h_W(z)&=\int_0^{\theta_c}{k_{\rm XPS}\over\sqrt{2}\,z\,\theta_c}\,\cos\left(\theta+{\pi\over4}\right)\,\diff{\theta}
 +\int_{\theta_w}^{\theta_W}{k_{\rm PIR}\over z-D_w}\,{\cos\theta\over{\theta_W-\theta_w}}\,\diff{\theta}\cr
 &={k_{\rm XPS}\over\sqrt{2}\,z\,\theta_c}\,\left[\sin\left(\theta_c+{\pi\over4}\right)-\sin{\pi\over4}\right]
  +{k_{\rm PIR}\over z-D_w}\,\left[{\sin\theta_W-\sin\theta_w\over{\theta_W-\theta_w}}\right]\cr
 \theta_c=\arctan&{D_w-z\over D_w}
 \quad\theta_w=\arctan{D_w\over z-D_w}
 \quad\theta_W=\max\left(\theta_w,~{\arctan{L-D_w\over z-D_w}}\right)\cr}\eqdef{eq:h_W}$$

 Forced air flows parallel to the long dimension on two sides, but
 flows into the windward side and away from the leeward side.  Air
 heated by the windward side reduces heat transfer from the
 test-surface; air heated by the test-surface suppresses heat transfer
 from the leeward side.  Hence, the model excludes windward and
 leeward forced convection.
 The total forced convective conductance of the flow-parallel foam
 wedges is calculated by integrating $h_W(z)$ in series (reciprocal of
 the sum of reciprocals, which is also the $\ell^{-1}$-norm) with the
 local surface conductance $k\,\Nus(\Rex)/L$, where $\Nus(\Rex)$ is
 the pierced-laminar convection calculated by applying
 the~$\Nutol\to\Nut$ transform~\eqref{eq:smooth-local-convection} to
 the pierced-laminar convection Formula~\eqref{eq:pierced-convection}:
$$\eqalignno{U_W&=\int_0^{D_w}\int_0^L\left\|{h_W(z), {k\,\Nus(\Rex)\over L}}\right\|_{-1}\diff{x}\,\diff{z}&\eqdef{eq:U_W}\cr}$$

 The natural convection flow from the vertical faces is upward,
 perpendicular to the horizontal forced flow; hence, the forced
 convection $U_W$ and the natural convective conductance
 $k\,L\,L'\,\Nuolq/L'=k\,L\,\Nuolq$ combine as the $\ell^2$-norm
 (introduced in \ref{Combining Transfer Processes}).  The resulting
 mixed convective conductance is in mild competition with the side
 radiative conductance, $U_R=\epsilon_W\,\epsilon_{wt}\,h_R\,L\,D_w$;
 they combine as the $\ell^{\sqrt{2}}$-norm:
$$\eqalign{
 U_S(u)&=2\,\left\|{U_R,~\|{U_W,k\,L\,\Nuolq}\|_2}\right\|_{\sqrt{2}}
        +2\,\|{U_R,~{k\,L\,\Nuolq}}\|_{\sqrt{2}}\cr}\eqdef{eq:U_S}$$

 Each of the four side's natural convective conductance is $\Nuolq$
 vertical plate Formula~\eqref{eq:vertical}
 from Jaffer~\cite{thermo3010010},
 with characteristic length $L'=D_{\rm Al}+\sqrt{2}\,\varepsilon$,
 where the metal slab thickness $D_{\rm Al}\approx19.4{\rm~mm}$.

$$\eqalign{&\Nuolq=\left\|{{\Nuzq}\over2}~,~{\Nuzq^{4/3}\over8\,\root3\of{2}}\left[{\Ra\over\Xi(\Pra)}\right]^{1/3}\right\|_{1/2}\cr
  &\Xi(\Pra)=\left\|1~,~{0.5\over\Pra}\right\|_{\sqrt{1/3}}~\,\quad\Nuzq={8^{5/4}\over\pi^2}}\eqdef{eq:vertical}$$

\subsection{Measurement Methodology}
 The measurement methodology employed is unusual.  Instead of waiting
 until the plate reaches thermal equilibrium, the plate is heated to
 15~K above ambient, heating stops, the fan runs at the designated
 speed, and convection cools the plate.  All of the sensor readings
 are captured each second during the 102 minute
 process, \tabref{tab:dynamic quantities} lists the dynamic physical
 quantities measured each second.  \tabref{tab:computed quantities}
 lists computed quantities.  Both $U_S(u)$ and
 $\{\epsilon_{\rm{Al}}\,\epsilon_{wt}\,h_R\,A\}$ are subtracted from
 the combined heat flow.  The mean of $\hol(u,t)$ over the time
 interval in which~$\Delta{T}$ drops by half (or exceeds 6142~s total
 time) is the result from that experiment.

\medskip
\centerline{\bf\tabdef{tab:dynamic quantities}\quad{Dynamic quantities}}
\moveright 0.25\hsize\vbox{\settabs 8\columns
\toprule
\+{\bf Symbol} &{\bf Units} & {\bf Description}&&\cr
\midrule
\+$\omega$ &r/min & fan rotation rate&&\cr
\+$T_F$ &K & ambient air temperature&&\cr
\+$T_P$ &K & plate temperature&&\cr
\+$T_B$ &K & back surface temperature&&\cr
\+$P$ &Pa & atmospheric pressure&&\cr
\+$\Phi$ &Pa/Pa & air relative humidity&&\cr
\bottomrule
}
\medskip

\centerline{\bf\tabdef{tab:computed quantities}\quad{Computed quantities}}
\moveright 0.1875\hsize\vbox{\settabs 8\columns
\toprule
\+ {\bf Symbol} &{\bf Units} &{\bf Description} &&&\cr
\midrule
\+ $h_R$ &$\rm W/(m^2K)$ &radiative surface conductance &&&\cr
\+ $U_S(u)$ &W/K &side radiative and convective conductance &&&\cr
\+ $\hol(u,t)$ &$\rm W/(m^2K)$ &convective surface conductance &&&\cr
\bottomrule
}

\subsection{Heat Balance}
Collecting into $U_T(u)$ Formula~\eqref{eq:U_T(u)} those terms which have a
factor of temperature difference $\overline{T_P}-\overline{T_F}$,
Formula~\eqref{eq:heat-balance} is the heat balance equation of the
plate during convective cooling:
$$\eqalignno{U_T(u)&=U_S(u)+\{\hol(u)\,A\}+\{\epsilon_{\rm{Al}}\,\epsilon_{wt}\,h_R\,A\}&\eqdef{eq:U_T(u)}\cr
  0&=U_T(u)\,\left[\overline{T_P}-\overline{T_F}\right]+U_I\left[\overline{T_P}-\overline{T_B}\right]+
  C_{pt}\,{\diff{\overline{T_P}}\over \diff{t}}&\eqdef{eq:heat-balance}\cr}$$

The plate and ambient temperatures are functions of time $t$.
Determined experimentally during heating, the temperature group-delay
through the 2.54~cm block of insulation between the slab and back
sheet is~110~s:
$$\overline{T_P}(t)={U_T(u)\,\overline{T_F}(t)+U_I\,\overline{T_B}(t-110{\rm~s})
-C_{pt}\,[\diff{\overline{T_P}(t)}/\diff{t}]
  \over U_T(u)+U_I}\eqdef{eq:delay}$$

To compute Nusselt number $\Nuol=\hol\,L/k$, equation~\eqref{eq:delay} is
solved for the $\{\hol(u,t)\,A\}$ term from equation~\eqref{eq:U_T(u)}.
$$\eqalignno{\zeta(u,t)&=-U_I\,\left[\overline{T_P}(t)-\overline{T_B}(t-110{\rm~s})\right]&\eqdef{eq:eta(u,t)}\cr
  \{\hol(u,t)\,A\}&={\zeta(u,t)-C_{pt}\,\left[\overline{T_P}(t)-\overline{T_P}(t')\right]/[t-t']
         \over \overline{T_P}(t)-\overline{T_F}(t)}
        -\{\epsilon_{\rm{Al}}\,\epsilon_{wt}\,h_R\,A\}-U_S(u)&\eqdef{eq:U_P(t)}}$$
 where $t'$ is the previous value of $t$.  In
 equations~\eqref{eq:eta(u,t)} and \eqref{eq:U_P(t)}, $\overline{T_P}(t)$,
 $\overline{T_F}(t)$, and $\overline{T_B}(t)$ are the 15-element
 cosine averages of plate and fluid temperatures (centered at
 time~$t$).

\subsection{Measurement Uncertainty}
Following Abernethy, Benedict, and Dowdell~\cite{Abernethy1985}, the
final steps in processing an experiment's data are:
\orderedlist

 \li Using equation~\eqref{eq:U_P(t)}, calculate the sensitivities of
 convected power $\hol\,A\,\Delta{T}$ per each parameter's
 average over the measurement time-interval;

 \li multiply the absolute value of each sensitivity by its estimated
 parameter bias to yield component uncertainties;

 \li calculate combined bias uncertainty as the root-sum-squared (RSS)
 of the component uncertainties;

 \li calculate the RSS combined measurement uncertainty as the RSS of
 the combined bias uncertainty and twice the product of the rotation
 rate sensitivity and variability.

\endorderedlist

\vfill\eject

 \tabrefs{tab:20160830T024010}
 and \tabrefn{tab:20230712T011216} list the sensitivity, bias,
 and uncertainty for each component contributing more than 0.20\%
 uncertainty for the 3~mm and 1~mm roughness plates, respectively.
 \figrefs{fig:convect-vs-theory-3} and \figrefn{fig:convect-vs-theory-1} show
 the measurements relative to the present theory for rough flow
 and turbulent flow, respectively.

\medskip
\centerline{\hfil{\bf\tabdef{tab:20160830T024010}\quad{Estimated measurement uncertainties, bi-level 3mm roughness at $\Rey = 59593$.}}}
\vbox{\settabs 7\columns
\toprule
\+\hfil{\bf Symbol}&\hfil{\bf Nominal}&\hfil{\bf Sensitivity}&\hfil{\bf Bias}&\hfil{\bf Uncertainty}&{\bf Component}&&\cr
\midrule
\+\hfil      $\Delta T$&\hfil    9.47K  &\hfil $+12.2$\%/K  &\hfil   0.10K  &\hfil   1.22\%&       LM35C differential&\cr
\+\hfil             $P$&\hfil    101kPa &\hfil $+0.0009$\%/Pa &\hfil   1.5kPa &\hfil   1.28\%& MPXH6115A6U air pressure&\cr
\+\hfil        $C_{pt}$&\hfil   4.69kJ/K&\hfil $+0.024$\%/(J/K)&\hfil     47J/K&\hfil   1.13\%&   plate thermal capacity&\cr
\+\hfil          $\eta$&\hfil   0.401   &\hfil $+180$\%   &\hfil   0.014   &\hfil   2.52\%&   anemometer calibration&\cr
\+\hfil     $\varsigma$&\hfil   6.00mm  &\hfil $+11285$\%/m  &\hfil   100um  &\hfil   1.13\%&              post height&\cr
\+&&&&\hfil    3.49\%&combined bias uncertainty&\cr
\+\hfil{\bf Symbol}&\hfil{\bf Nominal}&\hfil{\bf Sensitivity}&\hfil{\bf Variability}&\hfil{\bf Uncertainty}&{\bf Component}&&\cr
\midrule
\+\hfil        $\omega$&\hfil     905r/min&\hfil $+0.081$\%/(r/min)&\hfil    5.2r/min&\hfil   0.43\%&        fan rotation rate&\cr
\+&&&&\hfil    3.60\%&RSS combined uncertainty&\cr
\bottomrule

}

\medskip
\centerline{\hfil{\bf\tabdef{tab:20230712T011216}\quad{Estimated measurement uncertainties, bi-level 1mm roughness at $\Rey = 55935$.}}}
\vbox{\settabs 7\columns
\toprule
\+\hfil{\bf Symbol}&\hfil{\bf Nominal}&\hfil{\bf Sensitivity}&\hfil{\bf Bias}&\hfil{\bf Uncertainty}&{\bf Component}&&\cr
\midrule
\+\hfil      $\Delta T$&\hfil    10.2K  &\hfil $+11.7$\%/K  &\hfil   0.10K  &\hfil   1.17\%&       LM35C differential&\cr
\+\hfil             $P$&\hfil  100.0kPa &\hfil $+0.0008$\%/Pa &\hfil   1.5kPa &\hfil   1.26\%& MPXH6115A6U air pressure&\cr
\+\hfil        $C_{pt}$&\hfil   4.24kJ/K&\hfil $+0.028$\%/(J/K)&\hfil     42J/K&\hfil   1.17\%&   plate thermal capacity&\cr
\+\hfil          $\eta$&\hfil   0.340   &\hfil $+195$\%   &\hfil   0.003   &\hfil   0.66\%&   anemometer calibration&\cr
\+\hfil           $u_u$&\hfil   6.381   &\hfil $+2.44$\%   &\hfil   0.100   &\hfil   0.24\%&diffuser airflow upper bound&\cr
\+\hfil           $L_T$&\hfil   8.34mm  &\hfil $+9365$\%/m  &\hfil   100um  &\hfil   0.94\%&              post length&\cr
\+\hfil           $L_m$&\hfil   3.57mm  &\hfil $+454$\%/m  &\hfil   500um  &\hfil   0.23\%&   side metal strip width&\cr
\+\hfil $\epsilon_{rs}$&\hfil   0.040   &\hfil $+20.4$\%   &\hfil   0.010   &\hfil   0.20\%&  test-surface emissivity&\cr
\+\hfil $\epsilon_{wt}$&\hfil   0.900   &\hfil $+9.05$\%   &\hfil   0.025   &\hfil   0.23\%&   wind-tunnel emissivity&\cr
\+&&&&\hfil    2.44\%&combined bias uncertainty&\cr
\+\hfil{\bf Symbol}&\hfil{\bf Nominal}&\hfil{\bf Sensitivity}&\hfil{\bf Variability}&\hfil{\bf Uncertainty}&{\bf Component}&&\cr
\midrule
\+\hfil        $\omega$&\hfil   1.03kr/min&\hfil $+0.065$\%/(r/min)&\hfil    2.5r/min&\hfil   0.16\%&        fan rotation rate&\cr
\+&&&&\hfil    2.46\%&RSS combined uncertainty&\cr
\bottomrule

}

\medskip

\vbox{\settabs 2\columns
\+\hfill\figscale{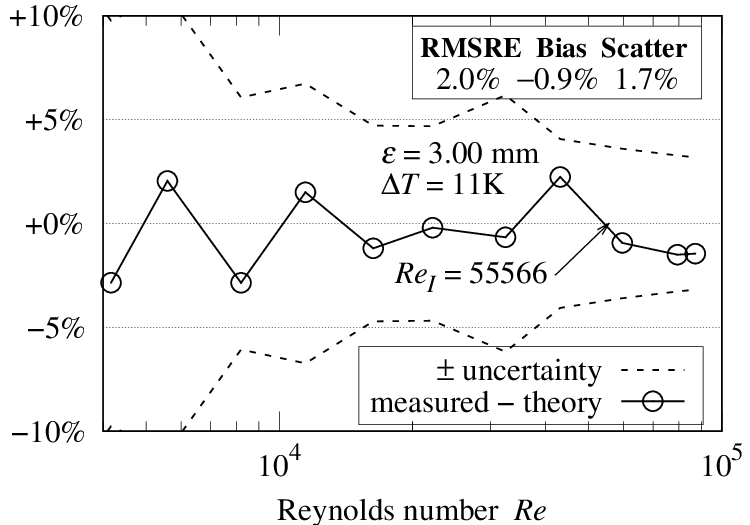}{234pt}\hfill
 &\hfill\figscale{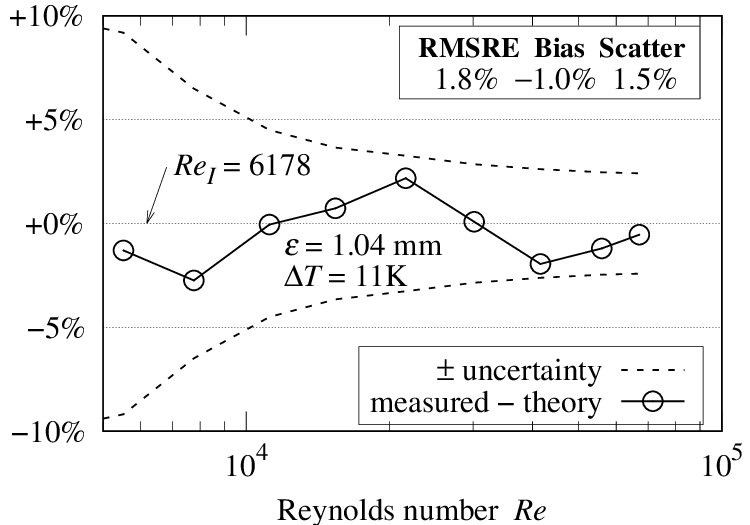}{234pt}\hfill&\cr
\+\hfill{\bf\figdef{fig:convect-vs-theory-3}\quad Measured versus theory $\varepsilon=3$~mm}\hfill
 &\hfill{\bf\figdef{fig:convect-vs-theory-1}\quad Measured versus theory $\varepsilon=1$~mm}\hfill&\cr
}

\subsection{Details}
  Documentation, photographs, electrical schematics, and software
  source-code for the apparatus, as well as calibration and
  measurement data are available from:

  {\tt{http://people.csail.mit.edu/jaffer/convect}}

\vfill\eject

\section{Appendix B: Thermal and Transport Properties of Humid Air}

  Wexler~\cite{Wexler1976} approximates the (partial) pressure of
  saturated water vapor as:
$$\eqalign{P_v=\exp\bigl(&-0.63536311\times10^{4}/T +0.3404926034\times10^{2}\cr
               &-0.19509874\times10^{-1}\,T +0.12811805\times10^{-4}\,T^2\bigr)\cr}$$
  where $T$ is absolute temperature in Kelvins.

  An ideal gas has density~$\rho=P\,M/[\overline{R}\,T]$, where the
  gas constant $\overline{R}=8.314\rm~J/(kg\cdot mol)$.  Air with
  relative humidity~$0<\phi<1$, modeled as a mixture of dry air and
  water vapor, has density:
$$\rho={M_a\,[P-\phi\,P_v]+M_v\,\phi\,P_v\over\overline{R}\,T}$$
  where the molar masses of air and water are
  $M_a=28.97\times10^{-3}$~kg/mol and
  $M_v=18.0153\times10^{-3}$~kg/mol.

  Tsilingiris~\cite{TSILINGIRIS20081098}, approximates the specific
  heats of dry air and water vapor:
$$\eqalign{c_{pa}=&+1034-0.2849\,T+0.7817\times10^{-3}\,T^2\cr
                  &-0.4971\times10^{-6}\,T^3+0.1077\times10^{-9}\,T^4\cr
           c_{pv}=&+1869-0.2578\,[T-273.15]+1.941\times10^{-2}\,[T-273.15]^2\cr
           c_p=&{c_{pa}\,[1-\chi_p]\,M_a+c_{pv}\,\chi_p\,M_v\over[1-\chi_p]\,M_a+\chi_p\,M_v}\qquad
 \chi_p=\phi\,P_v/P\cr}$$

  Morvay and Gvozdenac~\cite{Toolbox6}, approximate the viscosity of
  air and water vapor as:
$$\eqalign{\mu_a=&+0.40401\times10^{-6}+0.074582\times10^{-6}\,T-5.7171\times10^{-11}\,T^2\cr
                 &+2.9928\times10^{-14}\,T^3-6.2524\times10^{-18}\,T^4\cr
           \mu_v=&{10^{-6}\,\gamma^{-1/2}\over0.0181583+\gamma\,[0.0177624+\gamma\,[0.0105287-\gamma\,0.0036744]]}}$$
  where $\gamma=647.27/T$.
  They combine these into dynamic viscosity $\mu$ using absolute
  humidity $\chi$, the mass ratio of water vapor to air.
$$\chi={M_v\,\phi\,P_v\over M_a\,[P-\phi\,P_v]}\qquad r_m=M_a/M_v\qquad\chi_m=\chi\,r_m$$
$$\Phi(r_m,r_\mu)=\left[1+{\sqrt{r_\mu}\over\root4\of{r_m}}\right]^2\sqrt{1\over8\,[1+r_m]}$$
$$\mu={\mu_a\over1+\Phi(r_m,\mu_a/\mu_v)\,\chi_m}+{\mu_v\over1+\Phi(1/r_m,\mu_v/\mu_a)/\chi_m}$$

  Morvay and Gvozdenac~\cite{Toolbox6} approximates the thermal
  conductivity of water vapor at $t=T-273.15$:
$$\eqalign{k_v=&+1.74822\times10^{-2}+7.69127\times10^{-5}\,{t}-3.23464\times10^{-7}\,{t}^2\cr
               &+2.59524\times10^{-9}\,{t}^3-3.17650\times10^{-12}\,{t}^4\cr}$$

  Kadoya, Matsunaga and Nagashima~\cite{doi:10.1063/1.555744}
  approximates the thermal conductivity of dry air as:
$$\eqalign{k_a=0.0259778\,\bigl(&+0.239503\,T_r+0.00649768\,\sqrt{T_r}+1.0-1.92615/T_r\cr
        &+2.00383/T_r^2-1.07553/T_r^3+0.229414/T_r^4\cr
        &+0.402287\,\rho_r+0.356603\,\rho_r^2-0.163159\,\rho_r^3\cr
        &+0.138059\,\rho_r^4-0.0201725\,\rho_r^5\bigr)\cr}$$
  where
$$\rho_r={P/314.3\over287.058\,T}\qquad T_r={T\over132.5}$$

  Both Morvay and Gvozdenac~\cite{Toolbox6} and
  Tsilingiris~\cite{TSILINGIRIS20081098} develop the combined
  thermal conductivity $k$ as:
$$k={k_a\over1+\Phi(r_m,r_\mu)\,\chi_m}+{k_v\over1+\Phi(1/r_m,1/r_\mu)/\chi_m}\qquad
     r_\mu=\mu_a/\mu_v$$

\vfill\eject
\bye